\documentclass[11pt,sort&compress]{elsarticle}

\makeatletter
  \long\def\pprintMaketitle{\clearpage
  \iflongmktitle\if@twocolumn\let\columnwidth=\textwidth\fi\fi
  \resetTitleCounters
  \def\baselinestretch{1}%
  \printFirstPageNotes
  \begin{center}%
 \thispagestyle{pprintTitle}%
   \def\baselinestretch{1}%
    {\large\bf\@title}\par\vskip5pt
    \normalsize\elsauthors\par\vskip5pt
    \footnotesize\itshape\elsaddress\par\vskip10pt
    \end{center}%
  \gdef\thefootnote{\arabic{footnote}}%
  }
\makeatother

\newcommand\blfootnote[1]{%
  \begingroup
  \renewcommand\thefootnote{}\footnote{#1}%
  \addtocounter{footnote}{-1}%
  \endgroup
}

\usepackage{amsmath}
\usepackage{amsfonts}
\usepackage{graphicx}
\usepackage{xspace}
\usepackage{epsf}
\usepackage{setspace}
\usepackage{listings}
\usepackage{float}
\usepackage{abstract}
\usepackage{outlines}
\usepackage{tikz}
\usepackage{algpseudocode}
\usepackage{algorithm}
\usepackage{longtable}
\usepackage{booktabs}
\usepackage{subcaption}

\usepackage[]{siunitx}
\usepackage[]{xcolor}

\usepackage{bm}
\newcommand{\ve}[1]{\bm{#1}}

\newcommand{\bb}{\ve{b}}
\newcommand{\bu}{\ve{u}}

\newcommand{\bn}{\ve{n}}

\newcommand{\bq}{\ve{q}}

\newcommand{\bx}{\ve{x}}
\newcommand{\be}{\ve{e}}
\newcommand{\bg}{\ve{g}}
\newcommand{\bh}{\ve{h}}
\newcommand{\bs}{\ve{s}}
\newcommand{\bl}{\ve{l}}

\newcommand{\bD}{\ve{D}}

\newcommand{\bS}{\ve{S}}

\newcommand{\bF}{\ve{F}}
\newcommand{\bI}{\ve{I}}

\newcommand{\bT}{\ve{T}}
\newcommand{\bOmega}{\ve{\Omega}}

\newcommand{\cO}{\mathcal{O}}
\newcommand{\cT}{\mathcal{T}}
\newcommand{\bcL}{\boldsymbol{\mathcal{L}}}

\newcommand{\dd}{\text{d}}
\newcommand{\ddd}{\dd R \dd \dot{R}}

\newcommand{\Rdot}{\dot{R}}
\newcommand{\Rddot}{\ddot{R}}

\newcommand{\bxi}{\boldsymbol{\xi}}

\newcommand{\hw}{\widehat{w}}

\newcommand{\hR}{\widehat{R}}
\newcommand{\hRdot}{\widehat{\dot{R}}}

\newcommand{\mom}{\mu}
\newcommand{\bmom}{\boldsymbol{\mu}}
\newcommand{\vbmom}{\vec{\bmom}}

\newcommand\Rey{\mbox{\text{Re}}\xspace}
\newcommand\Ma{\mbox{\text{Ma}}\xspace}

\newcommand{\cL}{\mathcal{L}}

\newcommand{\bzero}{\boldsymbol{0}}
\newcommand{\APC}{\mathrm{APC}}

\newcommand{\tr}{\operatorname{tr}}
\newcommand{\diag}{\operatorname{diag}}

\definecolor{lightblue}{rgb}{0.63, 0.74, 0.78}
\definecolor{seagreen}{rgb}{0.18, 0.42, 0.41}
\definecolor{orange}{rgb}{0.85, 0.55, 0.13}
\definecolor{silver}{rgb}{0.69, 0.67, 0.66}
\definecolor{rust}{rgb}{0.72, 0.26, 0.06}
\definecolor{purp}{RGB}{68, 14, 156}

\colorlet{lightrust}{rust!50!white}
\colorlet{lightorange}{orange!25!white}
\colorlet{lightlightblue}{lightblue}
\colorlet{lightsilver}{silver!30!white}
\colorlet{darkorange}{orange!75!black}
\colorlet{darksilver}{silver!65!black}
\colorlet{darklightblue}{lightblue!65!black}
\colorlet{darkrust}{rust!85!black}
\colorlet{darkseagreen}{seagreen!85!black}

\usepackage{microtype}

\usepackage[margin=1in]{geometry}

\usepackage{stmaryrd}
\SetSymbolFont{stmry}{bold}{U}{stmry}{m}{n}

\usepackage{hyperref}
\hypersetup{
  colorlinks=true,
}

\usepackage[labelfont=bf,font=footnotesize]{caption}

\usepackage[nameinlink]{cleveref}
\crefname{lstlisting}{listing}{listings}
\Crefname{lstlisting}{Listing}{Listings}
\crefname{equation}{}{}

\usepackage[parfill]{parskip}

\usepackage{xparse}
\DeclareDocumentCommand{\diff}{O{} m}{
	\frac{\mathrm{d} #1}{\mathrm{d}#2}
}
\DeclareDocumentCommand{\difftwo}{O{} m}{
	\frac{\mathrm{d}^2 #1}{\mathrm{d}#2^2}
}
\DeclareDocumentCommand{\pdiff}{O{} m}{
	\frac{\partial #1}{\partial #2}
}
\DeclareDocumentCommand{\pdifftwo}{O{} m}{
	\frac{\partial^{2} #1}{\partial #2^{2}}
}
\DeclareDocumentCommand{\integral}{O{} O{} m O{x}}{
	\int_{#1}^{#2} #3\ \mathrm{d}#4
}

\usepackage{multirow}
\usepackage{pifont}
\newcommand{\cmark}{\color{seagreen}{\ding{52}}}
\newcommand{\xmark}{\color{rust}{\ding{55}}}
\usepackage{tabularx}

\journal{}

\begin{document}

\hypersetup{
  linkcolor=darkrust,
  citecolor=seagreen,
  urlcolor=darkrust,
  pdfauthor=author,
}

\begin{frontmatter}

\title{{\large\bfseries MFC 5.0: An exascale many-physics flow solver}}

\author[gtcse]{Benjamin~Wilfong\corref{cor1}}
\author[gtcse]{Henry~A.~Le~Berre\corref{cor1}}
\author[gtcse]{Anand~Radhakrishnan\corref{cor1}}
\author[gtcse]{Ansh~Gupta}
\author[gtcse]{Daniel~J.~Vickers}
\author[wpi]{Diego~Vaca-Revelo}
\author[gtcse]{Dimitrios~Adam}
\author[gtae]{Haocheng~Yu}
\author[ct]{Hyeoksu~Lee}
\author[ct]{Jose~Rodolfo~Chreim}
\author[brown]{Mirelys~Carcana~Barbosa}
\author[ct]{Yanjun~Zhang}
\author[illinoisme]{Esteban~Cisneros-Garibay}
\author[wpi]{Aswin~Gnanaskandan}
\author[brown]{Mauro~Rodriguez~Jr.}
\author[ornl]{Reuben~D.~Budiardja}
\author[hpe]{Stephen~Abbott}
\author[ct]{Tim~Colonius}
\author[gtcse,gtae,gtme]{Spencer~H.~Bryngelson}
\ead{shb@gatech.edu}

\cortext[cor1]{Equal contribution}

\address[gtcse]{School of Computational Science \& Engineering, Georgia Institute of Technology, Atlanta, GA 30332, USA\vspace{-0.15cm}}
\address[wpi]{Mechanical and Materials Engineering, Worcester Polytechnic Institute, Worcester, MA 01609, USA\vspace{-0.15cm}}
\address[gtae]{Daniel Guggenheim School of Aerospace Engineering, Georgia Institute of Technology, Atlanta, GA 30332, USA\vspace{-0.15cm}}
\address[ct]{Department of Mechanical and Civil Engineering, California Institute of Technology, Pasadena, CA 91125, USA\vspace{-0.15cm}}
\address[brown]{School of Engineering, Brown University, Providence, RI 02912, USA\vspace{-0.15cm}}
\address[illinoisme]{Mechanical Science \& Engineering, University of Illinois at Urbana--Champaign, Urbana, IL 61820, USA\vspace{-0.15cm}}
\address[ornl]{Oak Ridge National Laboratory, Oak Ridge, TN 37830, USA\vspace{-0.15cm}}
\address[hpe]{Hewlett Packard Enterprise, Bloomington, MN 55435, USA\vspace{-0.15cm}}
\address[gtme]{George~W.~Woodruff School of Mechanical Engineering, Georgia Institute of Technology, Atlanta, GA 30332, USA}

\date{}

\end{frontmatter}

\blfootnote{
\noindent Code available at: \url{https://github.com/MFlowCode/MFC}
}

\begin{abstract}
    Many problems of interest in engineering, medicine, and the fundamental sciences rely on high-fidelity flow simulation, making performant computational fluid dynamics solvers a mainstay of the open-source software community.
    Previous work
    MFC~3.0 was made a published, documented, and open-source solver via Bryngelson et al. \textit{Comp.\ Phys.\ Comm.} (2021) with numerous physical features, numerical methods, and scalable infrastructure.
    MFC~5.0 is a significant update to MFC~3.0, featuring a broad set of well-established and novel physical models and numerical methods, as well as the introduction of GPU and APU (or superchip) acceleration.
    We exhibit state-of-the-art performance and ideal scaling on the first two exascale supercomputers, OLCF~Frontier and LLNL El~Capitan.
    Combined with MFC's single-accelerator performance, MFC achieves exascale computation in practice, and achieved the largest-to-date public CFD simulation at 200 trillion grid points as a 2025 ACM Gordon Bell Prize finalist.
    New physical features include the immersed boundary method, $N$-fluid phase change, Euler--Euler and Euler--Lagrange sub-grid bubble models, fluid-structure interaction, hypo- and hyper-elastic materials, chemically reacting flow, two-material surface tension, magnetohydrodynamics (MHD), and more.
    Numerical techniques now represent the current state-of-the-art, including general relaxation characteristic boundary conditions, WENO variants, Strang splitting for stiff sub-grid flow features, and low Mach number treatments.
    Weak scaling to tens of thousands of GPUs on OLCF~Summit and Frontier and LLNL El~Capitan achieves efficiencies within 5\% of ideal to over 90\% of their respective system sizes.
    Strong scaling results for a 16-times increase in device count show parallel efficiencies over 90\% on OLCF~Frontier.
    MFC's software stack has undergone further improvements, including continuous integration, which ensures code resilience and correctness through over 300 regression tests; metaprogramming, which reduces code length while maintaining performance portability; and code generation for computing chemical reactions.
\end{abstract}

{
\textit{Program title}:  MFC 5.0 \\
\textit{Developer's repository link}:   \url{https://github.com/MFlowCode/MFC} \\
\textit{Licensing provisions}:   MIT license (MIT) \\
\textit{Programming language}:   Fortran08 and Python \\
\textit{Journal Reference of previous version}:  doi: 10.1016/j.cpc.2020.107396 \\
\textit{Does the new version supersede the previous version?}:  Yes \\
\textit{Reasons for the new version}:  An abundance of new features and capabilities that are challenging to convey outside of an academic article. \\
\textit{Nature of problem}:  Simulation of compressible flows requires careful physical model selection and treatment of spatial derivatives to keep solutions physically consistent and free of numerical artifacts. The associated methods should be high-order accurate to reduce computational cost. \\
\textit{Solution method}:  The present software uses multiple physical models and numerical schemes for the treatment of multi-phase, multi-component, chemically reacting, and magnetohydrodynamic flows.
Additional physical effects and models are included.
The numerical method is based on the finite volume method.
}

\section{Introduction}
MFC was introduced to the literature and open source community via \citet{bryngelson21}, which described the software design and features of MFC~3.0, a CPU-based compressible multi-component flow code (MFC) for the 5-equation diffuse interface methods of \citet{kapila2001} and \citet{allaire2002} and the six equation diffuse interface model of \citet{saurel2009}.
MFC~3.0 included other features that are discussed throughout this manuscript.
Since MFC~3.0's release, the MFC community has added a broad set of physical models, numerical methods, GPU offloading capabilities, and performance optimizations.
This paper provides a holistic overview of MFC's development over the past four years. 
We present physical models, numerical methods, validity, software robustness and testing, and performance on diverse architecture sets at scale on exascale machines and beyond.

\subsection{The history of MFC}

The MFC codebase archaeology dates back to the mid-2000s.
At this time, MFC was an unnamed solver developed by Eric~Johnsen.
This code represented a multi-component flow using the diffuse interface method, as described in \citet{johnson2008} and \citet{johnsen2006implementation}.
Following this effort, Vedran~Coralic performed a major code rewrite and adorned MFC with its name.
These features adorned MFC with a multi-dimensional numerical reconstruction treatment and are described in \citet{Coralic2014}.
Jomela~Meng added cylindrical coordinate systems and corresponding simulations of high-speed shock--droplet interaction~\citep{meng2016}.
Kazuki~Maeda added an Euler--Lagrange representation of sub-grid bubble dynamics~\citep{maeda2018eulerian} and acoustic wave generation~\citep{Maeda2017}.
Kevin~Schmidmayer and Spencer~Bryngelson added the 5-equation model of \citet{kapila2001} and the 6-equation model of \citet{saurel2009}.
These efforts were conducted in the research group of Tim~Colonius at the California~Institute~of~Technology.
Spencer~Bryngelson later led a restructuring of MFC at the Georgia Institute of Technology, which included additional modeling and numerical features. 
This effort culminated in the open-sourcing of MFC~3.0 in 2020 and is largely described in \citet{bryngelson21}.
Modeling, method, and software scalability and portability capabilities were added in subsequent years, which are discussed in this manuscript.

\subsection{New MFC capabilities}

To make high-fidelity CFD a practical instrument for today's many-physics problems, a solver must span liquids, gases, and solids (at least) and accommodate complex geometries and broad separations of time and length scales.
At the same time, this should be accomplished efficiently on modern accelerators, which serve as the backbone of current supercomputing centers.
MFC~5.0 is architected around that premise.
The new capabilities unify established and novel models.
Each of these capabilities is accompanied by state-of-the-art numerics and GPU/APU offloading, allowing a single code to transition from low-Mach, geometry-rich flows to shock- and detonation-dominated regimes at exascale.
In this sense, the new capabilities were crafted to establish an infrastructure for credible, end-to-end simulations across engineering, medicine, and the sciences.

MFC now includes more capable physical models and numerical methods, software resistance through continuous integration and deployment, code coverage testing, and computation acceleration via GPU and APU devices.
With these, MFC conducted the largest publicly known CFD simulation at 200T grid points, or 1 quadrillion degrees of freedom~\citep{wilfongGB25}, and was an ACM 2025 Gordon~Bell Prize Finalist.
Physical features include six new models for phase change, non-polytropic sub-grid bubble dynamics, treatment for elastic materials, chemical reactions and combustion, and surface tension.
An immersed boundary method that supports complex geometries from analytical level sets or STL files.
New numerical methods include generalized characteristic boundary conditions, improved shock capturing with TENO and WENO-Z constructions, Strang splitting for stiff sub-grid dynamics, and treatment for low-Mach number flow.
Significant software additions have made MFC a portable and performant solver on CPUs, GPUs, and APUs from various vendors, including Intel, AMD, and NVIDIA.
Decreases in runtime are, in part, enabled via metaprogramming and static code generation.


\subsection{Use of MFC}

MFC has a history of being used for early access programs on flagship supercomputers.
At the time of writing, these programs include JSC~JUPITER (JUREAP) and the LLNL El~Capitan and OLCF~Frontier final and early access systems, including LLNL~Tioga (AMD~MI300A testbed) and OLCF Spock and Crusher (AMD~MI100 and MI250X testbeds).
Commodity cluster use of MFC is broad, including the ACCESS-CI~\citep{access-ci} (formerly XSEDE~\citep{Towns2014-po}) clusters (PSC Bridges~\citep{PSCBridges} and Bridges2~\citep{Bridges2}, SDSC Comet~\citep{Comet2015} and Expanse~\citep{Expanse2021}, Purdue Anvil~\citep{Anvil2022}, NCSA Delta~\citep{Delta} and DeltaAI~\citep{DeltaAI}, TACC Stampede1--3~\citep{Stampede1,Stampede2,Stampede3}, among others) university clusters, cloud computer systems (Intel's AI cloud~\citep{IntelAICloud}), and AMD and Cray's internal systems, among numerous others.
MFC is a SPEChpc benchmark candidate, which comprises software that evaluates the performance of supercomputers~\citep{spechpc}; it is also used as part of the OLCF Test Harness for the performance and robustness of OLCF~Frontier~\citep{wilfong-hpctests}.

\subsection{Manuscript structure}

In \cref{sec:baseCapabilities}, we introduce MFC~5.0, describing the code and methods it builds upon while connecting it to MFC~3.0.
\Cref{sec:underlyingMethodology} discusses established numerical methods that MFC builds upon.
New features follow in \cref{sec:updatedCapabilities}, including physical models, numerical methods, and software tooling.
We show example MFC simulations in \cref{sec:examples}.
A discussion of other open source flow solvers with some similar features in \cref{sec:otherSolvers}.
\Cref{sec:Conclusion} summarizes the manuscript and MFC~5.0.

\section{Multiphase models}
\label{sec:baseCapabilities}

\subsection{Multi-component treatment}

MFC uses reduced versions of the Baer--Nunziato model~\cite{ANDRIANOV2004} and diffuse interface methods to model multiphase flow.
The core of the methodology, as described in this subsection, remains unchanged from MFC~3.0~\citep{bryngelson21} and readers are directed there for a more fundamental understanding and detailed algorithmic strategy.
Changes to the fundamental diffuse interface methodology are limited, except for performance improvements, code testing strategies, and additional physical features that are now accommodated.
These additional features are discussed in \cref{sec:updatedCapabilities}.

\subsubsection{Five-equation models}\label{sec:5eqn}

For a two-fluid flow configuration, the Baer--Nunziato model reduces to the so-called five-equation model
\begin{align*}\label{e:five_eqn_transport}
    \pdiff[\alpha_i\rho_i]{t} + \nabla \cdot (\alpha_i\rho_i\bu) &= 0, \\
    \pdiff[\rho \bu]{t} + \nabla \cdot (\rho\bu \otimes \bu + p\bI - \bT^v) &= 0, \\
    \pdiff[\rho E]{t} + \nabla \cdot \left[ (\rho E + p)\bu - \bT^v \cdot \bu \right] &= 0, \\
    \pdiff[\alpha_i]{t} + \bu \cdot \nabla \alpha_i &= K \nabla \cdot \bu,
\end{align*}
of \citet{kapila2001} by assuming that both species share a velocity and pressure, and $i = 1,\dots,N_\mathrm{comp.}$ where $N_\mathrm{comp.}$ is the number of fluid components.
With this model, $\rho$, $\bu$, $p$, and $E$ are the mixture density, velocity, pressure, and total energy (per unit mass).
The $\alpha_i$ and $\rho_i$ are the volume fraction and density of component $i$.
Viscous terms are introduced via the mixture viscous stress tensor
\begin{gather}
    \bT^v = 2\eta (\bD - \tr(\bD)\bI/3),
\end{gather}
where $\eta$ is the mixture shear viscosity, and the strain rate tensor is
\begin{gather}
    \bD = (\nabla \bu + (\nabla \bu)^\top)/2.
\end{gather}
The mixture internal energy $e$ is
\begin{gather}
    e = Y_1 e(\rho_1 p) + Y_2 e(\rho_2 p),
\end{gather}
where $Y_i = \alpha_i\rho_i/\rho$ is the mass fraction of component $i$.
The model is closed using the mixture rules 
\begin{gather}
    1 = \sum_{i = 1}^{N}\alpha_i, 
    \quad 
    \rho = \sum_{i = 1}^N \alpha_i\rho_i, \quad \rho e = \sum_{i = 1}^N \alpha_i \rho_i e_i.
\end{gather}
For two components, $K$ is
\begin{gather}
    K = \frac{\rho_2c_2^2 + \rho_1c_1^2}{\rho_1c_1^2/\alpha_1 + \rho_2c_2^2/\alpha_2},
\end{gather}
where $c_i$ is phasic sound speed
\begin{gather}
    c_i = \sqrt{\gamma_i(p + \pi_{\infty,i})/\rho_i}.
\end{gather}
The term $K\nabla \cdot \bu$ ensures thermodynamic consistency and accounts for the expansion and compression of each species in mixture regions.
The $K\nabla \cdot \bu$ is necessary to model phenomena like spherical bubble collapse, but it is non-conservative and can lead to numerical instabilities~\cite{schmidmayer2020}.
For some problems, $K\nabla \cdot \bu$ is not necessary, and the model reduces to that of  \citet{allaire2002}.
The differences between these models are discussed in \citet{schmidmayer2020}.

\subsubsection{Six-equation model}\label{sec:sixEqn}

The six-equation model of \citet{saurel2009} allows for pressure disequilibrium between phases and can be used to avoid numerical stability issues associated with the five-equation Kapila model while maintaining thermodynamic consistency.
For two fluids, the six-equation model is
\begin{align*}
    \pdiff[\alpha_i\rho_i]{t} + \nabla \cdot (\alpha_i\rho_i\bu) &= 0, \\
    \pdiff[\rho \bu]{t} + \nabla \cdot (\rho \bu \otimes \bu + p\bI - \bT^v) &= 0. \\
    \pdiff[\alpha_1\rho_1e_1]{t} + \nabla \cdot (\alpha_1 \rho_1 e_1 \bu) + \alpha_1 p_1 \cdot \nabla \bu &= -\mu p_\mathrm{I}(p_2 - p_1) - \alpha_1\bT_1^v:\nabla \bu, \\
    \pdiff[\alpha_2\rho_2e_2]{t} + \nabla \cdot (\alpha_1 \rho_2 e_2 \bu) + \alpha_2 p_2 \cdot \nabla \bu &= -\mu p_\mathrm{I}(p_1 - p_2) -\alpha_2\bT_2^v:\nabla \bu , \\
    \pdiff[\alpha_1]{t} + \bu \cdot \nabla \alpha_1 &= \mu(p_1 - p_2).
\end{align*}
Here, $\rho$ and $\bu$ are the mixture density and velocity, $\alpha_i$, $\rho_i$, $p_i$, $e_i$, and $\bT_i^v$ are the volume fraction, density, pressure, internal energy, and species viscous stress tensor of species $i$, and $\mu$ is a pressure relaxation parameter.
The interfacial pressure $P_\mathrm{I}$ is
\begin{gather}
    P_\mathrm{I} = \frac{z_2 p_1 + z_1 p_2}{z_1 + z_2},
\end{gather}
where $z_i = \rho_i c_i$ is the phasic acoustic impedance, and the phasic speed of sound $c_i$ is
\begin{gather}
    c_i = \sqrt{ \gamma(p_i + \pi_{\infty,i}) / \rho_i }.
\end{gather}
The model is closed using the same mixture rules of the five-equation model in \cref{sec:5eqn}.
Following \citet{schmidmayer2020}, the six-equation model in the limit of infinite relaxation can represent the same physical processes as the model of \citet{kapila2001}, but avoids numerical instability issues that arise from some interface capture schemes.

\subsection{Equation of state}

The five- and six-equation models are closed using the stiffened gas equation of state (EOS), which accurately models liquids and gases~\cite{menikoff1989}.
In its simplest form, the stiffened gas EOS relates the internal energy to the pressure and density of each species $i$ as
\begin{equation*}
    e_i = \frac{p_i + \gamma_i \pi_{\infty,i}}{(\gamma_i - 1)\rho_i},
    \label{e:eos}
\end{equation*}
where $e_i$, $p_i$, and $\rho_i$ are the internal energy, pressure, and density of species $i$.
Parameter $\gamma_i$ and the liquid stiffness $\pi_{\infty,i}$ can be tuned to represent different liquids and gases.

\subsection{Cylindrical coordinates}

Axisymmetric and cylindrical coordinates are implemented via the strategies of \citet{johnson2008} and \citet{meng2016}.
The singularity at $r \to 0$ is handled by placing the axis at a cell boundary such that the radius is defined at the cell center.
Then, the solution in cells 180 degrees apart from each other supports stencils for reconstruction.
Spectral filtering in the azimuthal direction relaxes the strict time-step restrictions induced by the small cells near the axis~\citep{mohseni2000}.

\section{Numerical methods}
\label{sec:underlyingMethodology}

We next describe the method used to solve for the advective and diffusive terms in the multi-component models of MFC.
The strategy employed here aligns with that of \citet{bryngelson21}, and the reader is directed to that work for further details.

\subsection{Finite volume method}

The models in MFC are solved using a finite volume method based on the framework of \citet{Coralic2014}.
The model equations take the form
\begin{equation}
    \pdiff[\bq]{t} + \pdiff[\bF^{(x)}(\bq)]{x}
        + \pdiff[\bF^{(y)}(\bq)]{y} +
        \pdiff[\bF^{(z)}(\bq)]{z} =
        \bm{s}(\bq) - \bm{h}(\bq) \nabla \cdot \bu,
    \label{eqn:discretization}
\end{equation}
where $\bq$ is the vector of cell-averaged conservative variables, $\bs$ and $\bh(\bq)$ are source terms, and $\bF^x,\ \bF^y,$ and $\bF^z$ are the fluxes in the $x$-, $y$- and $z$-directions.
\Cref{eqn:discretization} is integrated in space as
\begin{equation}
\begin{aligned}
    \pdiff[\bq_{i,j,k}]{t} = 
    \frac{1}{\Delta x_i}&\left(\bF^{(x)}_{i - 1/2,j,k} - \bF^{(x)}_{i + 1/2,j,k}\right)
    + \frac{1}{\Delta y_j}\left(\bF^{(y)}_{i, j- 1/2,k} - \bF^{(y)}_{i,j+1/2,k}\right) \\
    & + \frac{1}{\Delta z_k}\left(\bF^{(z)}_{i,j,k- 1/2} - \bF^{(z)}_{i,j,k+ 1/2}\right)
    + \bm{s}(q_{i,j,k}) - \bm{h}(\bq_{i,j,k})\left(\nabla\cdot \bu\right)_{i,j,k}.
\end{aligned}\label{eqn:rhs}
\end{equation}
The flux $\bF^x(\bq_{i + 1/2, j, k})$ is calculated at the center of the finite volume face, and the other coordinate directions follow in the same fashion.

\subsection{Shock and interface capturing}

\subsubsection{WENO Reconstructions}

The fluxes in finite-volume methods are calculated using the left and right states of each interface by solving the Riemann problem
\begin{align*}
    \bF^{(x)}_{i + 1/2, j, k} &= \bF^{(x)}\left(\bq^{\mathrm{(L)}}_{i + 1/2, j, k}, \bq^{\mathrm{(R)}}_{i + 1/2, j, k}\right), \\
    \bu^{(x)}_{i + 1/2, j, k} &= \bu^{(x)}\left(\bq^{\mathrm{(L)}}_{i + 1/2, j, k}, \bq^{\mathrm{(R)}}_{i + 1/2, j, k}\right),
\end{align*}
where superscripts $\mathrm{(L)}$ and $\mathrm{(R)}$ indicate the state variables reconstructed from the left- and right-hand sides of the finite volume interface.
WENO reconstructions obtain high-order spatial accuracy by considering a convex combination of lower-order reconstructions.
A $(2k-1)$-th order WENO reconstruction in the $x$-direction uses $k$ staggered candidate polynomials to reconstruct the state variable $f_{i + 1/2, j, k}$ with $f^\mathrm{(r)}_{i + 1/2, j, k}$ for $r = 0,1,\dots,k-1$.
The weighted sum of a set of crafted candidate polynomials gives the reconstructed state variables.
Fifth-order accuracy is maintained through mapped weights, as described by \citet{hendrick2005}.

\subsubsection{Approximate Riemann solver}

The Riemann problem at each finite volume interface is solved using either the Harten--Lax--van~Leer (HLL) or the Harten--Lax-van~Leer contact (HLLC) approximate Riemann solver~\cite{Toro1997}.
The HLL approximate Riemann solver admits three constant states separated by two discontinuities with different wave speeds.
The wave speeds are estimated using the left and right state variables $\bq^{\mathrm{(L)}}_{i + 1/2, j, k}$ and $\bq^{\mathrm{(R)}}_{i + 1/2, j, k}$.
The HLL flux satisfies the Rankine--Hugoniot conditions that ensure the conservation of mass, momentum, and energy across the discontinuity.
A shortcoming of the HLL approximate Riemann solver is that it does not account for the contact discontinuity in the region between the two wave speeds (the so-called ``star'' region).
As a result, the HLL approximate Riemann solver has more numerical dissipation than the HLLC approximate Riemann solver, which accounts for this third contact discontinuity.
Continuity of normal velocity and pressure is imposed across the contact discontinuity, which enables computations of the star states and the resulting HLLC flux.
Once computed, the HLL and HLLC fluxes are used to calculate the right-hand side of \cref{eqn:rhs}.
Identical calculations are performed in the $x$- and $z$-directions.

\subsection{Boundary conditions}

Extrapolation and Characteristic~\cite{thompson1990} boundary conditions are implemented using buffer regions outside the domain to support the stencils of the finite volume reconstructions and provide information outside of the domain for characteristic boundary conditions.
Subsonic and supersonic free-stream and wall boundary conditions are implemented to handle various simulation configurations.
Simple boundary conditions remain a mainstay of MFC, which were established in \citet{bryngelson21}.
These boundary conditions include symmetry, slip and no-slip walls, prescription of inlet conditions along partial boundary regions, and first-order accurate extrapolation for Neumann-like boundary conditions.

\subsection{Time integration}

The conservative variables are advanced in time using high-order total variation diminishing (TVD) and strong stability preserving (SSP) explicit Runge--Kutta time steppers~\cite{Gottlieb1998}.
First- and second-order accurate time-steppers Runge--Kutta are also available.

\subsection{Convergence}

The convergence of the existing numerical methods with the five- and six-equation models is verified using a one-dimensional two-component advection problem with periodic boundary conditions.
The computational domain spans the interval $[0, 1]$ with a uniform pressure and velocity equal to one.
The volume fraction and density of each component are defined as $\alpha_i = \alpha_i\rho_i = 0.5 + \left(-1\right)^i\sin\left(2\pi x\right)$ and sum to one everywhere in the domain.

\begin{figure}
    \centering
    \includegraphics{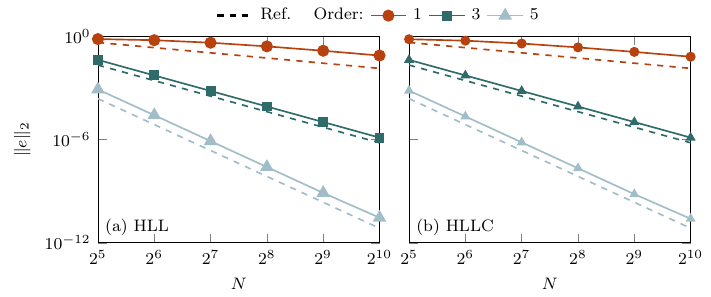}
    \caption{Convergence results for a 1D 2-component advection problem with the (a) HLL and (b) HLLC approximate Riemann solvers.}
    \label{fig:oldConvergence}
\end{figure}

\Cref{fig:oldConvergence} shows the RMS error after five periods of advection for $1^{st}$, $3^{rd}$, and $5^{th}$ order accurate reconstructions using the HLL and HLLC approximate Riemann solvers.
Convergence results can be reproduced using the automated scripts provided in the \lstinline{examples/1D_convergence} example case.

\section{Updated capabilities}
\label{sec:updatedCapabilities}

This section is organized by the role each additional physical, numerical, or performance capability plays in the overarching goal of providing an end-to-end state-of-the-art simulation platform for a broad range of flow problems.
First, we broaden physical coverage from MFC~3.0 (see \citet{bryngelson21}) to capture the couplings that govern realistic flows (immersed boundaries for complex geometry; N-fluid phase change; Euler--Euler and Euler--Lagrange bubble models; fluid–structure interaction; chemically reacting mixtures; two-material surface tension; MHD/RMHD).
We equip the solver with the numerical machinery required to preserve accuracy and robustness across regimes (general characteristic boundary conditions, WENO-Z/TENO reconstructions, Strang splitting for stiff sub-grid dynamics, and low-Mach treatments).
Lastly, we enhance the software's performance and reproducibility, including portable GPU/APU offload, metaprogramming and code generation, continuous integration with over 500 regression tests, and extreme-scale I/O.
So, these methods run consistently from a single accelerator to tens of thousands of devices.

\subsection{Physical features}

\subsubsection{Immersed boundary method}\label{sec:IBM}

Fluid-solid interactions with rigid bodies are simulated using the Ghost-Cell Immersed Boundary Method with an implementation similar to \citet{tseng2003ghost}.
This method imposes Neumann boundary conditions for the pressure, densities, and volume fractions, as well as a no-slip or slip boundary condition for the velocity on the immersed boundary.

In MFC~5.0, the solid geometries are parsed.
Each cell is characterized as being either within a fluid or a solid region.
For each immersed boundary, the level set field $\boldsymbol{\varphi}$ is calculated.
This field stores the vector from each cell to the closest point on the immersed boundary.
The ghost points, image points, and interpolation coefficients are calculated and used to enforce the appropriate boundary conditions.
These are computed before simulation execution and follow the algorithm below.
\begin{algorithm}[H]
\caption{Calculation of level sets.}
\label{alg:levelset}
\begin{algorithmic}
\For{each immersed boundary $\Gamma_i$}
    \For{each cell}
        \If{$\Gamma_i$ is defined as a simple geometry} \Comment{i.e: Sphere, Cuboid, etc.}
            \State Calculate the level set with an analytical function.
        \ElsIf{$\Gamma_i$ is defined by an STL/OBJ file}
            \State Calculate the level set using the procedure described in \cref{sec:stl+ibm}.
        \EndIf
    \EndFor
\EndFor
\end{algorithmic}
\end{algorithm}
The algorithm that applies the boundary conditions follows.
MFC first parses the STL or OBJ geometry file (or level set).
We determine whether each finite volume is in a solid, immersed region by examining its surface normals.
We use \cref{alg:levelset} to compute the level set field for each immersed boundary.
This field stores the vector from each finite volume center to the closest point on the immersed boundary.
If each cell center in the solid regions is within three cells of a fluid cell, we characterize it as a ghost point.
The fluid properties at the image points are interpolated from surrounding grid cells.
We follow \cref{alg:GPandIP} to compute the ghost points, image points, and the interpolation coefficients.
Finite volumes in an immersed solid region are excluded from interpolation.
Following \cref{alg:GP_treatment}, we compute $q_{\mathrm{(ip)}} = q_{\mathrm{(gp)}}$ for each fluid property $q$ with a Neumann boundary condition and set the velocity based on its boundary conditions.

The independence of ghost point treatment allows the algorithm to be readily parallelized.
The computational cost of applying boundary conditions across the immersed body is negligible compared to the rest of the flow simulation.

\begin{algorithm}[H]
\caption{Determining Ghost Points, Image Points, and Interpolation Coefficients}
\label{alg:GPandIP}
\begin{algorithmic}
\For{each cell}
    \If{the cell is in a solid region and it is within three cells of the fluid region}
        \State{Add cell to the list of ghost points}
    \EndIf
\EndFor

\For{each ghost point}
    \State{$\bx_{\mathrm{(ip)}} = \bx_{\mathrm{(gp)}} + 2\boldsymbol{\varphi}(\bx_{\mathrm{(gp)}})$} \Comment{$\bx_{\mathrm{(ip)}}$ is the image point position, $\bx_{\mathrm{(gp)}}$ is the ghost point position}
    \State{Determine the four grid cells surrounding the image point}
    \For{each surrounding grid cell $\bg_j(\bx_{\mathrm{(ip)}})$}
        \If{$\bg_j(\bx_{\mathrm{(ip)}})$ is in the fluid region}
            \State{$d = \lVert \bx_{\mathrm{(ip)}} - \bg_j(\bx_{\mathrm{(ip)}})\rVert_2$} \Comment{Distance from image point to grid cell}
            \State{$c_j(\bx_{\mathrm{(ip)}}) = 1/d^2$}
        \Else
            \State{$c_j(\bx_{\mathrm{(ip)}}) = 0$}
        \EndIf
    \EndFor

    \State{$c(\bx_{\mathrm{(ip)}}) = c(\bx_{\mathrm{(ip)}})/\sum_j c_j(\bx_{\mathrm{(ip)}})$} \Comment{Normalize the interpolation coefficients}
\EndFor
\end{algorithmic}
\end{algorithm}

\begin{algorithm}[H]
\caption{Ghost point treatment at each timestep.}
\label{alg:GP_treatment}
\begin{algorithmic}
\For{each ghost point}
    \For{each fluid property $q$}
        \State $q(\bx_{\mathrm{(ip)}}) = \sum q(\bg_i(\bx_{\mathrm{(ip)}})) \cdot c_i(\bx_{\mathrm{(ip)}}) $
        \If{$q$ is velocity}
            \If{velocity boundary condition is no-slip}
                \State $\bu(\bx_{\mathrm{(gp)}}) = 0$
            \ElsIf{velocity boundary condition is slip}
                \State $\bu(\bx_{\mathrm{(gp)}}) = \bu(\bx_{\mathrm{(ip)}}) - (\hat{\bn} \cdot \bu(\bx_{\mathrm{(ip)}}))\hat{\bn}$
            \EndIf
        \Else \Comment{The fluid property has a Neumann boundary condition}
            \State{$q(\bx_{\mathrm{(gp)}}) = q(\bx_{\mathrm{(ip)}})$}
        \EndIf
    \EndFor
\EndFor
\end{algorithmic}
\end{algorithm}

\subsubsection{Treatment of complex geometries} \label{sec:stl+ibm}

\begin{figure}[ht]
    \centering
    \includegraphics[trim=28 10 0 20, clip]{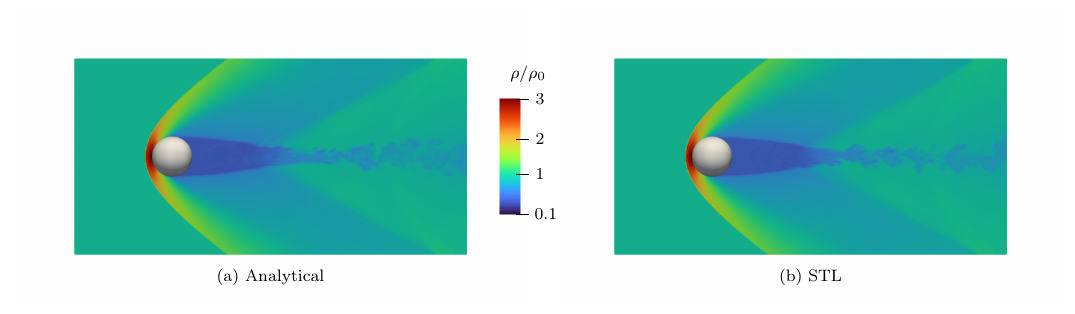}
    \caption{Steady density field ($\rho / \rho_0$) of Mach~2 helium flow over a sphere with (a) analytical level set and (b) STL-based level set.}
    \label{fig:ibm+stl}
\end{figure}

We use a ray-tracing algorithm to translate complex geometries in ASCII and stereolithography (STL) formats onto computational grids compatible with the immersed boundary method.
The ASCII STL format represents geometric models using a triangular mesh.
Each model surface is decomposed into a set of non-overlapping triangular facets.
Three vertices and a normal vector define each triangular facet.
Unlike simple geometries such as cubes and spheres, most complex geometric models do not have analytical solutions for their boundary normal vectors. Consequently, the level sets of complex geometries mentioned in \cref{sec:IBM} also lack analytical solutions.
We need to approximate the geometry's boundary and boundary normal vectors using the STL vertices located on the boundary.
We use an edge-manifoldness algorithm based on \citet{weiler1986topological} to group the boundary STL vertices that are used to determine the level sets and image points.
Coarse STL files are interpolated during ray tracing to achieve results consistent with those obtained from higher-resolution STL files.
The implementation is validated for a Mach~2 helium flow over a sphere with initial flow density $\rho_0$ and $\Rey = 6.5 \times 10^5$ in \cref{fig:ibm+stl}.

\subsubsection{Phase change: First order transition, liquid--vapor}

The phase change model for $N$ fluids is integrated into MFC implemented by introducing disequilibrium terms for pressure ($p$), temperature ($T$), and chemical potential ($g$) into $\bs(\bq)$ in \cref{eqn:discretization}.
This inclusion increases the system stiffness, making a fractional-step method essential~\citep{Strang1968,Zein2010,saurel2009}.
In the case of infinitely fast relaxation, the governing PDEs can be replaced with nonlinear, lower-dimensional, algebraic relations.
For example, for the $pT$-relaxation, we have \citep{flaatten2011solutions}
\begin{gather*}
     f(p) = \sum_{i=1}^N \alpha_i - 1 = \sum_{i=1}^N \frac{\gamma_i-1}{\gamma_i} \frac{\alpha_i\rho_i c_{p, i}}{\sum_{j=1}^N \alpha_j\rho_j c_{p, j}} \frac{ \rho e + p - \sum_{j=1}^N \alpha_j\rho_i e_{*, i} }{\sum_{i=1}^N \alpha_i\rho_i c_{p, i} } - 1 = 0,
\end{gather*}
and for the $pTg$-relaxation
\begin{gather*}
    F_1 \left( \alpha_1\rho_1, p \right) = A + \frac{B}{T} + C \ln( T ) + D\ln( p+\pi_{\infty,1} ) -\ln( p+\pi_{\infty,2} )  = 0,
\end{gather*}
and
\begin{align*}
    F_2 \left( \alpha_1\rho_1, p \right) = \rho e + p + \alpha_1\rho_1 &( e_{*, 2} - e_{*, 1} ) - \alpha_m\rho_m e_{*, 2}  \\ 
    - \sum_{i=3}^N \alpha_i\rho_i q_i + & T \big[\alpha_1\rho_1 ( c_{p,2} - c_{p,1} ) - \alpha_m\rho_m c_{p,2} - \sum_{i=3}^N \alpha_i\rho_i c_{p, i}\big] = 0,
\end{align*}
in which $T=T\left( \alpha_1\rho_1, p \right)$ is the relaxed temperature
\begin{gather*}
     T\left( \alpha_1\rho_1, p \right) = \left\{ \alpha_1\rho_1 \left[ \frac{c_{p,1}-c_{v,1}}{ p + \pi_{\infty,1} } - \frac{ c_{p,2} - c_{v,2} }{ p + \pi_{\infty,2} } \right] + \alpha_m\rho_m \left[ \frac{ c_{p,2} - c_{v,2} }{ p + \pi_{\infty,2} } \right] + \sum_{i=3}^N \left[ \alpha_i\rho_i \frac{ c_{p,i}- c_{v,i}}{ p + \pi_{ \infty, i } } \right] \right\}^{-1},
\end{gather*}
with $\alpha_m \rho_m = \alpha_1 \rho_1 + \alpha_2 \rho_2$, $e_{*, i}$ a reference parameter added to the stiffened gas EOS (see \cref{e:eos}) internal energy equation, and $c_{v,i}$ and $c_{p,i}$ the specific heat coefficient at constant volume and pressure, respectively, for species $i$.
Subscripts $(\, \cdot \,)_{1,2}$ indicate the reacting liquid and vapor.
The remaining subscripts are the inert phases.
$A$, $B$, $C$, and $D$ are fluid-dependent parameters~\citep{saurel2008}, reproduced here for clarity:
\begin{equation*}
    A = \frac{ c_{p,1} - c_{p,2} + e'_{*,2} - e'_{*,1} }{ c_{p,2} - c_{v,2} }, \quad B = \frac{ e_{*,1} - e_{*,2} }{ c_{p,2} -c_{v,2} }, \quad C = \frac{ c_{p,2} -c_{p,1} }{ c_{p,2} -c_{v,2} }, \quad D = \frac{ c_{p,1} -c_{v,1} }{ c_{p,2} -c_{v,2} },
\end{equation*}
in which $e'_{*,i}$ is a reference entropy parameter of the stiffened gas EOS for species $i$.
These strategies are compatible with five- and six-equation models and are solved via Newton's method.
Once the solution is obtained, the solver advances to the next time step.

To demonstrate the model capabilities, we compare numerical results to a set of shock tube experiments by \citet{simoesmoreira1999evaporation} (`chocked flow series'), focusing on the $pT\mu$-relaxation: metastable liquid n-dodecane on the left half of the tube is suddenly discharged into an almost-vacuum region of $\SI{1e-2}{\pascal}$, depressurizing it and initiating phase change.
Consequently, an evaporation wave propagates into the liquid with a steady front mean velocity $U_F$, which depends on the approximately constant temperature used for each run, with $T \in [ 180, 300 ]\,\si{\degreeCelsius}$.
$U_F$ is the ratio of the differences in momentum and density~\citep{saurel2005modelling} after (``a'') and before (``b'') the evaporation wave as
\begin{equation*}
    U_F = \frac{(\rho u)_a-(\rho u)_b}{\rho_a-\rho_b}.
    \label{eq:UF}
\end{equation*}

\begin{figure}[!ht]
	\centering
    \includegraphics{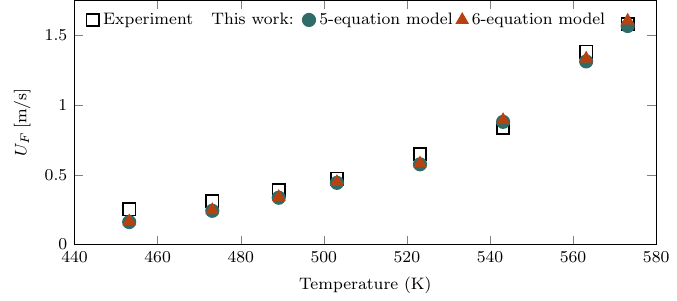}
	\caption{
        Comparison between the numerical and experimental values for the evaporation front velocity using the 5- and 6-equation models for the chocked flow series of n-dodecane tests~\citep{simoesmoreira1999evaporation}.
    }
	\label{fig:SimoesMoreira1D}
\end{figure}

\Cref{fig:SimoesMoreira1D} compares the numerical (5- and 6-equation models) and experimental results.
We observe that both numerical results match well, with some minor discrepancies noted.
These are suspected to be due to the sensitivity in the data obtained to calculate $U_F$, as the precise locations of ``a'' and ``b'' are unclear.
The numerical results quantitatively match the experimental data, with closer matching at larger $T$.

\subsubsection{Euler--Euler sub-grid bubble dynamics}

We use Euler--Euler sub-grid models for bubble dynamics, including the method of classes and the method of moments.
The method of classes is based on the ensemble phase-averaged equations~\citep{zhang1994}, which can be used to represent dispersions of radially oscillating bubbles as well as other particles~\citep{bryngelson19_IJMF,bryngelson23,Bryngelson2020,bryngelson19_whales}.
Note that the method of classes approach was included in MFC~3.0~\citep{bryngelson21}, though a statistical representation enabled via the moment method was not included, which we focus on here.

The void fraction of the dispersed phase $\alpha$ is assumed to be negligible compared to the liquid phase ($\alpha_l$) under dilute assumptions.
The bubbles are represented statistically via random variables $R$, $\Rdot$, and $R_o$, corresponding to the instantaneous bubble radius, time derivative, and equilibrium bubble radius.
The mixture-averaged pressure field is
\begin{gather*}
    p(\bx,t) = (1-\alpha) p_\ell +
	\alpha  \left(
		\frac{\overline{R^3 p_{bw} }}{\overline{R^3}} - \rho \frac{ \overline{ R^3 \dot{R}^2 }}{ \overline{R^3} }
	\right),
    \label{e:pressure}
\end{gather*}
where $p_{bw}$ is the associated bubble wall pressure and $p_\ell(\bx,t)$ is the liquid pressure according to the modified stiffened-gas EOS~\citep{menikoff89}:
\begin{gather*}
	\Gamma_\ell p_\ell + \Pi_{\infty,\ell} =
    \frac{1}{1-\alpha} \left( E - \frac{1}{2} \rho \bu^2 \right).
	\label{e:SEOS}
\end{gather*}
The bubble pressure $p_b$ follows the polytropic assumption, and the bubble wall pressure $p_{bw}$ is
\begin{gather*}
    p_{bw} = p_o\left(\frac{R_o}{R}\right)^{3\gamma} - \frac{4 \mu \dot{R}}{R} - \frac{2\sigma}{R}.
\end{gather*}
The bubble number density per unit volume $n_\mathrm{bub.}(\bx,t)$ is conserved as
\begin{gather*}
    \frac{\partial n_\mathrm{bub.} }{\partial t } + \nabla \cdot ( n_\mathrm{bub.} \bu ) = 0.
    \label{e:consn}
\end{gather*}
For the spherical bubbles considered here, $n_\mathrm{bub.}$ is related to the void fraction $\alpha$ via
\begin{gather*}
    \alpha(\bx,t) = \frac{4}{3} \pi { \overline{R^3} } n_\mathrm{bub.}(\bx,t),
    \label{e:ndf}
\end{gather*}
and so the void fraction $\alpha(\bx,t)$ transports as
\begin{gather*}
    \frac{\partial \alpha }{\partial t } +
        \bu \cdot \nabla \alpha =
	3 \alpha \frac{ \overline{R^2 \Rdot }}{ \overline{R^3} },
    \label{e:alpha}
\end{gather*}
The right-hand side represents the change in void fraction associated with bubble growth and collapse.
The over-barred terms appearing in the above equations denote average quantities of the bubble dispersion.
These are averaged over $N_\mathrm{bin}$ bins of $R_o$ using a log-normal probability distribution function (pdf).
Hybrid schemes and fast integration techniques have been developed for polydisperse cases that otherwise require large bin counts~\citep{charalampopoulos21,bryngelson_levin26}.
We implement the Rayleigh--Plesset and Keller--Miksis equations~\cite{plesset77} for the radial bubble dynamics.
The Keller--Miksis dynamics are
\begin{gather*}
    R \ddot{R} \left(1-\frac{\Rdot}{c}\right) + \frac{3}{2} \Rdot^2\left(1-\frac{\Rdot}{3c}\right) = \frac{p_{\mathrm{bw}} - p_l}{\rho_l}\left(1+\frac{\Rdot}{c}\right) + \frac{R\dot{p}_\mathrm{bw}}{\rho c}.
    \label{e:rpe}
\end{gather*}
The method of classes incorporates stochasticity in $R_o$ using a log-normal probability density function (pdf).
However, modeling complex bubbly flow phenomena requires the expansion of the set of stochastic variables, including instantaneous bubble variables.
The inaccuracy of the polytropic assumption can also limit its scope of application.
The method of moments is employed for this purpose, utilizing a population balance equation (PBE) to govern the probability density function (pdf) of the bubbles.
The pdf $f(R, \Rdot, R_o)$ governs the moments $\mom_{lmn}$ as
\begin{gather*}
    \mom_{lmn} = \overline{R^l \Rdot^m R_o^n} =
        \int_\Omega R^l \Rdot^m R_o^n f(R,\Rdot,R_o) \, \ddd \dd R_o,
    \label{e:raw}
\end{gather*}
The moments are evaluated using a conditional inversion procedure that follows \citet{Fox2018}.
The conditional probability density function $f(R,\dot{R}|R_o)$ without coalescence or breakup effects is governed by the PBE
\begin{gather*}
    \frac{\partial f}{\partial t} +
	\frac{\partial}{\partial R} (f \Rdot ) +
	\frac{\partial}{\partial \Rdot} (f \ddot{R} ) = 0.
    \label{e:master}
\end{gather*}

The evolution of the raw moments $\vbmom$ is obtained by integrating the population balance equation.
The set of transport equations for $\vbmom$ are
\begin{gather}
    \frac{\partial n_\mathrm{bub.} \mom_i}{\partial t} + \nabla \cdot (n_\mathrm{bub.} \mom_i \bu) =
    n_\mathrm{bub.} {\dot\mom_i} =
    n_\mathrm{bub.} g_i,
    \label{e:transport}
\end{gather}
which are enumerated over each moment and corresponding right-hand side of the column vectors $\vbmom$ and $\bg$, denoted as $\mom_i$ and $g_{lmn}$.
So,
\begin{gather*}\label{e:rhs}
    g_{lmn} = l \mom_{l-1,m+1,n} + m \iiint_\Omega \Rddot  R^l \Rdot^{m-1} R_o^n f(\vbmom) \, \ddd \dd R_o,
\end{gather*}
and $\Omega = \Omega_{R} \times \Omega_{\Rdot} \times \Omega_{R_o} = (0,\infty) \times (-\infty,\infty) \times (0,\infty)$~\citep{Bryngelson2020}.
The integral is computed via quadrature nodes obtained by the inversion procedure.

The pdf is conditioned on $R_o$ as
\begin{gather*}
    f(R,\Rdot,R_o) = f(R,\Rdot|R_o) f(R_o),
\end{gather*}
and the raw moments are
\begin{gather*}
    \mom_{lmn}
        \equiv \int_{\Omega_{R_o}} f(R_o) R_o^m \mom_{lm}(R_o) \, \dd R_o
        \approx \sum_{i=1}^{N_{R_o}} w_i \hR_{o,i}^n \, \mom_{lm}(\hR_{o,i}),
    \label{e:full}
\end{gather*}
where $N_{R_o}$ represents the polydispersity of the bubble population.
The conditional moments $\mom_{lm}$ are evaluated using a conditional hyperbolic inversion procedure (CHyQMOM).
The CHyQMOM algorithm is described in~\citet{Fox2018}, and a detailed discussion of its application to bubble dynamics can be found in \citet{bryngelson23}.
The total moments follow as
\begin{gather*}
    \mom_{lmn} =
        \sum_{i=1}^{N_{R_o}}
        w_i \hR_{o,i}^n
        \sum_{j=1}^{N_R} \sum_{k=1}^{N_{\Rdot}} \left[ \hw_{j,k} \hR_j^l \hRdot_k^m \right]_{\hR_{o,i}},
\end{gather*}
These moments evaluate the right-hand side of the moment evolution equation and the ensemble-phase averaged terms.
The evaluation of $\overline{R^3 p_{bw}}$ requires the accounting of statistics of $p_b$.
Without polytropic assumptions, this requires including additional internal variables in the PBE.
The ODEs for $p_b$ and $m_v$ are initialized using an isothermal assumption and evolved at the quadrature nodes, keeping computation costs low.
This strategy assumes their probability densities to be Dirac delta functions centered at the quadrature nodes for $R$ and $\dot{R}$.
The ODEs for $p_b$ and $m_v$ follow \citet{ando09} as
\begin{gather*}
    \dot{p}_b = \frac{3 \gamma_b}{R} \left(- \Dot{R} p_b + R_v T_{bw} \Dot{m_v} + \frac{\gamma_b - 1}{\gamma_b} k_{bw} \frac{\partial T}{\partial r}\Bigr|_{\substack{r=R}} \right)
    \quad \text{and} \quad
    \dot{m}_v = \frac{D \rho_{bw}}{1 - \chi_{vw}} \frac{\partial \chi_{vw}}{\partial r}\Bigr|_{\substack{r=R}}.
    \label{e:pbmv}
\end{gather*}

\subsubsection{Euler--Lagrange sub-grid bubble dynamics}\label{s:el-bubbles}

The Euler--Lagrange model for sub-grid bubble dynamics is based on the volume-averaged equations of motion and describes the dynamics of a mixture of dispersed bubbles in a compressible liquid.
Mixture properties $\overline{(\cdot)}$ are defined as $\overline{(\cdot)}=(1-\alpha)(\cdot)_l+\alpha(\cdot)_g$, where $\alpha$ is the volume fraction of the gas or void fraction, and the subscripts $l$ and $g$ denote the liquid and gas phase, respectively.
Assuming zero slip-velocity between the liquid and gas phase and applying the volume averaging, yields the following set of inhomogeneous equations
\begin{equation}
    \begin{aligned}
    \frac{\partial \rho_l}{\partial t}+\nabla \cdot\left(\rho_l \bu_l\right) & =\frac{\rho_l}{1-\alpha}\left[\frac{\partial \alpha}{\partial t}+\bu_l \cdot \nabla \alpha\right], \\
    \frac{\partial\left(\rho_l \bu_l\right)}{\partial t}+\nabla \cdot\left(\rho_l \bu_l \otimes \bu_l+p \mathcal{I}-\cT_l\right) & =\frac{\rho_l \bu_l}{1-\alpha}\left[\frac{\partial \alpha}{\partial t}+\bu_l \cdot \nabla \alpha\right]-\frac{\alpha \nabla \cdot\left(p \mathcal{I}-\cT_l\right)}{1-\alpha}, \\
    \frac{\partial E_l}{\partial t}+\nabla \cdot\left(\left(E_l+p\right) \bu_l-\cT_l \cdot \bu_l\right) & =\frac{E_l}{1-\alpha}\left[\frac{\partial \alpha}{\partial t}+\bu_l \cdot \nabla \alpha\right]-\frac{\alpha \nabla \cdot\left(p \bu_l-\cT_l \cdot \bu_l\right)}{1-\alpha}.
    \end{aligned}
    \label{eqn:EL}
\end{equation}
Here, $\cT_l$ is the viscous stress tensor of the liquid host.
The advantage of the above form of equations is that the left-hand side of the equation can be seen as the conservation equations for the liquid phase (the details of which are explained in \cref{sec:5eqn}), and the right-hand side part can be seen as source terms that carry the effect of the bubbles in the host liquid.

The volumetric oscillations of the sub-grid bubbles due to pressure variations in the surrounding medium are modeled using the Keller--Miksis equation of \cref{e:rpe}.
The Lagrangian field equation is integrated using a fourth-order accurate Runge--Kutta scheme with adaptive time stepping.

\begin{figure}[ht]
    \centering
    \includegraphics[scale=1]{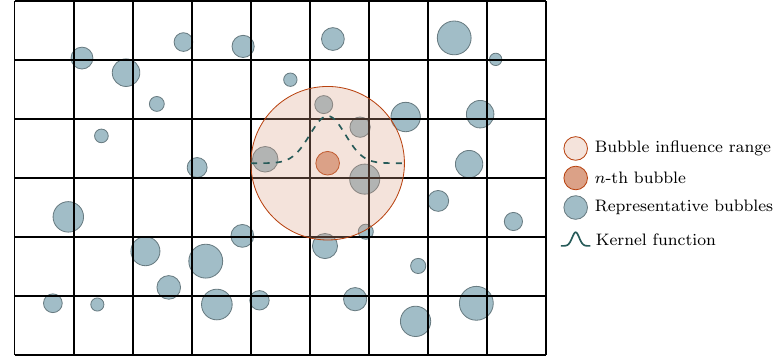}
    \caption{
        Volume spreading of the $n$-th bubble for the void fraction computations using a Gaussian kernel of characteristic radial extent (not to scale).
    }
    \label{fig:smearingBubble}
\end{figure}

MFC communicates the bubble sizes to the carrier-flow solver via the local void fraction $\alpha$.
This coupling is achieved by computing an effective void fraction derived from the contribution of each bubble to its surrounding computational cells, which is illustrated in \cref{fig:smearingBubble}.

The bubble volume is smeared in the continuous field of the void fraction in the mixture at its coordinate $\bx_n$ using a regularization kernel $\delta$ as
\begin{gather}
    \alpha(\bx)=\sum_{n=1}^{N_\mathrm{bub}} V_n \delta = \left( \frac{4}{3} \pi R_n^3 \right) \delta,
\end{gather}
where $N_\mathrm{bub}$ is the number of bubbles, and $V_n$ is the volume of bubble $n$.
We use the continuous, second-order, truncated Gaussian function for the kernel.
With $\alpha$, all the terms on the right-hand side of \cref{eqn:EL} can be computed.
More details on the numerical implementation can be found in \citet{maeda2018eulerian} and \citet{fuster2011modelling}.

\begin{figure}[ht]
    \centering
    \includegraphics[scale=1]{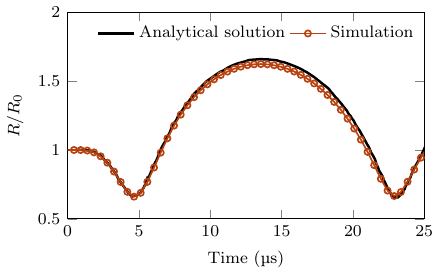}
    \caption{
        Evolution of an isolated bubble in response to a single cycle of a sinusoidal pressure wave.
        The analytical solution of the Keller--Miksis equation was reported by \citet{maeda2018eulerian}.
    }
    \label{fig:el_oscillatingBub}
\end{figure}

The Euler--Lagrange model is validated against the analytical solution of the Keller–-Miksis equation reported by \citet{maeda2018eulerian}.
In this test, we consider a single gas bubble suspended in a water tank.
The bubble has an initial radius of \SI{50}{\micro\meter} and is positioned at the center of the fluid domain $(0, 0, 0)$.
It is subjected to a planar sinusoidal acoustic wave with an amplitude of \SI{0.2}{\mega\pascal} and a frequency of \SI{150}{\kilo\hertz}.
To prevent acoustic reflections at the domain boundaries, we use the non-reflective boundary conditions.
The temporal evolution of the bubble radius, shown in \cref{fig:el_oscillatingBub}, is compared with the analytical Keller--Miksis solution.
The results exhibit excellent agreement, yielding an RMSE error of \SI{2.76}{\percent}.

\begin{figure}[ht]
    \centering
    \includegraphics[scale=1]{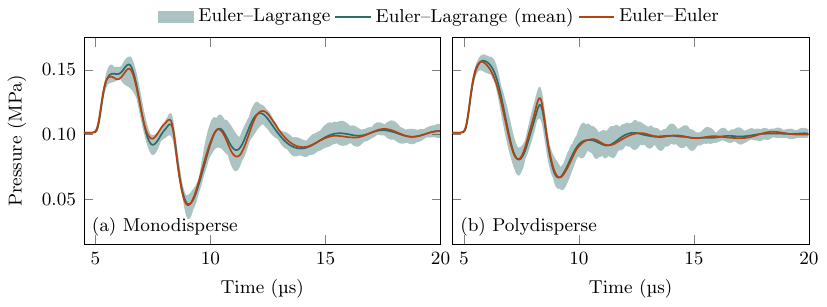}
    \caption{
        Pressure profiles at the origin of the bubble screen using the Euler--Lagrange and Euler--Euler subgrid models. Mono and polydisperse (log-normally distributed with $\sigma_p = 0.3$) bubble clouds are tested as labeled.
    }
    \label{fig:ee-el_bubbleScreen}
\end{figure}

To evaluate the Euler--Euler subgrid model, we simulate the interaction between a dilute bubble screen and a planar sinusoidal acoustic wave.
The results from the Euler--Euler formulation are compared directly with those obtained using our Euler--Lagrange model.
The computational domain spans $x \in [\SI{-20}{\milli\meter}, \SI{20}{\milli\meter}]$ and $y, z \in [\SI{-2.5}{\milli\meter}, \SI{2.5}{\milli\meter}]$, with a bubble cloud centered at the origin and an initial void fraction of $\alpha_0 = 4 \times 10^{-5}$.
In the monodisperse case, all bubbles are initialized with an identical radius of \SI{10}{\micro\meter}.
In contrast, in the polydisperse case, the bubble radii follow a log-normal distribution centered at \SI{10}{\micro\meter} with a shape parameter of $\sigma_p = 0.3$.
A single \SI{0.1}{\mega\pascal}, \SI{300}{\kilo\hertz} acoustic wave is introduced at $x = \SI{-7.5}{\milli\meter}$ and propagates in the positive $x$-direction.
Simulations use a uniform 3D grid ($400 \times 50 \times 50$) with \SI{100}{\micro\meter} spacing, providing fifty cells per wavelength.
For the volume-averaged model, 40 realizations are performed to ensure statistical reliability, while the ensemble-averaged model resolves 21~bins in bubble size space.
\Cref{fig:ee-el_bubbleScreen} shows that both models produce closely matching pressure responses at the cloud center, with RMSE values below \SI{2.1}{\percent}.
The shaded region represents the variability range observed across all Euler--Lagrange realizations.

\subsubsection{Fluid--elastic structure interaction}

Fluid--structure interaction with hypoelastic materials is modeled by adding the elastic shear stress $\bT^{e}$ to the Cauchy stress tensor of the five-equation and six-equation models described in \cref{sec:5eqn} and \cref{sec:sixEqn}, respectively.
The model relies on transforming strains to strain rates using the Lie objective temporal derivative \cite{Guilaume2015,guillaume2016}.
The strain rates are computed using high-order accurate central finite differences.
The mixture energy is also updated to include the elastic contribution, which gives
\begin{gather*}
    E = e + \frac{\lVert\bu\rVert^2}{2} + \frac{\bT^e:\bT^e}{4\rho G},
\end{gather*}
where $G$ is the shear modulus of the medium.
The third term on the right-hand side is the hypoelastic energy term.
The Lie objective temporal derivative of the elastic stresses
\begin{gather*}
    \dot\bT^e = \frac{D \bT^e}{D t} - \bl\cdot\bT^e - \bT^e\cdot\bl^{\top} + \bT^e\tr(\bD) ,
\end{gather*}
where $\bl$ is the velocity gradient, and for an isotropic Kelvin--Voigt material $\dot{\bT}^e = 2G\bD^{d}$~\cite{rodriguez2019} are used to derive a time evolution equation for the elastic stresses.
Here, the deviatoric component of the velocity gradient is $\bD^{d} = \bD - \mathrm{tr}(\bD)\bI/3$.
Appending this evolution equation to the five-equation model gives the five-equation model with hypoelasticity
\begin{align*}
    \pdiff[\alpha_i\rho_i]{t} + \nabla \cdot (\alpha_i\rho_i\bu) &= 0, \\
    \pdiff[\rho \bu]{t} + \nabla \cdot (\rho \bu \otimes \bu + p\bI - \bT^e) &= \nabla \cdot \bT^v, \\
    \pdiff[\rho E]{t} + \nabla \cdot \left[ (\rho E + p)\bu - \bT^e \cdot \bu \right] &= \nabla \cdot (\bT^v \cdot \bu), \\
    \pdiff[\alpha_i]{t} + \bu \cdot \nabla \alpha_i &= K \nabla \cdot \bu, \\
    \pdiff[(\rho \bT^e)]{t} + \nabla \cdot \left( \rho\bT^e \otimes \bu \right) &= \bS^e,
\end{align*}
where $\bS^e$ is the elastic source term:
\begin{gather*}
    \bS^e = \rho \left(\bl\cdot \bT^e + \bT^e\cdot \bl^{\top} - \bT^e \tr(\bD) + 2G\bD^d \right).
\end{gather*}
Details of the hypoelastic model implementation can be found in \citet{spratt2024}.

Fluid--structure interaction with hyperelastic materials is modeled with an evolution equation for the reference map using the reference map technique (RMT) of \cite{kamrin2012}.
The gradient of the reference map $\bxi$, the inverse of the deformation gradient $\bF$,
\begin{gather*}
    \bF = \left(\nabla \bxi\right)^{-1},
\end{gather*}
is computed using a high-order central finite-difference scheme.
The evolution equation of the reference map is combined with the conservation of mass equation to obtain a conservative form,
\begin{gather*}
    \frac{\partial \left(\rho \bxi\right)}{\partial t} + \nabla \cdot \left(\rho \bxi \otimes \bu\right) = \bzero,
\end{gather*}
and solved with the system of equations in \cref{sec:5eqn,sec:sixEqn}.
The deformation gradient $\bF$, the left Cauchy--Green strain $\bb$,
\begin{gather*}
    \bb = \bF\bF^{\top},
\end{gather*}
and the elastic deviatoric contribution to the Cauchy stress tensor $\bT^e$ is computed.
For a compressible neo-Hookean material model, the elastic deviatoric part of the Cauchy stress tensor is
\begin{gather*}
    \bT^e = \frac{G}{J} \left(\bb - \frac{1}{3}\operatorname{tr}(\bb) \bI \right),
\end{gather*}
where $J$ is the determinant of $\bb$ and $\bI$ is the identity tensor and the hyperelastic energy is $e^e = G \left(I_b-3\right) / 2$, where $I_b$ is the first invariant of $\bb$.
Further details of the hyperelastic model implementation can be found in \citet{ctr24_mrj-mcb}.

\subsubsection{Chemical reactions and combustion}

High-fidelity simulations of reacting flows are critical to designing efficient propulsion systems.
Such simulations require models that account for the effects of chemical reactions on the flow.
There are two major differences with respect to simulations of inert flow.
First, the mixture composition varies locally due to chemical reactions, and as a result, fluid properties, such as viscosity, also change.
This effect must be accounted for to achieve acceptable accuracy; it is typically accomplished using transport equations for individual species.
Second, because chemical reactions release heat, temperature changes can become substantial; therefore, variable thermodynamic properties are necessary for accuracy.
As we expand on this, we incorporate these effects through detailed models in the gas phase.

The gas phase is a mixture of $N_\mathrm{species}$ species with mass fractions $\{ Y_{m} \}_{m = 1}^{N_\mathrm{species}}$.
Without molecular transport, these evolve as
\begin{gather*}\label{e:massfrac_transport}
    \frac{\partial \rho_{g} Y_m}{\partial t} + \frac{\partial \rho_{g} u_i Y_m}{\partial x_i} = W_{(m)} \dot{\omega}_m,
    \quad
    m = 1, \dots, N_\mathrm{species},
\end{gather*}
where $\rho_{g}$ is the density of the gas phase, $W_{m}$ is the molecular weight of the $m^{\mathrm{th}}$ species (with parenthesized subscript obviating Einstein summation) and $\dot{\omega}_m$ its net production rate by chemical reactions---the chemical source term.

Next, we provide detailed expressions for the chemical source term.
Their implementation uses code generation and is discussed in \cref{sss:pyro}.
We simulate gas-phase combustion by integrating \cref{e:massfrac_transport} along with the equations for conservation of momentum and total energy density \cref{e:five_eqn_transport}.
Plans to extend this capability to multi-phase combustion with liquid or solid fuels entail the implementation of detailed expressions for phase changes~\cite{ref:Aier2020,ref:Ren2021,ref:Zeng2002}.

The net production rate in \cref{e:massfrac_transport} is
\begin{equation}
    \dot{\omega}_{m} = \sum_{n = 1}^{N_\mathrm{species}}(\nu_{m n}^{\prime\prime} - \nu_{m n}^{\prime})R_{n},
    \quad m = 1, \dots, N_\mathrm{species},
\end{equation}
where $\nu_{mn}^{\prime},\,\nu_{mn}^{\prime\prime}$ are the forward and reverse stoichiometry of the $m^{\mathrm{th}}$ species in the $n^{\mathrm{th}}$ reaction, and $R_{n}$ the net reaction rate of progress. By the law of mass action,
\begin{equation}\label{e:rxn_rate}
    R_{n} = k_{n}(T)\Bigg[ \prod_{j = 1}^{N}\Bigg(\frac{\rho_{g} Y_{j}}{W_{j}}\Bigg)^{\nu_{jn}^{\prime}} - \frac{1}{K_{n}(T)}\prod_{k = 1}^{N}\Bigg(\frac{\rho_{g}Y_{k}} {W_{k}}\Bigg)^{\nu_{kn}^{\prime\prime}} \Bigg]
\end{equation}
where $k_{n}$ and $K_{n}$ are the rate coefficient and the equilibrium constant. Expressions for the rate coefficient are reaction-dependent but conventionally take a modified Arrhenius form
\begin{equation}
    k_{n}(T) = A_{n}T^{b_{n}}\exp( -T_{a,n}/T ),
\end{equation}
where $A_{n},\, b_{n},\, T_{a,n}$ are the pre-exponential factor, temperature exponent, and activation temperature of the $n^{\mathrm{th}}$ reaction. The equilibrium constant in \cref{e:rxn_rate} is determined from standard equilibrium thermodynamics; detailed expressions can be found elsewhere~\cite{ref:cisneros2022,ref:alkhateeb2009}.

As stated, one must consider the effect of the mixture thermodynamics on combustion.
For the equation of state, we assume a mixture of ideal gases, so
\begin{gather}
    p = \rho R_{u}T / W,
\end{gather}
where $R_{u}$ is the universal gas constant and
\begin{equation}
    W = \Bigg(\sum_{m = 1}^{N_\mathrm{species}}\frac{Y_{m}}{W_{m}}\Bigg)^{-1}
\end{equation}
is the mixture molecular weight.
The temperature follows from
\begin{equation}\label{e:combust_temp}
    e_{g} - \sum_{m = 1}^{N_\mathrm{species}}e_{m}(T)\,Y_{m} = 0,
\end{equation}
where $e_{g}$ is the gas-phase internal energy per unit mass (obtained from \cref{e:five_eqn_transport}), and $\{ e_{m}(T) \}_{m=1}^{N_\mathrm{species}}$ the species internal energies.
These are
\begin{equation}\label{e:species_energy}
  e_{m}(T) = \frac{\hat{h}_{m}(T) - R_{u}T}{W_{m}},\qquad m =
  1,\dots,N_\mathrm{species},
\end{equation}
where $\{ \hat{h}_{m}(T) \}_{m = 1}^{N_\mathrm{species}}$ are the species molar enthalpies, modeled using NASA polynomials~\cite{ref:mcbride2002}:
\begin{equation}\label{e:nasa_poly}
  \frac{\hat{h}_{m}}{R_{u}T} = \frac{\hat{C}_{0}}{T} + \sum_{r =
    1}^{5}\frac{\hat{C}_{r}}{r}T^{r-1},
\end{equation}
with fit coefficients $\{ \hat{C}  \}_{r = 0}^{5}$ determined from experiments and collision integrals~\cite{ref:Gardiner1984}.
The fitting procedure for \cref{e:nasa_poly} includes heats of formation, so, in concert with \cref{e:combust_temp,e:species_energy}, it accounts for combustion heat release.
In practice, to obtain the temperature, \cref{e:combust_temp} is solved iteratively using a standard Newton method.

\begin{figure}[htbp]
    \centering
    \includegraphics{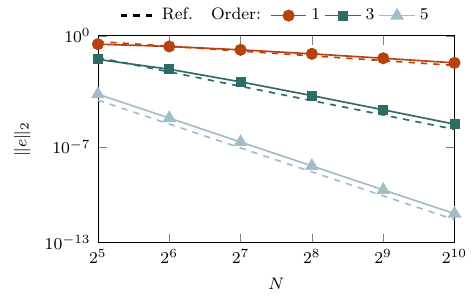}
    \caption{Convergence results for a 1D 2-component advection problem with constant specific heats.}
    \label{fig:chemConvergence}
\end{figure}

The order accuracy is verified via a test problem: the advection of a two-species, calorically perfect mixture.
The flow is initialized with a sinusoidal density and temperature profile, as well as spatially uniform species mass fractions.
The errors compared to a reference solution are in \cref{fig:chemConvergence}, which shows the expected order of accuracy for all numerical methods.

\begin{figure}
    \centering
    \includegraphics{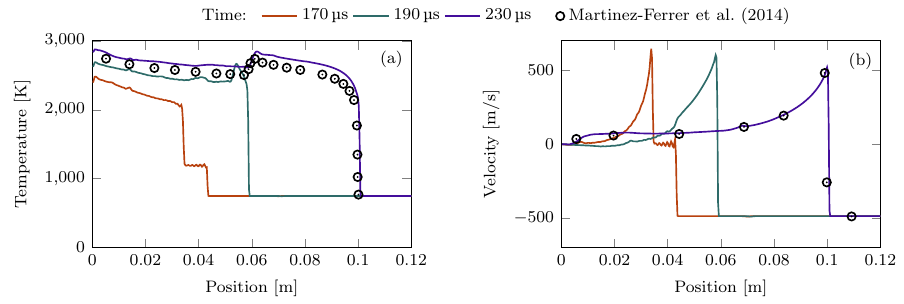}
    \caption{(a) Temperature and (b) velocity profiles of a one-dimensional reactive shock tube compared to the results of \citet{martinez2014}.}
    \label{fig:chemShock}
\end{figure}

To validate our implementation, we simulate a one-dimensional reacting shock tube and reproduce the published results of \citet{martinez2014}.
We uniformly discretize the domain of size $L = \SI{12}{\centi\meter}$ with 400 grid cells.
The left boundary represents a reflecting wall, while the right boundary is an outflow.
Reactions are modeled using the San Diego mechanism~\cite{ref:Saxena2006}.
The initial shock is left-going, reflects from the wall, and ignites the mixture.
The combustion wave and the shock coalesce into a detonation wave.
The comparison between our implementation and the results of \citet{martinez2014} is shown in \cref{fig:chemShock}.
The agreement is sufficient for validation, considering the uncertainties associated with different numerical methods and combustion mechanisms.

\subsubsection{Surface tension at diffuse material interfaces}

\begin{figure}
    \centering
    \includegraphics{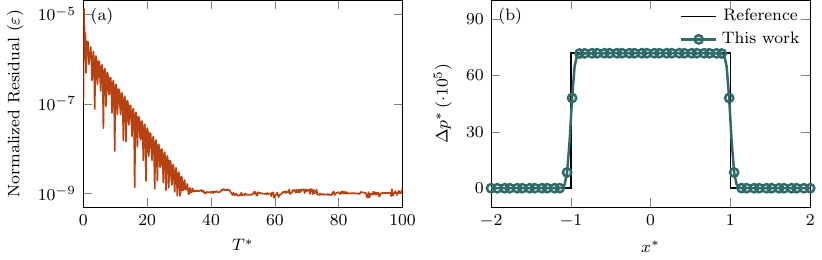}
    \caption{
        Validation results for the surface tension implementation using a Laplace pressure jump problem with a $128 \times 128$ grid.
        (a)~shows the normalized residual as a function of time; (b)~shows the pressure along the centerline for an $R^* = 1$ water column with a surface tension coefficient $\sigma^* = 7.2 \times 10^{-4}$.
    }
    \label{fig:STValidation}
\end{figure}

Surface tension for two-fluid flows is implemented using the model of \citet{schmidmayer2017}.
An advection equation for the color function $c$, which is $0$ in one fluid and $1$ in the other, is added to the system of equations.
This color function is then used to calculate the capillary stress tensor
\begin{gather*}
    \bOmega = -\sigma \left( \lVert\nabla c\rVert\bI -
    \frac{\nabla c \otimes \nabla c}{\lVert\nabla c\rVert}\right),
\end{gather*}
where $\sigma$ is the surface tension coefficient.
The capillary stress tensor defines the contribution of surface tension to the momentum and mixture energy equations.
The six-equation model with surface tension is
\begin{align*}
    \pdiff[\alpha_i\rho_i]{t} + \nabla \cdot (\alpha_i\rho_i \bu) &= 0, \\
    \pdiff[\rho \bu]{t} + \nabla \cdot (\rho \bu \otimes \bu + p\bI + \bOmega - \bT^v) &= 0, \\
    \pdiff[\alpha_1\rho_1e_1]{t} + \nabla \cdot (\alpha_1 \rho_1 e_1 \bu) + \alpha_1 p_1 \cdot \nabla \bu &= -\mu p_I(p_2 - p_1) - \alpha_1\bT_1^v:\nabla\bu, \\
    \pdiff[\alpha_2\rho_2e_2]{t} + \nabla \cdot (\alpha_1 \rho_2 e_2 \bu) + \alpha_2 p_2 \cdot \nabla \bu &= -\mu p_I(p_1 - p_2) - \alpha_2 \bT_2^v : \nabla \bu , \\
    \pdiff[(\rho E + \varepsilon_0)]{t} + \nabla \cdot \left((\rho E + \varepsilon_0 + P)\bu + (\bOmega - \bT^v) \cdot \bu\right) &= 0, \\
    \pdiff[\alpha_1]{t} + \bu \cdot \nabla \alpha_1 &= \mu(p_1 - p_2), \\
    \pdiff[c]{t} + \bu \cdot \nabla c &= 0,
\end{align*}
where $\varepsilon_0$ is the capillary mixture energy.
The six-equation model with surface tension is conservative, except for the pressure relaxation terms, and follows the second law of thermodynamics.
The implementation of surface tension is validated by resolving the Laplace pressure jump inside a liquid water column in air.
The theoretical pressure jump in two dimensions is given by $\Delta P = \sigma / R_\text{d}$ where $R_\text{d}$ is the radius of the water column.
\Cref{fig:STValidation} shows the normalized pressure residual
\begin{equation*}
    \varepsilon = \max \left[ \frac{\lvert P_{i,j}^N - P_{i,j}^{N-1}\rvert}{P_{i,j}^N} \right]
\end{equation*}
as a function of time and the pressure across the centerline for an $R^* = 1$ water column with a surface tension coefficient $\sigma^* = 7.2\times 10^{-4}$ for a $128 \times 128$ grid.
The pressure jump is resolved exactly, and the residual is time-stable.

\subsubsection{Magnetohydrodynamics}

Magnetohydrodynamics (MHD) governs the behavior of conducting fluids under the influence of magnetic fields.
The governing equations for ideal MHD are~\cite{dai1994extension}:
\begin{align*}
  \frac{\partial \rho}{\partial t} + \nabla \cdot \left(\rho \mathbf{u}\right) &= 0, \\
  \frac{\partial \left(\rho \mathbf{u}\right)}{\partial t} + \nabla \cdot \left[
    \rho \mathbf{u}\otimes\mathbf{u} + \left(p + \frac{1}{2}|\mathbf{B}|^2\right) \mathbf{I} - \mathbf{B}\otimes\mathbf{B}
  \right] &= 0, \\
  \frac{\partial E}{\partial t} + \nabla \cdot \left[
    \left(E + p + \frac{1}{2}|\mathbf{B}|^2\right)\mathbf{u} - \left(\mathbf{u} \cdot \mathbf{B}\right) \mathbf{B}
  \right] &= 0, \\
  \frac{\partial \mathbf{B}}{\partial t} + \nabla \cdot \left( \mathbf{u}\otimes\mathbf{B} - \mathbf{B}\otimes\mathbf{u} \right) &= 0.
\end{align*}
Here, $p$ is the thermal pressure and $\mathbf{B}$ is the magnetic field. The total energy density is $E=\rho e + (\rho |\mathbf{u}|^2 + |\mathbf{B}|^2)/2$.

The fast magnetosonic speed, which represents the maximum wave propagation speed, is

\begin{equation*}
c_f = \sqrt{\frac{1}{2} \left( c_s^2 + v_A^2 + \sqrt{ \left( c_s^2 + v_A^2 \right)^2 - 4 c_s^2 v_A^2 \cos^2\theta } \right)},
\end{equation*}
where $c_s$ is the sound speed and $v_A=\sqrt{|\mathbf{B}|^2/\rho}$ is the Alfvén speed.

\begin{figure}
    \centering
    \includegraphics{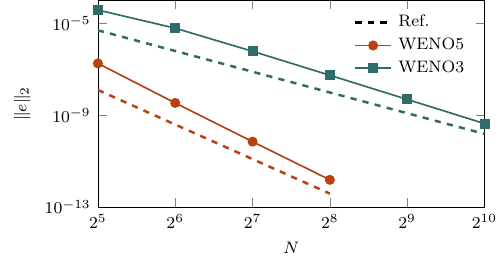}
    \caption{Convergence test for the smooth Alfvén‐wave problem~\citep{toth2000} using WENO3 and WENO5, showing errors decreasing at the expected rate.}
    \label{fig:mhd-convergence}
\end{figure}

A one-dimensional Brio--Wu MHD shock-tube validation of this implementation is shown in \cref{fig:mhd-rmhd-density}~(a), which closely matches the reference~\citep{miyoshi2005multi}.
The convergence test in \cref{fig:mhd-convergence} also shows the expected order of accuracy for the smooth Alfvén-wave problem~\citep{toth2000}, confirming the robustness of the implementation.

\subsubsection{Relativistic Magnetohydrodynamics}

Relativistic magnetohydrodynamics (RMHD) extends the classical MHD approach to high-energy density plasmas where fluid velocities approach the speed of light.
The governing equations can be expressed in conservative form using the conserved variable vector:
\begin{equation*}
\mathbf{U} = \begin{pmatrix}
  D \\[0.25em] \mathbf{m} \\[0.25em] \tau \\[0.25em] \mathbf{B}
\end{pmatrix}
=
\begin{pmatrix}
  \Gamma \rho \\[0.25em]
  \Gamma^2 \rho h \mathbf{u} + |\mathbf{B}|^2 \mathbf{u} - (\mathbf{u}\cdot\mathbf{B}) \mathbf{B} \\[0.25em]
  \Gamma^2 \rho h - p + \left[ |\mathbf{B}|^2 + |\mathbf{u}|^2 |\mathbf{B}|^2 - (\mathbf{B}\cdot\mathbf{u})^2 \right]/2 - \Gamma \rho \\[0.25em]
  \mathbf{B}
\end{pmatrix}.
\end{equation*}
Here, $\rho$ denotes the rest-mass density, $\mathbf{u}$ is the spatial 3-velocity, $\Gamma$ is the Lorentz factor, $p$ is the thermal pressure, and $h = 1 + e + p/\rho$ is the specific enthalpy.
We adopt natural units, where the speed of light is $c$ and defined by reference as $c=1$.
The rest-mass contribution is subtracted from the total energy density to reduce numerical errors in the non-relativistic limit.

The governing equations are~\cite{mignone2006hllc}:
\begin{align*}
\frac{\partial D}{\partial t} + \nabla \cdot \left( D \mathbf{u} \right) &= 0, \\
\frac{\partial \mathbf{m}}{\partial t} + \nabla \cdot \left[ \mathbf{m} \otimes \mathbf{u}+\left(p+\frac{1}{2}\left(\frac{|\mathbf{B}|^2}{\Gamma^2}+(\mathbf{u} \cdot \mathbf{B})^2\right)\right) \mathbf{I}-\frac{\mathbf{B} \otimes \mathbf{B}}{\Gamma^2}-(\mathbf{u} \cdot \mathbf{B}) \mathbf{u} \otimes \mathbf{B} \right] &= 0, \\
\frac{\partial \tau}{\partial t} + \nabla \cdot \left( \mathbf{m} - D\mathbf{u} \right) &= 0, \\
\frac{\partial \mathbf{B}}{\partial t} + \nabla \cdot \left( \mathbf{u}\otimes \mathbf{B} - \mathbf{B}\otimes \mathbf{u} \right) &= 0.
\end{align*}

Recovery of the primitive variables from these highly nonlinear conservative variables requires an iterative scheme.
In this work, we solve a single nonlinear equation using the Newton--Raphson method~\cite{mignone2006hllc}.

\begin{figure}
    \centering
    \includegraphics{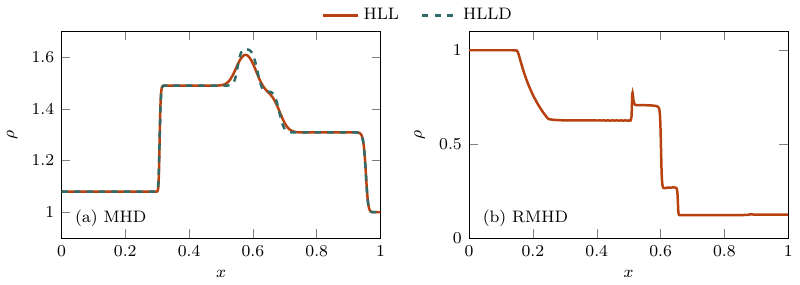}
    \caption{Density profiles for (a) ideal MHD and (b) RMHD shock-tube problems.
    Panel (a) follows \citet{miyoshi2005multi}~(their Fig.~5), and both solvers align closely with the reference solution.
    Panel (b) follows \citet{mignone2006hllc}~(their Fig.~3), and shows similarly tight agreement; the RMHD case here uses WENO3 rather than the reference's second-order scheme.
    Only density is shown; other variables behave comparably.}
    \label{fig:mhd-rmhd-density}
\end{figure}

To validate the implementation, we consider a relativistic one-dimensional RMHD shock-tube problem based on \citet{mignone2006hllc}; the resulting density profile in Figure~\ref{fig:mhd-rmhd-density} (b) aligns closely with the reference, as do the other variables (not shown).

\subsection{Numerical methods}

\subsubsection{General characteristic boundary conditions}

The priority for time-dependent boundary conditions for hyperbolic equations is to ensure a smooth egress of the outgoing artifacts with minimal impact on the interior solution.
For this purpose, the non-reflecting characteristic boundary conditions~\cite{thompson1990} are used, which perform a characteristic decomposition of the governing equation at the boundary.
Ignoring the contribution of the viscous and transverse terms, this is
\begin{gather*}
    \frac{\partial q_c}{\partial t} + R_x \Lambda_x L_x \frac{\partial q_c}{\partial x} = 0,
\end{gather*}
where $q_c$ is the set of conservative variables, $\Lambda_x = \diag(\lambda_j) = \diag(u - c, u, u, u, u + c)$ are the eigenvalues of the Jacobian, and $R_x$ and $L_x$ are the right and left eigenvectors.
The contribution of the characteristics can be further decomposed into incoming and outgoing waves, depending on the sign of the eigenvalue, as
\begin{align*}
    \lambda_j^{i} = \lambda_j - n_x |\lambda_j|
    \quad \text{and} \quad
    \lambda_j^{o} = \lambda_j + n_x |\lambda_j|,
\end{align*}
with $n_x = 1$ for the right boundary and $n_x = -1$ for the left boundary.
The non-reflecting characteristic boundary conditions subsequently ignore the contribution of the incoming wave to get
\begin{gather*}
\frac{\partial q_c}{\partial t} = - R_x \Lambda^o_x L_x \frac{\partial q_c}{\partial x}.
\end{gather*}
In many applications, the primitive variables $q_p$ outside the computational domain are not known, necessitating a more general approach.
The evolution equation for the conservative variables at the boundary can be cast into primitive form as
\begin{gather*}
    \frac{\partial q_c}{\partial t} = -  R_x \Lambda_x L_x \frac{\partial q_c}{\partial x} = -  R_x \Lambda_x L_x P \frac{\partial q_p}{\partial x},
\end{gather*}
where $P = \partial_{q_p}q_c$ is the conversion Jacobian matrix.
The characteristics $\bcL = \Lambda_x L_x P \partial_x q_p$ are
\begin{align*}
    \cL_1 &= (u - c) \left(\frac{\partial p}{\partial x} - \rho c \frac{\partial u}{\partial x} \right), \quad
    \cL_2 = u \left(c^2 \frac{\partial \rho}{\partial x} - \frac{\partial p}{\partial x} \right), \\
    \cL_3 &= u \frac{\partial v}{\partial x}, \quad
    \cL_4 = u \frac{\partial w}{\partial x}, \quad
    \cL_5 = (u + c)\left(\frac{\partial p}{\partial x} + \rho c \frac{\partial u}{\partial x}\right),
\end{align*}
with $v$ and $w$ being the transverse velocities.
The generalized characteristic boundary conditions make use of the solution outside the domain at ghost points ($q_p^{\mathrm{(g)}}$) to incorporate the contribution of the incoming characteristics~\cite{pirozzoli2013generalized}.
The characteristics $\bcL$ in this case are
\begin{gather*}
    \bcL =
        - \Lambda^o_x L_x \frac{\partial q_c}{\partial x} - \Lambda^i_x L_x \frac{\partial q_c}{\partial x} =
        - \Lambda^o_x L_x \frac{\partial q_c}{\partial x} + n_x \lambda_x^i L_x P \frac{(q_p^{\mathrm{(g)}} - q_p)}{\Delta}
\end{gather*}
where $\Delta$ is the grid spacing between the boundary ($q_p$) and the ghost point ($q_p^{\mathrm{(g)}}$).
In practice, $\Delta$ is set to be the grid spacing between the boundary and the first interior point for accuracy and stability.
For a subsonic outflow at the right boundary, coefficients $\cL^i_2$ through $\cL^i_5$ are set to $0$ as these correspond to outgoing waves.
The incoming characteristic $\cL^i_1$ is calculated as
\begin{gather*}
    \cL_1^i = (u - c)((p^{\mathrm{(g)}} - p) - \rho c (u^{\mathrm{(g)}} - u)),
\end{gather*}
where $p^{\mathrm{(g)}}$ and $u^{\mathrm{(g)}}$ are the pressure and normal velocity in the exterior of the domain.

One can extend the 1D approach implemented here to treat strong oblique waves.
Roughly, one would add a dependence on tangential wavenumbers at the boundary, following the oblique non-reflecting formulations of \citet{Giles1990} and the 3D NSCBC extensions by \citet{LodatoDomingoVervisch2008}.
Impedance in the time domain can be enforced via a stable filter~\citep{TamDong1996} and time-domain impedance~\citep{Jaensch2016}, resulting in a non-local time closure that is efficient.

\subsubsection{WENO-Z, TENO, and high-order non-uniform reconstruction}

The WENO-Z scheme~\cite{borges2008improved} enhances classical WENO methods by improving their spectral properties while being faster than WENO-M. The $(2k-1)$-th order WENO-Z weight $\omega_r^{(\mathrm{Z})}$ for the $r$-th stencil is computed as
\begin{gather*}
    \omega_r^{(\mathrm{Z})}=\frac{\alpha_r}{\sum_{i=0}^{k-1} \alpha_i}, \quad \alpha_r=d_r\left(1+\left(\frac{\tau}{\beta_r+\epsilon}\right)^q\right), \quad \tau=\left|\beta_0-\beta_{k-1}\right|,
\end{gather*}
where $d_r$ are the ideal weights, $\tau$ is the global smoothness indicator, and the parameter $q$ (typically set to $1$ for fifth-order reconstruction) influences the convergence rate at the critical points and the spectral properties. This formulation applies to schemes of order 3, 5, and 7.

The Targeted Essentially Non-Oscillatory (TENO) scheme further reduces numerical dissipation by employing a stencil selection process similar to the ENO method~\cite{fu2016family}.
A $K$-th order accurate TENO scheme uses $K-2$ candidate stencils of increasing width.
Its equations are given by
\begin{gather*}
    \omega_r^{(\mathrm{T})}=\frac{d_r\delta_r}{\sum_{i=0}^{K-3} d_i\delta_i},
    \quad
    \delta_r =
    \begin{cases}
        0, & \text { if } \chi_r<C_T \\ 
        1, & \text { otherwise }
    \end{cases},
    \quad
    \chi_r=\frac{\gamma_r}{\sum_{i=0}^{K-3} \gamma_i},
    \quad
    \gamma_r = d_r\left(1+\frac{\tau}{\beta_r+\epsilon}\right)^6,
\end{gather*}
where $C_T$ is a smoothness threshold for stencil activation and $\tau$ is the WENO-Z parameter.

The explicit expressions for the seventh-order WENO reconstruction in our code are omitted here due to their considerable length.
Instead, we present a generalized approach for reconstructing a $(2k-1)$-order scheme from cell averages on non-uniform grids.
For each stencil, the reconstruction polynomial is obtained using the Lagrange interpolation method~\cite{shu1997essentially}:
\begin{gather*}
    p(x) = \sum_{j=0}^{k-1} \hspace{0.5em} \underbrace{
        \Delta x_j \sum_{m=j+1}^k
        \left( \frac{
            \sum_{\substack{l=0 \\ l \neq m}}^k \prod_{\substack{q=0 \\ q \neq m,\, l}}^k \left(x - x_{q-\frac{1}{2}}\right)
        }{
            \prod_{\substack{l=0 \\ l \neq m}}^k \left(x_{m-\frac{1}{2}} - x_{l-\frac{1}{2}}\right)
        } \right)
    }_{\mbox{$c_j(x)$}} \hspace{0.5em} f_j.
\end{gather*}
Here, $x_j$ denotes cell-center location, $x_{j+1/2}$ the cell-boundary location, and $\Delta x_j := x_{j+1/2} - x_{j-1/2}$ is the cell width. For each stencil $r = 0, 1, \dots, k-1$, the $k$-cell stencil spans the interval from $x_{-1/2}$ to $x_{k-1/2}$.
Consequently, the overall $(2k-1)$-cell stencil is centered at $x_{k-r-1}$, and the right flux is $p_r:=p(x_{k-r-1/2})$.

To reduce computational cost, we rewrite $p(x)$ in terms of the differences $\Delta f_j := f_{j+1}- f_j$, so that $p(x) = \sum_{j=0}^{k-2}\hat{c}_j(x)\Delta f_{j}+ f_0$.
For flux evaluation at $x_{k-r-1/2}$, the coefficients $\hat{c}_{r,j} := \hat{c}_j(x_{k-r-1/2})$ for each $\Delta f_j$ and stencil $r$ are precomputed at the start of the program.

The smoothness indicator for stencil $r$ is defined by
\begin{equation*}
    \beta_r =\sum_{l=1}^{k-1} \Delta x_{k-r-1}^{2 l-1} \int_{x_{k-r-\frac{3}{2}}}^{x_{k-r-\frac{1}{2}}}
    \left(\frac{\partial^l p(x)}{{\partial x}^l}\right)^2 \dd x.
\end{equation*}
Rewriting this in terms of the difference formulation yields
\begin{equation*}
    \beta_r =\sum_{j=0}^{k-2} \sum_{m=j}^{k-2} \hspace{0.5em} \underbrace{
        \left(2-\delta_{j m}\right)\sum_{l=1}^{k-1} \Delta x_{k-r-1}^{2 l-1} \int_{x_{k-r-\frac{3}{2}}}^{x_{k-r-\frac{1}{2}}}
            \frac{\partial^l \hat{c}_j(x)}{{\partial x}^l} \frac{\partial^l \hat{c}_m(x)}{{\partial x}^l} \dd x
    }_{\mbox{$\hat{b}_{r,j,m}$}} \hspace{0.5em} \Delta  f_j \Delta  f_m,
\end{equation*}
where $\delta_{jm}$ is the Kronecker delta.
The coefficients $\hat{b}_{r,j,m}$ for each cross term $\Delta f_j \Delta f_m$ and stencil $r$ are precomputed.
Reformulating in terms of $\Delta f$ reduces the number of terms from $k$ to $k-1$ for $p_r$, and from $k(k+1)/2$ to $(k-1)k/2$ for $\beta_r$, thereby reducing the computational cost.

\begin{figure}
    \centering
    \includegraphics{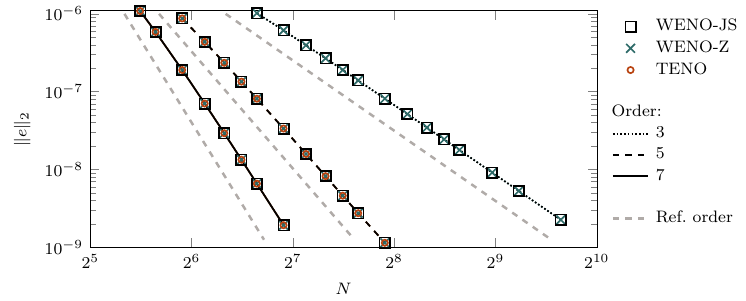}
    \caption{Convergence results for a 1D linear advection problem with WENO-JS, WENO-Z, and TENO schemes at orders of accuracy 3, 5, and 7.}
    \label{fig:WENO-convergence}
\end{figure}

The convergence study in \cref{fig:WENO-convergence} demonstrates that WENO-Z and TENO recover their intended design orders, consistent with the original WENO-JS for 3rd-, 5th-, and 7th-order reconstruction.

\subsubsection{Strang splitting for stiff sub-grid dynamics}

Sub-grid bubbles often exhibit rapid dynamics, necessitating a small time step for stable computation of the bubble source term.
When the time scale of the background flow significantly exceeds that of the sub-grid bubbles, fully resolving the temporal dynamics becomes computationally prohibitive.
To address this, we implement the Strang splitting scheme, which separates the time integration of the background flow and the sub-grid bubbles~\cite{strang1968construction}.
The implementation supports a single resolved phase with sub-grid bubbles governed by
\begin{gather*}
    \frac{\partial \bq}{\partial t} = \nabla \cdot \bF(\bq) + \bs(\bq).
\end{gather*}
The Strang splitting scheme integrates the equation over one time step by integrating three sub-equations for $\bq^{*}$, $\bq^{**}$, and $\bq^{***}$ as
\begin{alignat*}{3}
    \frac{\partial \bq^{*}}{\partial t} &= \bs(\bq), \quad &t& \in [0, \Delta t/2], \quad &\bq&^{*}(0) = \bq^n, \\
    \frac{\partial \bq^{**}}{\partial t} &= - \nabla \cdot \bF(\bq^{**}), \quad &t& \in [0, \Delta t], \quad &\bq&^{**}(0) = \bq^{*}(\Delta t/2), \\
    \frac{\partial \bq^{***}}{\partial t} &= \bs(\bq^{***}), \quad &t& \in [0, \Delta t/2],  \quad &\bq&^{***}(0) = \bq^{**}(\Delta t), \quad \bq^{n+1} = \bq^{***}(\Delta t/2).
\end{alignat*}

To solve for $\bq^{*}$ and $\bq^{***}$, where the stiff bubble source term is evaluated, we use a third-order accurate embedded Runge--Kutta scheme with adaptively controlled step sizes~\cite{hairer93solving}.
The step size, $h$, is updated based on the estimated error, $\be$, calculated as
\begin{align*}
    \bq^{n+1} &= \bq^{n} + \frac{h}{6} \left( k_1 + k_2 + 4 k_3 \right), \\
    \hat{\bq}^{n+1} &= \bq^{n} + \frac{h}{8} \left( 3 k_1 + 3 k_2 + 2 k_3 \right), \\
    \be &= \bq^{n+1} - \hat{\bq}^{n+1} = -\frac{5}{24} h \left( k_1 + k_2 - 2 k_3 \right),
\end{align*}
where $k_1 = \partial \bq^n/\partial t$, $k_2 = \partial \bq^{(1)} / \partial t$, and $k_3 = \partial \bq^{(2)} / \partial t$.
Here, $\bq^{n+1}$ represents a solution of the third-order TVD Runge--Kutta scheme, and $\hat{\bq}^{n+1}$ is a 2nd-order accurate solution obtained using the same $k_1$, $k_2$, and $k_3$ that requires minimal additional computational cost.
More details of the adaptive step size control algorithm can be found in \citet{hairer93solving}.

We demonstrate the utility of the operator splitting scheme with adaptive time stepping for a 0D simulation of bubble dynamics.
When a bubble is located in a high-pressure liquid, it collapses in radial size and rebounds (size increases) repeatedly, eventually damping due to viscous dissipation.
Here, we test our implementation via a $\boldsymbol{u}=\boldsymbol{0}$ case and initial pressure field such that the liquid pressure is $1427$-times larger than the bubbles' internal pressure $p_{b}$.
The bubbles are assumed to be monodisperse in size and air-filled with equilibrium radius $R_{\mathrm{eq}}=\SI{5}{\milli\meter}$.
The 0D simulations are implemented as spatially homogeneous 1D cases.

\begin{figure}
    \centering
    \includegraphics{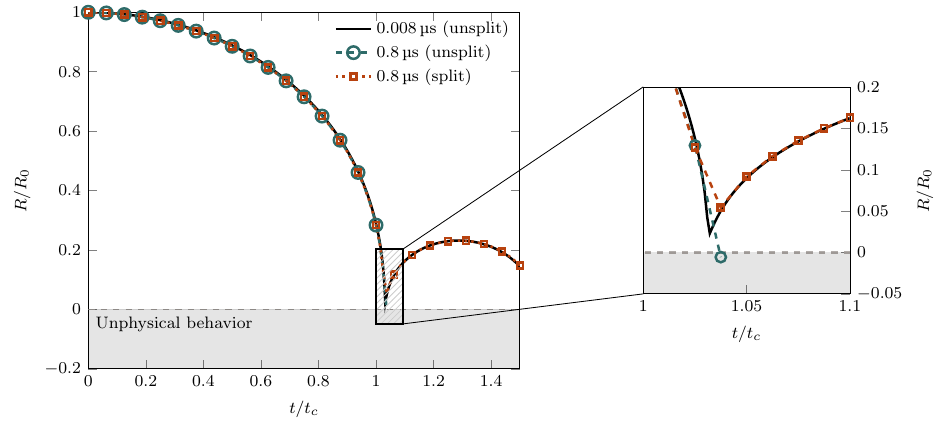}
    \caption{
    Evolution of bubble radius at pressure ratio $p_\infty/p_b = 1427$ with and without the Strang splitting scheme and adaptive time stepping.
    $t_c$ is the total collapse time~\cite{brennen1995}.
    The right plot is a magnified view of the boxed region near the rebound in the left plot.
    }
    \label{fig:bubcollapse}
\end{figure}

Figure \ref{fig:bubcollapse} shows the temporal evolution of bubble radius.
An unsplit scheme with time step size $\Delta t = \SI{0.008}{\micro\second}$ can resolve the collapse and rebound of bubbles.
However, $\Delta t = \SI{0.8}{\micro\second}$ is insufficiently small to resolve the collapse event and leads to simulation failure.
The failure occurs when the bubble radius becomes negative due to the time step sizes that are too large for the bubble radial velocity.
The use of an operator-splitting scheme with adaptive time stepping enables the simulation to represent the rapid collapse and rebound with a time step size of $\Delta t = \SI{0.8}{\micro\second}$.

\subsubsection{Low-Mach number treatment}

Godunov-type Riemann solvers, including the HLLC Riemann solver, are known to lose their accuracy at low Mach numbers due to numerical dissipation in the discrete solution~\cite{guillard1999behaviour}.
To address this limitation, we use two numerical treatments based on the approaches of \citet{thornber2008improved} and \citet{chen2022anti}, allowing users to select their preferred scheme.

\citet{thornber2008improved} proposed a simple modification to the reconstructed velocities of the left and right states as
\begin{equation*}
    \begin{aligned}
        u_{\mathrm{L}}^* &= \frac{u_{\mathrm{L}} + u_{\mathrm{R}}}{2} + z \frac{u_{\mathrm{L}} - u_{\mathrm{R}}}{2}, \\
        u_{\mathrm{R}}^* &= \frac{u_{\mathrm{L}} + u_{\mathrm{R}}}{2} + z \frac{u_{\mathrm{R}} - u_{\mathrm{L}}}{2}, \\
        z &= \mbox{min}\left( \mbox{max} \left(\frac{\lvert u_{\mathrm{L}} \rvert}{c_{\mathrm{L}}}, \frac{\lvert u_{\mathrm{R}} \rvert}{c_{\mathrm{R}}} \right), 1 \right),
    \end{aligned}
\end{equation*}
where the modified velocities $u_{\mathrm{L}}^*$ and $u_{\mathrm{R}}^*$ replace the original velocities $u_{\mathrm{L}}$ and $u_{\mathrm{R}}$ in the calculation of the signal velocity and fluxes.
This strategy ensures the correct scaling of the pressure and density fluctuations in the low Mach number regime.

On the other hand, \citet{chen2022anti} identified a source term, $p_d$, which is responsible for the wrong scaling at the low Mach number limit.
To address this, they proposed an anti-dissipation pressure correction (APC) term, which is incorporated into the flux calculation.
For the HLLC Riemann solver, the terms $p_d$ and APC are
\begin{equation*}
    \begin{aligned}
        p_d &= \frac{\rho_{\mathrm{L}} \rho_{\mathrm{R}} (S_{\mathrm{L}} - u_{\mathrm{L}}) (S_{\mathrm{R}} - u_{\mathrm{R}}) (u_{\mathrm{R}} - u_{\mathrm{L}})}{\rho_{\mathrm{R}} (S_{\mathrm{R}} - u_{\mathrm{R}}) - \rho_{\mathrm{L}} (S_{\mathrm{L}} - u_{\mathrm{L}})}, \\
        \APC_{\rho u} &= (z - 1) p_d \left( n_x, n_y, n_z \right)^\top, \\
        \APC_{\rho E} &= (z - 1) p_d S_*, \\
        z &= \mbox{min}\left( \mbox{max} \left(\frac{\lvert u_{\mathrm{L}} \rvert}{c_{\mathrm{L}}}, \frac{\lvert u_{\mathrm{R}} \rvert}{c_{\mathrm{R}}} \right), 1 \right).
    \end{aligned}
\end{equation*}
Here, $\APC_{\rho u}$ and $\APC_{\rho E}$ represent the correction terms for the momentum and energy equations, respectively, and $n_x$, $n_y$, and $n_z$ denote the components of the unit normal vector.
Then, the corrected HLLC flux is
\begin{gather*}
    \bF^{\mathrm{HLLC+APC}} = \begin{cases}
        \bF\left(\bq^{\mathrm{(L)}}\right) & S_{\mathrm{L}} \ge 0, \\
        \bF^{(*\mathrm{L})} + \bm{APC} & S_{\mathrm{L}} \le 0 \le S_*, \\
        \bF^{(*\mathrm{R})} + \bm{APC} & S_* \le 0 \le S_{\mathrm{R}}, \\
        \bF\left(\bq^{(\mathrm{R})}\right) & 0 \ge S_{\mathrm{R}}.
    \end{cases}
\end{gather*}

\begin{figure}
    \centering
    \includegraphics{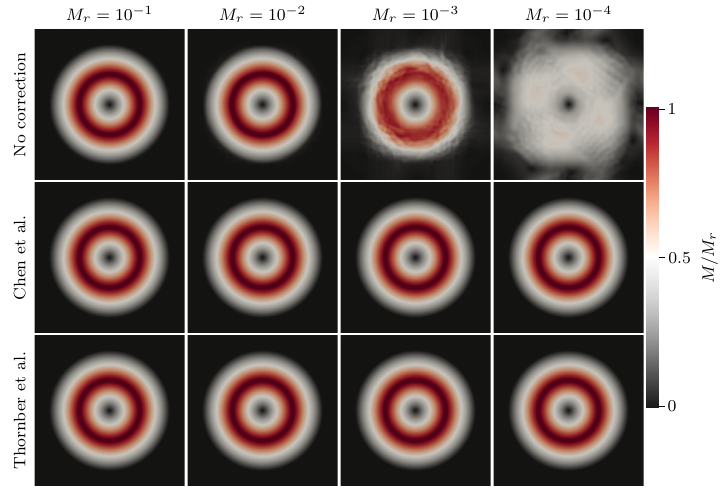}
    \caption{
        Ratio of local Mach number, $M$, to reference Mach number, $M_r$, for the Gresho vortex at $t/t_r = 1$ with $M_r = 10^{-1}$, $10^{-2}$, $10^{-3}$ and $10^{-4}$: (top row) without correction; (middle row) \citet{chen2022anti}; (bottom row) \citet{thornber2008improved}.
    }
    \label{fig:gresho}
\end{figure}

The HLLC Riemann solver with correction schemes is validated with the 2D Gresho vortex, which is an exact solution to the incompressible Euler equations~\citep{gresho1990}.
The angular velocity and pressure are given by
\begin{equation*}
    u_{\phi} = u_r \left\{
    \begin{array}{ll}
        r/R, & 0 \le r < R, \\
        2 - r/R, & R \le r \le 2R, \\
        0, & 2R \le r,
    \end{array}
    \right.
\end{equation*}
\begin{equation*}
    p = p_r + \rho_r u_r^2 \left\{
    \begin{array}{ll}
        \left(r/R\right)^2 / 2, & 0 \le r < R, \\
        \left(r/R\right)^2 / 2  + 4 \left( 1 - \left(r/R\right) + \ln\left(r/R\right)\right), & R \le r \le 2R, \\
        -2 + 4 \ln 2, & 2R \le r,
    \end{array}
    \right.
\end{equation*}
where $r = \sqrt{x^2 + y^2}$ is the radial coordinate, $u_r = 2\pi R M_r$ and $M_r =  u_r / \sqrt{\gamma \left(p_r / \rho_r + u_r^2/2\right)}$ are reference velocity and reference Mach number, respectively.
The corresponding reference time scale is $t_r = 2\pi R/u_r = 1/M_r$, which is the time for one vortex turn around.
Since the velocity and pressure fields are in equilibrium, the flow is expected to be time-stationary.
\Cref{fig:gresho} shows the local Mach number at $t/t_r = 1$ at four reference Mach numbers: $M_r = 10^{-1}$, $10^{-2}$, $10^{-3}$, and $10^{-4}$.
The standard HLLC approximate Riemann solver shows a loss of accuracy for smaller $M_r$.
In contrast, the HLLC solver with correction schemes preserves the flow field to visual accuracy for all tested Mach numbers.

\subsubsection{HLLD Riemann Solver}

For the MHD equations, the HLLD (Harten--Lax--van~Leer--Discontinuities) Riemann solver~\cite{miyoshi2005multi} builds on the HLL framework to resolve five of the seven characteristic waves, thereby reducing numerical dissipation and improving accuracy.

In this approach, the Riemann fan is divided into four intermediate states $\mathbf{U}_\mathrm{L}^*$, $\mathbf{U}_\mathrm{L}^{**}$, $\mathbf{U}_\mathrm{R}^{**}$, and $\mathbf{U}_\mathrm{R}^*$.
These states are separated by two Alfvén waves ($S_\mathrm{L}^*$ and $S_\mathrm{R}^*$) and a middle contact discontinuity ($S_\mathrm{M}$).
The fastest left and right waves ($S_\mathrm{L}$ and $S_\mathrm{R}$) form the outer boundaries of the entire Riemann fan.

Similar to the HLL solver, $S_\mathrm{L}$ and $S_\mathrm{R}$ are approximated by
\begin{equation*}
    S_\mathrm{L} = \min(u_\mathrm{L} - c_{f,\mathrm{L}}, u_\mathrm{R} - c_{f,\mathrm{R}})
    \quad \text{and} \quad
    S_\mathrm{R} = \max(u_\mathrm{L} + c_{f,\mathrm{L}}, u_\mathrm{R} + c_{f,\mathrm{R}}).
\end{equation*}

The HLLD solver enforces two key constraints: normal velocity and total pressure remain constant across the Riemann fan:
\begin{gather*}
    u_\mathrm{L}^* = u_\mathrm{L}^{**} = u_\mathrm{R}^{**} = u_\mathrm{R}^* = S_\mathrm{M}, \\
    p_{\mathrm{T},\mathrm{L}}^* = p_{\mathrm{T},\mathrm{L}}^{**} = p_{\mathrm{T},\mathrm{R}}^{**} = p_{\mathrm{T},\mathrm{R}}^* = p^*_\mathrm{T}.
\end{gather*}
The middle wave speed is calculated as:
\begin{equation*}
    S_\mathrm{M} = \frac{(S_\mathrm{R} - u_\mathrm{R})\rho_\mathrm{R} u_\mathrm{R} - (S_\mathrm{L} - u_\mathrm{L})\rho_\mathrm{L} u_\mathrm{L} - p_{\mathrm{T},\mathrm{R}} + p_{\mathrm{T},\mathrm{L}}}{(S_\mathrm{R} - u_\mathrm{R})\rho_\mathrm{R} - (S_\mathrm{L} - u_\mathrm{L})\rho_\mathrm{L}}.
\end{equation*}
For the outer intermediate states, we first compute the density:
\begin{equation*}
    \rho_\alpha^* = \rho_\alpha \frac{S_\alpha - u_\alpha}{S_\alpha - S_\mathrm{M}},
\end{equation*}
where $\alpha = \mathrm{L}$ or $\mathrm{R}$.
For brevity, explicit expressions for tangential velocities ($v_\alpha^*$, $w_\alpha^*$) and tangential magnetic fields ($B_{y,\alpha}^*$, $B_{z,\alpha}^*$), as well as the energy computation, are not shown here but are available in \citet{miyoshi2005multi}.
The Alfvén wave speeds are determined by:
\begin{equation*}
    S_\mathrm{L}^* = S_\mathrm{M} - \frac{|B_x|}{\sqrt{\rho_\mathrm{L}^*}}, \quad S_\mathrm{R}^* = S_\mathrm{M} + \frac{|B_x|}{\sqrt{\rho_\mathrm{R}^*}}.
\end{equation*}

For inner states, the tangential velocity is:
\begin{equation*}
    v^{**} = \frac{\sqrt{\rho_\mathrm{L}^*}v_\mathrm{L}^* + \sqrt{\rho_\mathrm{R}^*}v_\mathrm{R}^* + (B_{y,\mathrm{R}}^* - B_{y,\mathrm{L}}^*) \, \text{sign}(B_x)}{\sqrt{\rho_\mathrm{L}^*} + \sqrt{\rho_\mathrm{R}^*}}.
\end{equation*}
The corresponding expressions for $w^{**}$, $B_y^{**}$, and $B_z^{**}$ in the inner states, along with the associated energy calculation, follow a similar pattern and can be found in \citet{miyoshi2005multi}.

The numerical flux is chosen based on the region of the Riemann fan that contains the cell interface:
\begin{equation*}
    \mathbf{F}_\mathrm{HLLD} =
    \begin{cases}
    \mathbf{F}_\mathrm{L} & \text{if } S_\mathrm{L} > 0, \\
    \mathbf{F}_\mathrm{L}^* & \text{if } S_\mathrm{L} \leq 0 \leq S_\mathrm{L}^*, \\
    \mathbf{F}_\mathrm{L}^{**} & \text{if } S_\mathrm{L}^* \leq 0 \leq S_\mathrm{M}, \\
    \mathbf{F}_\mathrm{R}^{**} & \text{if } S_\mathrm{M} \leq 0 \leq S_\mathrm{R}^*, \\
    \mathbf{F}_\mathrm{R}^* & \text{if } S_\mathrm{R}^* \leq 0 \leq S_\mathrm{R}, \\
    \mathbf{F}_\mathrm{R} & \text{if } S_\mathrm{R} < 0.
    \end{cases}
\end{equation*}

To validate the HLLD implementation, we use the standard one-dimensional MHD Brio--Wu shock-tube problem from \citet{miyoshi2005multi}.
The density profile in Figure~\ref{fig:mhd-rmhd-density} (a) closely matches the reference, as do the other variables (not shown).

\subsection{Software and high-performance computing}

\subsubsection{APU and GPU offloading with OpenACC and OpenMP}\label{subsub:openacc}

\begin{table}
    \renewcommand\tabularxcolumn[1]{>{\centering\arraybackslash}m{#1}}
    \centering
    \caption{Compiler support for GPU offload to Nvidia hardware, AMD hardware, and for unified shared memory mode on AMD hardware.
    NVHPC SDK~25.9, CCE~19.0.0, and AMD Flang Preview~7.0.5 are used to populate this table.
    Support for both the AMD and CCE compilers is provided to enable AMD GPU offloading when CCE is not available.}
    {
    \begin{tabularx}{\textwidth}{@{}rXXXXXXX@{}}
    \toprule
    & \multicolumn{3}{c}{\textbf{OpenMP}} & \multicolumn{3}{c}{\textbf{OpenACC}} \\
    \cmidrule(lr){2-4} \cmidrule(lr){5-7}
    \textbf{Compiler} & \textbf{NVIDIA GPU} & \textbf{AMD A/GPU} & \textbf{AMD USM} & \textbf{NVIDIA GPU} & \textbf{AMD A/GPU} & \textbf{AMD USM} \\
    \midrule
    AMD   & \xmark & \cmark & \cmark & \xmark & \xmark & \xmark \\
    CCE   & \cmark & \cmark & \cmark & \cmark & \cmark & \xmark \\
    NVHPC & \cmark & \xmark & \xmark & \cmark & \xmark & \xmark \\
    \bottomrule
    \end{tabularx}
    }
    \label{tab:compilerSupport}
\end{table}

Vendor portable GPU offloading on AMD and NVIDIA hardware is implemented using OpenACC~\cite{Wienke2012} and OpenMP~\cite{OpenMP}.
OpenACC and OpenMP are directive-based tools that enable developers to generate GPU kernels by adding directive clauses around regions of parallelism identified in the code.
\Cref{tab:compilerSupport} summarizes the current state of compiler support for OpenMP and OpenACC on NVIDIA and AMD GPUs, and for unified shared memory mode on AMD hardware.
OpenACC is the preferred offloading tool in MFC due to its superior performance over OpenMP on NVIDIA and AMD hardware~\cite{wilfongGB25}.
OpenMP is used as a fallback when unified shared memory (USM) is required on AMD hardware.
Support for AMD compilers is provided for using AMD GPUs when CCE is not available.
Compilers like GNU~14 and Flang~18 support OpenACC, though their relative immaturity at the time of writing makes them insufficient to support MFC.

Support for OpenACC and OpenMP is implemented through a single source code version using the Fortran preprocessor Fypp~\cite{fypp} (additional uses of Fypp in MFC are described in \cref{sss:metaprogramming}).
\Cref{lst:kernel} shows an example of a typical GPU kernel in MFC.
Parallel regions are wrapped with the \texttt{GPU\_PARALLEL\_LOOP} and \texttt{END\_GPU\_PARALLEL\_LOOP} macros, which expand to OpenACC or OpenMP directives depending on the selected offloading tool.
The outer loops over \texttt{j}, \texttt{k}, and \texttt{l} iterate through $\cO(100)$ elements each, corresponding to the grid cells in the MPI subdomain.
The inner loop over \texttt{i} iterates through each equation in the current model.
For two-phase flow problems, this loop spans $\cO(1)$ elements.
When additional features like hypo- and hyperelasticity or chemistry are added to the problem, the extent of this inner loop can be $\cO(10)$.

\begin{lstlisting}[caption={Directive setup for a typical MFC OpenACC kernel.
    The Fypp macros \texttt{GPU\_PARALLEL\_LOOP} and \texttt{END\_GPU\_PARALLEL\_LOOP} expand to OpenACC or OpenMP directives depending on the selected offloading tool.
    },
    label={lst:kernel},
    xleftmargin = 0mm,
    float,
    escapeinside={(*}{*)}
    ]
$:GPU_PARALLEL_LOOP(collapse=3, private='[...]')
do l = 0, p ! Third coordinate direction
    do k = 0, n ! Second coordinate direction
        do j = 0, m ! First coordinate direction
            $:GPU_LOOP(parallelism='[seq]')
            do i = 1, num_PDEs
                ! Core kernel here
                ! O(100) arithmetic operations
            end do
        end do
    end do
end do
$:END_GPU_PARALLEL_LOOP
\end{lstlisting}

\Cref{lst:kernelACC} shows the OpenACC directives generated by the Fypp macros in \cref{lst:kernel}.
OpenACC describes parallelism using terminology in which gangs, workers, and vectors refer to blocks, warps, and threads in CUDA notation.
By default, when OpenACC encounters a \texttt{parallel loop} clause, it distributes the workload across gangs, leaving each gang to utilize a single vector.
Appending the \texttt{gang vector} clause to the \texttt{parallel loop} tells the compiler to split the work across multiple gangs with a fixed vector length, which more efficiently uses available resources and increases execution speed.
Appending \texttt{collapse(3)} to the three outer loops instructs the compiler to collapse those three loops and select the optimal gang and vector sizes based on the current problem and architecture.
Our tests show that serialization of the innermost loop \texttt{i} increases performance, in part due to its relatively small range.

\Cref{lst:kernelMP} shows the OpenMP directives generated for the same kernel by the Fypp macros in \cref{lst:kernel}.
OpenMP offloads parallel regions to the GPU when it encounters an \texttt{omp target} directive.
Adding \texttt{teams} after \texttt{omp target} instructs the compiler to launch a league of teams, which typically map to CUDA thread blocks on NVIDIA GPUs.
The combined \texttt{teams distribute parallel do} construct then further decomposes the iteration space: \texttt{distribute} partitions iterations across teams (i.e., across blocks), and \texttt{parallel do} partitions iterations across threads within each team.
The \texttt{defaultmap} clauses specify how scalars, aggregates, allocatables, and pointers should be implicitly mapped to the device, since OpenMP requires more explicit data-mapping semantics than OpenACC.
Adding \texttt{collapse(3)} collapses the three nested loops into a single iteration space, improving load balance and work distribution across teams and threads, just as in OpenACC.

\begin{centering}
    \addtocounter{lstlisting}{0}
    \renewcommand{\thelstlisting}{\arabic{lstlisting}a}
    \begin{minipage}[t]{0.47\textwidth}
         \begin{lstlisting}[caption={OpenACC directives generated by Fypp macros.},
            label={lst:kernelACC},
            xleftmargin = 0mm,
            escapeinside={(*}{*)}
            ]
!$acc parallel loop vector gang collapse(3) &
!$acc default(present) private(...)
do l = 0, p       ! Z-direction
  do k = 0, n     ! Y-direction
    do j = 0, m   ! X-direction
        !$acc loop seq
        do i = 1, num_PDEs
            ! Core kernel
            ! O(100) - O(1000) operations
        end do
    end do
  end do
end do
!$acc end parallel loop
        \end{lstlisting}
    \end{minipage}
    \hfill
    \begin{minipage}[t]{0.47\textwidth}
        \renewcommand{\thelstlisting}{\arabic{lstlisting}b}
        \addtocounter{lstlisting}{-1}
        \begin{lstlisting}[caption={OpenMP directives generated by Fypp macros.},
            label={lst:kernelMP},
            xleftmargin = 0mm,
            escapeinside={(*}{*)}
            ]
!$omp target teams distribute parallel do &
!$omp simd defaultmap(firstprivate:scalar) &
!$omp defaultmap(tofrom:aggregate) &
!$omp defaultmap(present:allocatable) &
!$omp defaultmap(present:pointer) &
!$omp collapse(3) private(...)
do l = 0, p       ! Z-direction
  do k = 0, n     ! Y-direction
    do j = 0, m   ! X-direction
        !$omp loop bind(thread)
        do i = 1, num_PDEs
            ! Core kernel
            ! O(100) - O(1000) operations
        end do
    end do
  end do
end do
!$omp end target teams distribute parallel do
        \end{lstlisting}
        \renewcommand{\thelstlisting}{\arabic{lstlisting}}
    \end{minipage}
\end{centering}

MFC also uses NVIDIA and AMD's vendor-provided libraries \texttt{cuTensor} and \texttt{hipBLAS} to perform hardware-optimized array reshaping for coalescing memory before the most expensive kernels.
Memory coalescence before the WENO reconstruction and approximate Riemann solver kernels results in a ten-fold speedup due to the increased throughput of high-bandwidth memory (HBM).
\texttt{cuTENSOR}~\citep{cuTENSOR} performs similarly to fully collapsed OpenACC loops on NVIDIA hardware, providing only marginal speedup.
\texttt{hipBLAS}~\citep{hipBLAS} performs array transposes approximately seven times faster than fully collapsed OpenACC loops.
NVIDIA and CCE's \texttt{cuFFT}~\citep{cuFFT} and \texttt{hipFFT}~\citep{hipFFT} perform fast Fourier transforms on GPU devices, and \texttt{FFTW}~\citep{FFTW} is used for CPU-based simulations.
Code enclosed by OpenACC \texttt{host\_data use\_device} and \texttt{end host\_data} or OpenMP \texttt{target data} and \texttt{end target data} clauses are executed on the GPU using these vendor-provided libraries, reducing the need for hardware-specific optimizations.

\subsubsection{Performance on various CPU, GPU, and APU architectures}

MFC~5.0 has been benchmarked on a broad range of CPUs, GPUs, and APUs (also known as superchips).
\Cref{t:serialperf} reports the single-device performance in terms of its grind time, which is computed as time per grid point, PDE, and right-hand side evaluation.
The grind times are calculated for one GPU, APU, or CPU socket.
The benchmarks were performed using a similar strategy to that employed in published testbed reports~\citep{elwasif}.
These quantities are for a typical, representative compressible multi-component problem: 3D, inviscid, 5-equation model problem with two advected species (8~PDEs) and 8M grid points (158-cubed 3D uniform grid).
The equation is solved via fifth-order accurate WENO finite volume reconstruction and the HLLC approximate Riemann solver.
This case is in the MFC source code under \lstinline{examples/3D_performance_test}.
One can execute it via
\begin{lstlisting}[keywordstyle={}]
./mfc.sh run examples/3D_performance_test/case.py -t pre_process simulation --case-optimization -n <num_cores_or_gpus> -j <num_build_threads>
\end{lstlisting}
This command can be executed for CPU cases.
It builds an optimized version of the code for this case and then executes it.
For benchmarking GPU devices, one will likely want to use
\lstinline[]{-n <num_gpus>}
where the usual case for single GPU device benchmarking leaves \lstinline[keywordstyle={}]{<num_gpus>} as unity.
Similar performance is also observed for other problem configurations, such as the Euler equations (4~PDEs).
All results are for the compiler that gave the best performance, including AOCC, Intel, GCC, CCE, and NVHPC.

CPU results may be performed on CPUs with more cores than reported in \cref{t:serialperf}; we report results for the best performance, given the full processor die, by checking performance for different core counts on that device.
CPU results are the best performance we achieved using a single socket (or die).
GPU results are for a single GPU device.
For single-precision (SP) GPUs, we performed computations in double-precision via compiler/software conversion; these numbers are not representative of single-precision computations.
AMD MI250X and MI300A devices have multiple graphics compute dies per socket; we report results for two GCDs for the AMD~MI250X and the entire APU (6~XCDs) for the AMD~MI300A.

{\footnotesize
\begin{longtable}{llll | llll}
    \caption{
        Observed grind time performance (time), reported as nanoseconds per grid point per PDE per right-hand side evaluation.
        Using more GPU devices or CPU sockets may not improve the grind time for the same problem, as the problem will be made smaller per marshaled device (and so grind time is expected to be approximately constant).
        \label{t:serialperf}
    }\\ 
        Hardware & Type & Usage & Time & Hardware & Type & Usage & Time  \\  \midrule
        NVIDIA GH200 & APU & 1 GPU & 0.32 & Intel Xeon 6740E & CPU & 92 cores & 4.2 \\ 
        NVIDIA H100 SXM5 & GPU & 1 GPU & 0.38 & NVIDIA A10 & GPU & 1 GPU & 4.3 \\ 
        NVIDIA H100 PCIe & GPU & 1 GPU & 0.45 & AMD EPYC 7713 & CPU & 64 cores & 5.0 \\ 
        NVIDIA B200$^\dagger$ & GPU & 1 GPU & 0.46 & Intel Xeon 8480CL & CPU & 56 cores & 5.0 \\ 
        AMD MI250X & GPU & 1 GPU & 0.55 & Intel Xeon 6454S & CPU & 32 cores & 5.6 \\
        AMD MI300A & APU & 1 APU & 0.57 & Intel Xeon 8462Y+ & CPU & 32 cores & 6.2 \\
        NVIDIA A100 & GPU & 1 GPU & 0.62 & Intel Xeon 6548Y+ & CPU & 32 cores & 6.6 \\ 
        NVIDIA V100 & GPU & 1 GPU & 0.99 & Intel Xeon 8352Y & CPU & 32 cores & 6.6 \\ 
        NVIDIA A30 & GPU & 1 GPU & 1.1 & Ampere Altra Q80-28 & CPU & 80 cores & 6.8 \\ 
        AMD EPYC 9965 & CPU & 192 cores & 1.2 & AMD EPYC 7513 & CPU & 32 cores & 7.4 \\ 
        AMD MI100 & GPU & 1 GPU & 1.4 & Intel Xeon 8268 & CPU & 24 cores & 7.5 \\ 
        AMD EPYC 9755 & CPU & 128 cores & 1.4 & AMD EPYC 7452 & CPU & 32 cores & 8.4 \\ 
        Intel Xeon 6980P & CPU & 128 cores & 1.4 & NVIDIA T4 & GPU & 1 GPU & 8.8 \\ 
        NVIDIA L40S & GPU & 1 GPU & 1.7 & Intel Xeon 8160 & CPU & 24 cores & 8.9 \\ 
        AMD EPYC 9654 & CPU & 96 cores & 1.7 & IBM Power10 & CPU & 24 cores & 10 \\ 
        Intel Xeon 6960P & CPU & 72 cores & 1.7 & AMD EPYC 7401 & CPU & 24 cores & 10 \\ 
        NVIDIA P100 & GPU & 1 GPU & 2.4 & Intel Xeon 6226 & CPU & 12 cores & 17 \\ 
        Intel Xeon 8592+ & CPU & 64 cores & 2.6 & Apple M1 Max & CPU & 10 cores & 20 \\ 
        Intel Xeon 6900E & CPU & 192 cores & 2.6 & IBM Power9 & CPU & 20 cores & 21 \\ 
        AMD EPYC 9534 & CPU & 64 cores & 2.7 & Cavium ThunderX2 & CPU & 32 cores & 21 \\ 
        NVIDIA A40 & GPU & 1 GPU & 3.3 & Arm Cortex-A78AE & CPU & 16 cores & 25 \\ 
        Intel Xeon Max 9468 & CPU & 48 cores & 3.5 & Intel Xeon E5-2650V4 & CPU & 12 cores & 27 \\ 
        NVIDIA Grace CPU & CPU & 72 cores & 3.7 & Apple M2 & CPU & \phantom{1}8 cores & 32 \\ 
        NVIDIA RTX6000 & GPU & 1 GPU & 3.9 & Intel Xeon E7-4850V3 & CPU & 14 cores & 34 \\ 
        AMD EPYC 7763 & CPU & 64 cores & 4.1 & Fujitsu A64FX & CPU & 48 cores & 63 \\ 
        \caption*{$^\dagger$Cross-module inlining not yet supported.}
\end{longtable}
}

\begin{figure}
    \centering
    \includegraphics{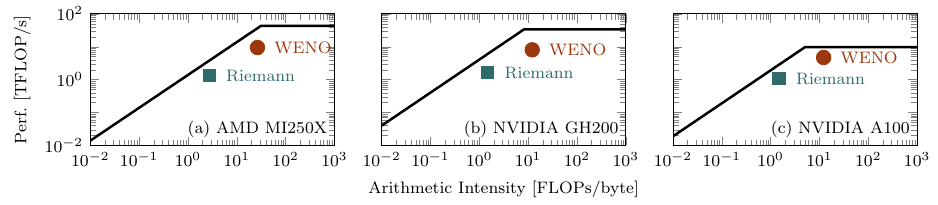}
    \label{fig:rooflines}
    \caption{
        The roofline performance of the fifth-order accurate WENO reconstruction and HLLC approximate Riemann solve kernels on the AMD~MI250X, NVIDIA~GH200, and NVIDIA~A100.
    }
\end{figure}

\Cref{fig:rooflines} shows the roofline performance for the most expensive kernels on NVIDIA and AMD accelerators.
At the time of writing, roofline analysis is not available on the MI300A~APU and therefore is not shown.
NVHPC~24.5 compilers and NVIDIA Nsight Compute profilers produce roofline results on NVIDIA devices, and CCE~18 with Omniperf profiles of the roofline on the AMD~MI250X GPU.
The WENO and approximate Riemann kernels achieve 37\% and 14\% of peak performance on the NVIDIA~A100.
On the NVIDIA~GH200, the WENO and approximate Riemann kernels achieve 24\% and 5\% of peak compute performance.
The AMD~MI250X achieves 22\% of peak compute performance in the WENO kernel and 3\% of peak performance in the approximate Riemann problem.

\begin{figure}
    \centering
    \includegraphics{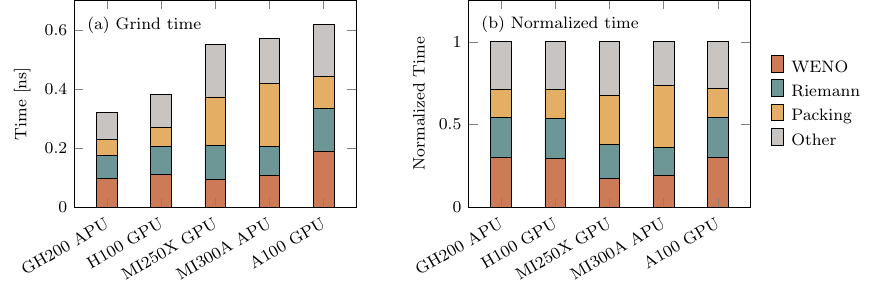}
    \caption{
        Grind time and normalized breakdown of the most expensive kernels, array packing, and all other computations on NVIDIA (GH200, H100, A100) and AMD (MI250X, MI300A) GPUs/APUs.
    }
    \label{fig:breakdown}
\end{figure}

The breakdown of time spent in the two most expensive kernels, time spent packing arrays, and time spent doing all other computations and communication is shown in \cref{fig:breakdown}.
The WENO and approximate Riemann kernels account for about 50\% of the total wall time on NVIDIA GPUs.
On AMD GPU devices, the time spent in the WENO and approximate Riemann kernels is similar to that of NVIDIA devices.
Still, these kernels account for about a third of the total wall time due to an increase in the time spent packing arrays for coalesced memory access.
The profiles show a high number of cache misses on the AMD~MI250X, which we expect is due to the smaller device caches compared to those of the shown NVIDIA device.

\subsubsection{Exascale capabilities and beyond}

\begin{figure}
    \centering
    \includegraphics{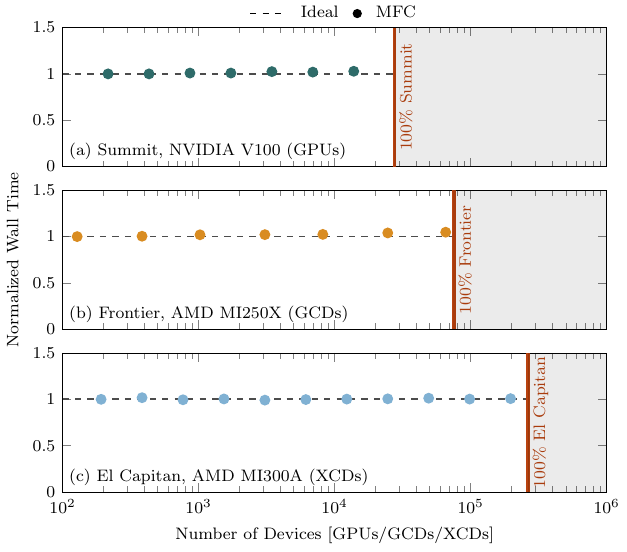}
    \caption{Weak scaling results on OLCF~Summit and Frontier and LLNL El~Capitan.
    MFC Scales from 128 to 13824 NVIDIA V100 GPUs with 97\% efficiency on Summit, from 128 to 65536 AMD MI250Xs (GCDs) with 95\% efficiency on Frontier, and 192 to 196608 AMD MI300As (XCDs) with 99\% efficiency on El~Capitan (XCDs).
    }
    \label{fig:weakScaling}
\end{figure}

MFC efficiently scales to tens of thousands of GPUs on the world's fastest computers.
\Cref{fig:weakScaling} shows the weak scaling performance of MFC on OLCF Summit (now retired) with NVIDIA V100 GPUs, OLCF Frontier with AMD MI250X GPUs, and LLNL El~Capitan with AMD MI300A APUs.
MFC scales to 13824 NVIDIA V100 GPUs on OLCF~Summit with 97\% efficiency, to 65536 AMD MI250X GCDs (32768 MI250X GPUs) on OLCF~Frontier with 95\% efficiency, and to 196608 AMD MI300As XCDs (32768 MI300A APUs) on LLNL El~Capitan.

The communication patterns in MFC lead to high strong scaling efficiencies.
Specifically, the communication involves only a small halo region around each uniformly sized domain decomposition chunk, and the relatively large amount of computation associated with each grid point.
The decomposed grid chunks, assigned one-to-one per MPI rank, are not expanded or contracted.
Since each computing device handles the same number of elements, the halo exchange cost is the same for each MPI rank, which also explains the ideal weak scaling behavior:
The number of domain decomposition chunks increases linearly with problem size, so there is no opportunity for weak scaling inefficiencies.

\begin{figure}
    \centering
    {\footnotesize
    \begin{tabular}{c c}
        \includegraphics{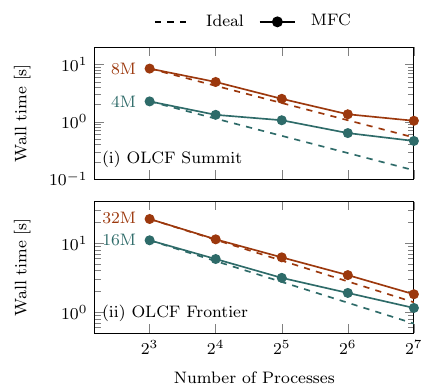} &
        \includegraphics{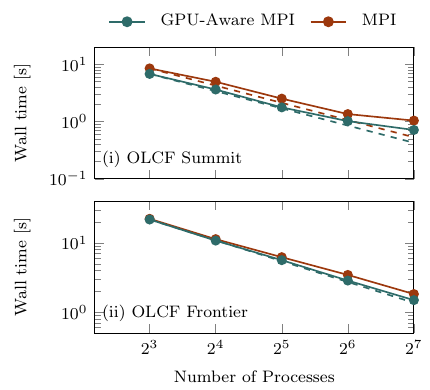} \\
        (a) Strong scaling & (b) GPU-aware MPI enhancement
    \end{tabular}
    }
    \label{fig:strongScaling}
    \caption{
        Strong scaling performance on OLCF~Summit and OLCF~Frontier.
        The figures on the left show the strong scaling performance for a species problem without using GPU-aware MPI.
        The quantities in (b) indicate the number of grid cells per device in the base case.
        The figures in (b) show the improvement in scaling performance when GPU-aware MPI is used.
        In (b), an 8M grid point case is used for benchmarking MFC on (i) Summit, and a 32M grid point case is used for (ii) Frontier.
        The number of processes is commensurate with Summit's NVIDIA~V100 GPUs or Frontier's AMD~MI250X GPUs.
    }
\end{figure}

\Cref{fig:strongScaling} shows the strong scaling performance of MFC on OLCF~Summit and OLCF~Frontier and compares weak scaling with and without GPU-aware MPI.
The problem sizes in \cref{fig:strongScaling}~(a) correspond to approximately 100\% and 50\% usage of available GPU memory per device in the base case.
In the base case, with 8M grid cells per device, strong scaling efficiencies of 79\% and 51\% are observed on OLCF~Summit when increasing the device count by a factor of 8 or 16 without GPU-aware MPI.
For the case with 32 million grid cells per device in the base case, strong scaling efficiencies of 81\% and 77\% when increasing the device count by a factor of 8 and 16 without GPU-aware MPI.
The higher efficiencies seen with the AMD MI250X on OLCF Frontier are partly due to the increased computation-to-communication ratio resulting from larger problem sizes and the improved internode communication on Frontier.
With GPU-aware MPI, the strong scaling efficiencies on OLCF~Summit increase to 84\% and 60\% when increasing the device count by a factor of 8 and 16.
On Frontier, the efficiencies are 96\% and 92\% when the device count is increased by the same factor.

\subsubsection{Parallel I/O at extreme scales}

Parallel I/O is implemented using the MPI I/O library to facilitate shared file parallel I/O and file-per-process parallel I/O.
Shared file parallel I/O was sufficient for handling up to 10K processes, but a significant slowdown in I/O times was observed when scaling to over 65K processes on OLCF Frontier.
For simulations that use over 10K processes, each process writes a file for each I/O operation.
To alleviate file system pressure resulting from metadata creation for tens of thousands of files, MFC accesses the file system in waves of 128 processes, separated by a fixed number of floating-point operations.

\subsubsection{Software resilience and continuous integration}

Each pull request to MFC goes through a suite of continuous integration actions.
The correctness, performance, and code quality are evaluated through this continuous integration suite.
MFC's test suite, comprising over 500 cases, is compiled and run using GNU and Intel compilers on macOS and Ubuntu operating systems, utilizing GitHub Actions and Runners.
The same tests are run using NVHPC on CPUs and NVIDIA GPUs, as well as with CCE on AMD GPUs via self-hosted runners.
All of these tests are compared to the same set of golden files to ensure that MFC's code provides the same results on CPUs and GPUs, regardless of compiler or operating system.
The coverage of the test suite is evaluated using Codecov to help ensure coverage of the test suite for additional (or removed) code.
Each pull request is benchmarked against the main MFC branch on CPUs and GPUs to prevent performance regressions; it also checks for compiler warnings and formatting to ensure code consistency.

\subsubsection{Metaprogramming}\label{sss:metaprogramming}

Metaprogramming is implemented using the Fortran preprocessor Fypp~\citep{fypp}.
Fypp is a Python-based preprocessor that inlines subroutines via directives and collapses repetitive code.
MFC uses Fypp for several tasks, including automating deep copies of derived types for allocation on the GPU device.
NVIDIA's NVHPC Fortran compiler automatically allocates derived types on the device, whereas CCE adheres more closely to the OpenACC technical specification (version~3.2 at the time of writing) and requires manual allocation.
If one holds many host variables, this process is cumbersome.

\Cref{lst:deepCopy} shows how Fypp is used to generate repetitive directives required for deep copying derived types.
Without metaprogramming, each component of a derived type would need to be copied individually.
The macro \texttt{@:SETUP\_VFs(var1,var2,\dots,varN)} is used after allocating the vector fields \texttt{var1,var2,\dots,varN} on the host.
This macro automatically traverses the derived type fields and inserts the correct enter data and create statements, ensuring that all vector fields and their subfields are allocated on the device with minimal manual coding.
This macro also performs deep copies, reducing the number of lines required to initialize the variables at runtime.

\begin{lstlisting}[
    caption={
        A Fortran+Fypp macro that manually deep copies derived types from the host to the device.
        In the loop, Fypp expands \texttt{@:SETUP\_VFs(var1, var2, \dots)} into repeated OpenACC or OpenMP directives.
        \texttt{\$:GPU\_ENTER\_DATA(copyin='[]'} reduces to \texttt{!\$acc enter data copyin} or \texttt{!\$omp target enter data copyin} depending on the desired GPU offload too.
        For each variable, the macro first copies the variable and its vector field array to the device.
        If the array is allocated, the loop iterates over its bounds, and if any scalar field pointers are associated, it performs further deep copies.
        This ensures a complete mirror of the host data structure on the GPU.
        },
        label={lst:deepCopy},
        xleftmargin = 0mm,
        float
        ]
#:def SETUP_VFs(*args)
block
    integer :: macros_setup_vfs_i
    #:for arg in args
        $:GPU_ENTER_DATA(copyin=('[' + arg + ']'))
        $:GPU_ENTER_DATA(copyin=('[' + arg + '%vf]'))
        if (allocated(${arg}$%vf)) then
            do macros_setup_vfs_i = lbound(${arg}$%vf, 1), ubound(${arg}$%vf, 1)
                if (associated(${arg}$%vf(macros_setup_vfs_i)%sf)) then
                    $:GPU_ENTER_DATA(copyin=('[' + arg + '%vf(macros_setup_vfs_i)]'))
                    $:GPU_ENTER_DATA(copyin=('[' + arg + '%vf(macros_setup_vfs_i)%sf]'))
                end if
            end do
        end if
    #:endfor
end block
#:enddef
\end{lstlisting}


Fypp simplifies memory management by wrapping \texttt{allocate} and \texttt{deallocate} into macros that allocate the variable and perform the necessary data movement required when using OpenACC or OpenMP for GPU offloading.
\Cref{lst:memManag} shows this abstraction.
A call such as \texttt{@:ALLOCATE(u, v, w)} expands to allocate the three variables and register them on the GPU device.
The code length is further reduced by using Fypp to abstract away repetitive code via preprocessor loops~\citep{radhakrishnan24}.

\begin{lstlisting}[caption={Fypp macros for automatically allocating and deallocating memory with the associated directives for GPU memory management.
    \texttt{\$:GPU\_ENTER\_DATA(create='[]')} reduces to \texttt{!\$acc enter data create} or \texttt{!\$omp target enter data create} and \texttt{\$:GPU\_EXIT\_DATA(delete='[]')} reduces to \texttt{!\$acc exit data delete} or \texttt{!\$omp target exit data delete} depending on the desired GPU offload tool.},
    label={lst:memManag},xleftmargin = 0mm,float]
#:def ALLOCATE(*args)
    #:set allocated_variables = ', '.join(args)
    allocate (${allocated_variables}$)
    #:set cleaned = strip_parenthesis(allocated_variables)
    #:set joined = ', '.join(cleaned)
    $:GPU_ENTER_DATA(create='[' + joined + ']')
#:enddef ALLOCATE

#:def DEALLOCATE(*args)
    #:set allocated_variables = ', '.join(args)
    $:GPU_EXIT_DATA(delete=('[' + allocated_variables + ']'))
    deallocate (${allocated_variables}$)
#:enddef DEALLOCATE
\end{lstlisting}



Metaprogramming sets problem parameters as constant parameters in case files.
This option is called \textit{case optimization}, enabled via the flag \texttt{--case-optimization} at run time.
Case optimization reduces register pressure by providing the compiler with the parameters of a configuration.
On CPUs, specifying the problem parameters at compile time results in a two-fold speedup of the code.
On GPUs, case optimization decreases run time by about a factor of ten.

\Cref{lst:fyppDec} demonstrates case optimization: when the option is enabled, configuration parameters are converted into compile-time constants (parameter).
Otherwise, they remain runtime variables.
This small change enables the compiler to perform constant folding and loop unrolling, thereby reducing register pressure and improving performance.
Specifically, variables are prescribed in a file with the Fypp code \texttt{\#:set var = val}.
This file is included in the Fortran source code; compile-time-known parameters are declared using the Fypp logic of \cref{lst:fyppDec}.

\begin{lstlisting}[
    caption={
        Fortran+Fypp code the declaration of configuration-specific parameters for compile-time optimization.
    }
    ,label={lst:fyppDec},
    xleftmargin = 0mm,
    float]
#:if CASE_OPTIMIZATION
    integer, parameter :: var = ${var}$
#:else
    integer :: var
#:endif
\end{lstlisting}

In summary, Fypp-style metaprogramming enables MFC to automate deep copies of complex data structures, simplify memory management with consistent macros, and facilitate compile-time specialization of problem parameters.
These techniques reduce boilerplate code, improve maintainability, and deliver performance improvement on CPUs and GPU devices.

\subsubsection{Code generation for efficient reactions and thermodynamics}\label{sss:pyro}

MFC now simulates chemically reacting flows.
We accommodate this feature via combustion routines that evaluate the chemical source terms and species thermodynamics.
Comprehensive libraries, such as Cantera, provide such routines.
However, these exist in a different compilation unit from the flow solver, so they cannot easily be offloaded via the OpenACC directive-based strategy of \cref{subsub:openacc}.
We address this challenge through code generation, producing a computational representation of thermochemistry at the same level of abstraction as MFC's compressible flow solver.
This strategy obviates the need to optimize comprehensive combustion libraries at link time.
MFC uses Pyrometheus~\cite{ref:cisneros2022,cisneros25} for code generation, generating thermochemistry code in OpenACC-decorated Fortran.

Pyrometheus is based on a symbolic representation of the combustion formulation that takes mechanism parameters from Cantera.
The symbolic representation is mapped to a user-specified target language.
For Fortran, the generated code is a module that can be easily integrated and offloaded with MFC.
The mapping process decorates the routines with OpenACC statements for GPU offloading, ensuring that the generated code does not require post-hoc modifications.
The generated code unrolls inner loops over species and reactions and prepares compile-time constants such as Arrhenius parameters.
This optimization meaningfully reduces kernel runtime~\cite{cisneros25}.
In a similar vein to case optimization of \cref{sss:metaprogramming}, Pyrometheus assigns compile-time-known bounds to the arrays in the generated code, enabling the optimization of register allocations and reducing runtime by approximately a factor of 5 for GPU kernels.

Pyrometheus is verified against Cantera, although it is thoroughly tested itself.
The testing suite is language-agnostic, so the generated Fortran code is as accurate as a C-based alternative.
The generated code is mechanism-specific, but because code generation is faster than simulation time, multiple mechanisms can be preprocessed.
For MFC, we generate code for established combustion mechanisms for hydrogen~\cite{ref:Saxena2006} and methane combustion~\cite{ref:gri30}, though we are not limited to these.

\section{Example simulations}
\label{sec:examples}

Example simulations using MFC~5.0 are included below.
Many of these could not be conducted using prior MFC releases, or would have been prohibitively expansive, as MFC~5.0 has improved distributed computation and compiler and non-NVIDIA GPU hardware support has been added with MFC~5.0.

\subsection{Shock--bubble-cloud interaction}\label{sec:example1}

The first example simulation shows the interaction between a cloud of $75$ randomly placed $\SI{3}{\milli\meter}$ bubbles and a shockwave in water.
The computational domain is $[-9D,9D]\times [0,10D] \times [0,10D]$ before grid stretching and is discretized as $(N_x, N_y, N_z) = (2000,1000,1000)$, or about 2B grid points.
The grid points are stretched away from the bubbles to avoid boundary effects via
\begin{equation*}
    x_\mathrm{stretch} = x + \frac{x}{a_x}\left[\log \left[ \cosh \left( \frac{a_x(x - x_a)}{L} \right) \right] + \log \left[ \cosh \left(\frac{a_x(x - x_b}{L}\right) \right] - 2\log\left[\cosh \left(\frac{a_x(x_b - x_a}{2L} \right) \right] \right],
\end{equation*}
where $a_x$ is the stretching parameter, $L$ is the domain length, and $x_a$ and $x_b$ control the stretching location.

\begin{figure}
    \centering
    \includegraphics[]{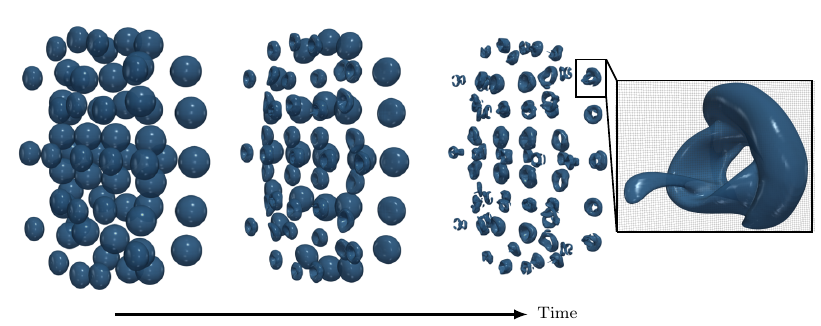}
    \caption{Shock-induced collapse of a cloud of 75 air bubbles in water.
    The $\alpha = 0.5$ contour is shown at increasing points in time from left to right.
    The magnified view shows the mesh resolution around one of the collapsing bubbles.}
    \label{fig:bubCloud}
\end{figure}

\Cref{fig:bubCloud} shows the $\alpha = 0.5$ isosurface at increasing points in time from left to right.
The detail view shows the grid resolution around one of the collapsing bubbles.
The initial condition has 100 uniform-sized grid points across each bubble diameter.
The simulation was performed using 1024~AMD~MI250X GCDs (512~AMD MI250X GPUs) on OLCF~Frontier and completed in approximately \SI{30}{\minute}.

\subsection{Shock--bubble interaction}\label{sec:example2}

\begin{figure}
    \centering
    \includegraphics[]{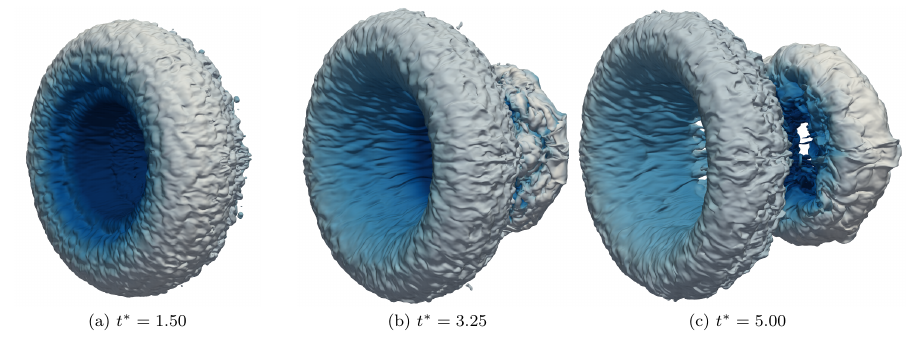}
    \caption{Shock-induced collapse of a helium bubble in air.
    The $\alpha = 0.5$ contour, representing the material interface, is shown at dimensionless times as labeled.
    The isosurface is colored by its velocity magnitude, with darker colors corresponding to higher velocities.
    The initial droplet has 640 grid cells across its diameter in each coordinate direction.}
    \label{fig:shockBubble}
\end{figure}

The second example simulation shows the interaction between a helium bubble in air impinged by a shock wave.
The computational domain has spatial extents $[0,20D]\times [-5D, 5D] \times [-5D, 5D]$ before grid stretching and is discretized with $(N_x, N_y, N_z) = (3200, 1600, 1600)$, which is over 8B grid points.
The domain boundaries are moved far away from the bubble to avoid boundary effects using grid stretching (described in \cref{sec:example1}).
\Cref{fig:shockBubble} shows the $\alpha = 0.5$ isosurface, which represents the interface between air and helium, at dimensionless times $t^* = tU/D = 1.5$, $3.25$, and $5.0$ for shock speed $U$ and bubble diameter $D$.
The isosurface is colored by velocity magnitude, with darker colors corresponding to higher velocities.
The bubble is resolved with 640~grid~cells in its initial diameter.
The simulation was performed using 144~NVIDIA~H200 GPUs in \SI{16}{\hour}.

\subsection{Taylor--Green vortex}\label{sec:example3}

\begin{figure}
    \centering
    \includegraphics[]{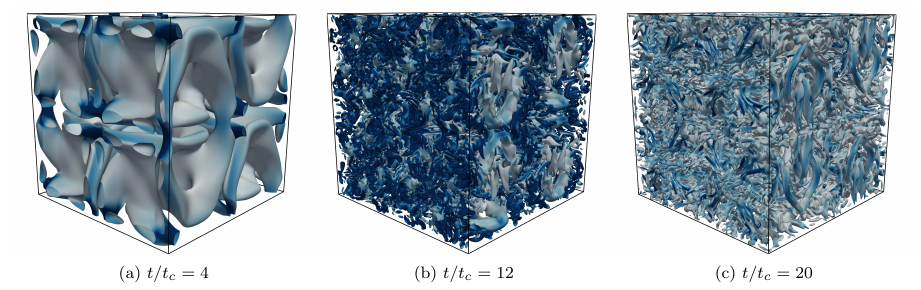}
    \caption{
        The isosurface with zero Q-criterion of a $\Rey = 1600$, $\Ma = 0.1$ Taylor--Green vortex at dimensionless times $t/t_c$ as labeled.
        The isosurface is colored by the vorticity magnitude, with darker colors corresponding to higher vorticity.
    }
    \label{fig:TGV}
\end{figure}

The third example simulation is a $\Rey = 1600$, $\Ma = 0.1$ Taylor--Green vortex.
The initial condition follows that of \citet{koen2013}.
The computational domain is a cube with side lengths of $L = 2\pi$ and is discretized with $N_x = N_y = N_z = 1600$, resulting in approximately 4 billion grid points.
\Cref{fig:TGV} shows the isosurface with zero Q-criterion colored by vorticity magnitude at dimensionless times $t/t_c = 4$, $12$, and $20,$ with darker colors corresponding to higher vorticity magnitudes.
The convective timescale $t_c$ is defined as $t_c = L/\left(c_0 \Ma\right),$ where $c_0$ is the free stream speed of sound.
The simulation was performed using 144 NVIDIA~H200 GPUs in \SI{28}{\hour}.

\subsection{Supersonic gas injection and plume dynamics}\label{sec:example4}

\begin{figure}
    \centering
    \includegraphics[width=0.5\textwidth]{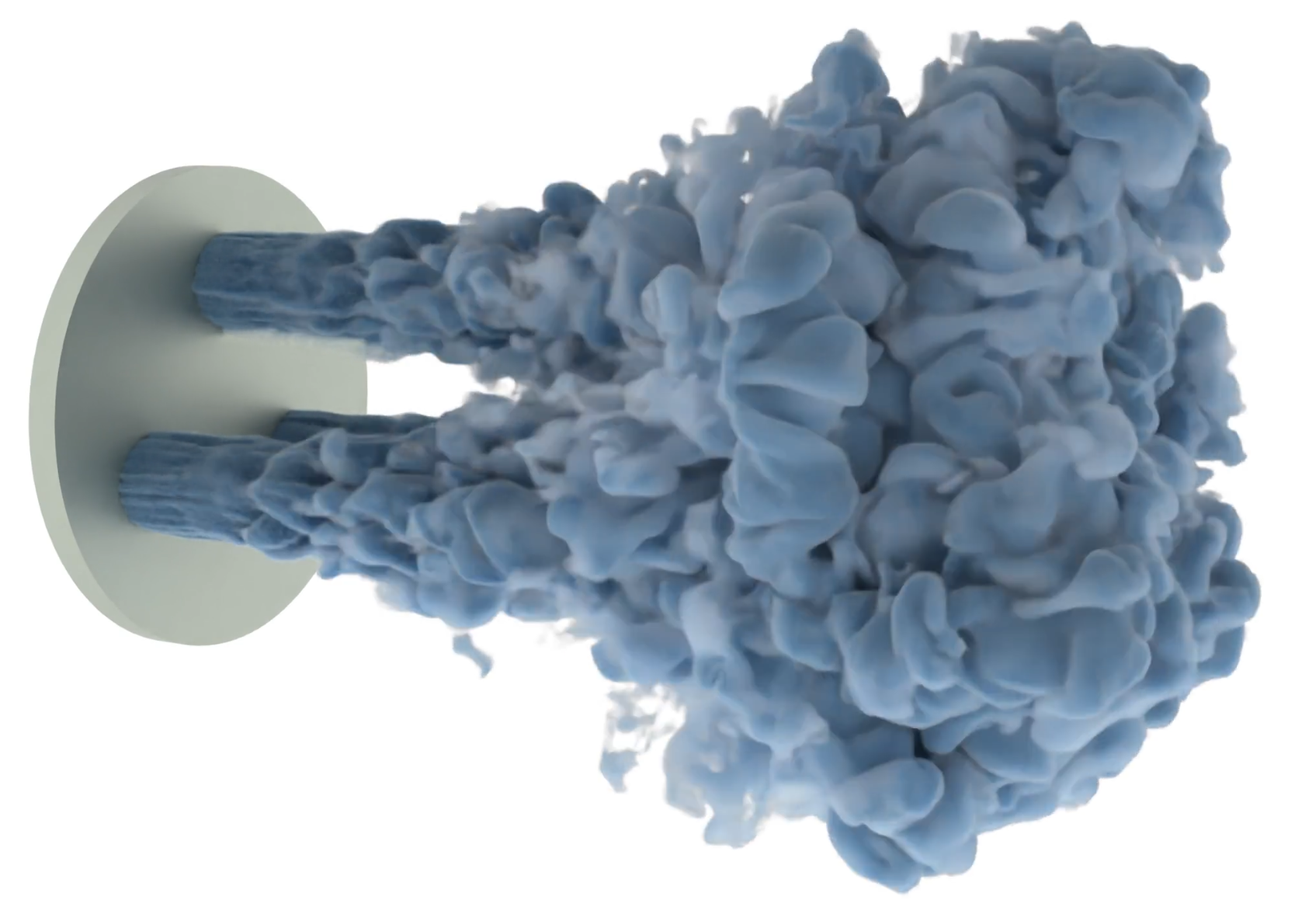}
    \caption{
        A rendering showing the injection and fluid dynamics of three $\Ma = 1.5$ jets into quiescent air.
    }
    \label{fig:jet}
\end{figure}

Three jets inject into the domain via Dirichlet boundary conditions.
The domain has size $[0,16D] \times [0,10D] \times [0,10D]$ where $D$ is the jet diameter and the first dimension is the mean flow direction (horizontal in \cref{fig:jet}).
The domain is discretized into a $600 \times 400 \times 400$ grid of cells.
The jets have an incoming velocity ($V_\mathrm{jet}$), pressure, and density that satisfy the Mach~1.5 normal shock conditions; they are organized in an equilateral triangle.
The jet Reynolds number is $\Rey = \rho V_\mathrm{jet} D/\mu = 2.5 \times 10^6$.

The simulation is executed until dimensionless time $40D/(\Ma \, c)$, where $\Ma$ is the Mach number and $c$ is the ambient speed of sound.
The last time step is visualized in \cref{fig:jet}.
The visualization shows a volume rendering of the injected fluid after it has meaningfully intruded into the domain.

\subsection{Validation and verification}\label{sec:vv}

The presented examples above are for demonstration of simulation capability and integrity.
The shock--bubble dynamics cases of \cref{sec:example1,sec:example2}, including the model and numerics, were validated at the scale of single-bubble dynamics by \citet{bryngelson21} and the cited art within.
The numerics themselves have been validated in prior works, such as \citet{johnsen2006implementation}.
The Taylor--Green vortex of \cref{sec:example3} is validated against standard turbulence statistics.
The jet case in \cref{sec:example4} was verified through a grid study, but is primarily presented as an example of computational readiness.

\section{Solvers with some shared capabilities}
\label{sec:otherSolvers}

Other open-source codes for CFD exist.
These and their relative trade-offs compared to MFC~5.0 are briefly discussed here.
URANOS-2.0 (\citet{devanna24}) is a modern Fortran code that uses OpenACC for GPU offloading to NVIDIA and AMD GPUs.
URANOS-2.0 incorporates some of MFC's shock-capturing capabilities and features compressible turbulence models.
STREAmS-2 (\citet{sathyanarayana2023}) is also a Fortran code for compressible flow, utilizing CUDA Fortran and hipFORT to offload computations to NVIDIA and AMD GPUs.
There is also an OpenMP port of STREAmS-2 that makes it portable to Intel GPUs.
AFiD-GPU of \citet{zhu2018} is similar in this context.
Neko (\citet{jansson2024}) is a particularly modern object-oriented solver that uses multiple levels of abstraction for offloading to GPUs as well as more extensive extensions for vector processors and FPGAs.
The above solvers have offloading capabilities similar to those of MFC, although their physical models and sophistication differ.
MFC offers a broad range of modeling capabilities and numerical methods for simulating multi-phase and multi-species flows. 

\section{Conclusion}
\label{sec:Conclusion}

Since MFC~3.0~\citep{bryngelson21}, the code has transitioned from a specialized research code to an exascale-capable framework for multiphysics and multiphase flows.
The implementation of physical models, numerical methods, software infrastructure, and high-performance computing tools embodies this transition.

Version~5.0 introduces six phase-change formulations for vapor--liquid systems and treatments of reacting flows.
The implementation extends to non-polytropic sub-grid bubble dynamics models capable of representing sophisticated bubble dynamic processes.
Hypo- and hyper-elastic material treatments are now available for solids under large strain rates.
A generalized surface tension framework is now included, which is both conservative and numerically robust in our studies.

Numerical advancements accompany MFC's extended feature set.
For shock-capturing, the implementation of TENO, WENO-Z, and higher-order reconstructions provides methods that are well-established to reduce the numerical dissipation errors inherent in conventional, lower-order schemes~\cite{shu1997essentially, leveque2002finite, borges2008improved, fu2016family}. 
The Strang-split particle solver with the adaptive time stepping algorithm was implemented to improve numerical efficiency for stiff sub-grid dynamics while ensuring second-order temporal accuracy~\cite{strang1968construction, hairer93solving}.
The known inaccuracy of Riemann solvers in the low-Mach number limit~\cite{guillard1999behaviour} is addressed with two implemented treatments that reduce numerical dissipation, following the approaches of Thornber et al.~\cite{thornber2008improved} and Chen et al.~\cite{chen2022anti}.
A ghost-cell immersed boundary method is now available.
This method supports complex 2D and 3D geometries that can be imported as level sets or STL files.

Software infrastructure improvements deliver performance portability across modern architectures.
Continuous integration pipelines enforce rigorous quality control through over 300 regression tests, including reproducibility checks across hardware platforms and compilers.

Exascale readiness is demonstrated through multiple performance tests.
OpenACC implementations and GPU-aware MPI throughout the codebase enable state-of-the-art strong scaling efficiency on AMD MI250X platforms.
Metaprogramming enables a speedup of nearly 10 times over the baseline.
Ideal weak scaling behavior is observed for current exascale machines, OLCF~Frontier and LLNL El~Capitan, which use AMD MI250X~GPUs and MI300A~APUs, and large-scale NVIDIA-based GPU machines like OLCF~Summit and LLNL~Lassen.

\section*{Declaration of competing interest}

The authors declare that they have no known competing financial interests or personal relationships that could have appeared to influence the work reported in this paper.

\section*{Statement of data availability}

MFC is available in perpetuity under the MIT license at \url{github.com/MFlowCode/MFC}.

\section*{Acknowledgments}

The authors gratefully acknowledge OLCF~Frontier Hackathons and Open~Hackathons, the fruitful support of numerous MFC contributors, hackathon mentors, and vendor experts, of whom there are too many to name individually.

SHB acknowledges support from the U.S. Department of Defense, Office of Naval Research under grant numbers N00014-22-1-2519 and N00014-24-1-2094, the Army Research Office under grant number W911NF-23-10324, the Department of Energy under DOE~DE-NA0003525 (Sandia~National~Labs, subcontract), the Oak Ridge Associated Universities (ORAU) Ralph E.\ Powe Junior Faculty Enhancement Award, and hardware gifts from NVIDIA and AMD.

MRJ acknowledges support from the U.S. Department of Defense under the DEPSCoR program Award No.\ FA9550-23-1-0485 (PM Dr. Timothy Bentley) and the U.S. National Science Foundation (NSF) under Grant No.\ 2232427.

AG acknowledges support from the U.S. National Science Foundation (NSF) under Grants CBET 2301721 and CBET 2301709.

Some computations were also performed on the Tioga, Tuolumne, and El~Capitan computers at Lawrence Livermore National Laboratory's Livermore Computing facility.
This research also used resources of the Oak Ridge Leadership Computing Facility at the Oak Ridge National Laboratory, which is supported by the Office of Science of the U.S.\ Department of Energy under Contract No.~DE-AC05-00OR22725 (allocation CFD154, PI Bryngelson).

This work used Delta and DeltaAI at the National Center for Supercomputing Applications and Bridges2 at the Pittsburgh Supercomputing Center through allocations PHY210084 and  PHY240200 (PI Bryngelson) from the Advanced Cyberinfrastructure Coordination Ecosystem: Services \& Support (ACCESS) program, which is supported by National Science Foundation grants \#2138259, \#2138286, \#2138307, \#2137603, and \#2138296.

\bibliography{main.bib}

@article{cisneros25,
  author = {Cisneros-Garibay, Esteban and {Le Berre}, H. and Bryngelson, Spencer H. and Freund, Jonathan B.},
  title = {{Pyrometheus: S}ymbolic abstractions for {XPU} and automatically differentiated computation of combustion kinetics and thermodynamics},
  journal = {arXiv preprint arXiv:2503.24286},
  year = {2025},
}

@article{johnsen2006implementation,
  title={Implementation of {WENO} schemes in compressible multicomponent flow problems},
  author={Johnsen, Eric and Colonius, Tim},
  journal={Journal of Computational Physics},
  volume={219},
  number={2},
  pages={715--732},
  year={2006},
  publisher={Elsevier}
}

@article{bryngelson19_whales,
  doi = {10.1121/10.0000746},
  year = {2020},
  volume = {147},
  number = {2},
  pages = {1126--1135},
  author = {Bryngelson, S. H. and Colonius, T.},
  title = {Simulation of humpback whale bubble-net feeding models},
  journal = {Journal of the Acoustical Society of America},
}

@inproceedings{elwasif,
  author = {Elwasif, W. and Bastrakov, S. and Bryngelson, S. H. and Bussmann, M. and Chandrasekaran, S. and Ciorba, F. and Clark, M. A. and Debus, A. and Godoy, W. and Hagerty, N. and Hammond, J. and Hardy, D. and Harris, J. A. and Hernandez, O. and Joo, B. and Keller, S. and Kent, P. and {Le Berre}, H. and Lebrun-Grandie, D. and MacCarthy, E. and Vergara, V. G. Melesse and Messer, B. and Miller, R. and Oral, S. and Piccinali, J.-G. and Radhakrishnan, A. and Simsek, O. and Spiga, F. and Steiniger, K. and Stephan, J. and Stone, J. E. and Trott, C. and Widera, R. and Young, J.},
  title = {Early application experiences on a modern {GPU}-accelerated {A}rm-based {HPC} platform},
  booktitle = {HPC Asia '23},
  series = {International Workshop on Arm-based HPC: Practice and Experience (IWAHPCE)},
  doi = {10.1145/3581576.3581621},
  address = {Singapore},
  year = {2023},
  pages = {35-49},
}

@article{bryngelson23,
  author = {Bryngelson, S. H. and Fox, R. O. and Colonius, T.},
  title = {Conditional moment methods for polydisperse cavitating flows},
  journal = {Journal of Computational Physics},
  year = {2023},
  volume = {477},
  pages = {111917}
}

@misc{IntelAICloud,
    title        = {{Intel Developer Cloud: AI-O}ptimized Supercomputing Platform},
    author       = {{Intel Corporation}},
    year         = 2024,
    note         = {Accessed: 2025-02-14},
    howpublished = {\url{https://console.cloud.intel.com/docs/index.html}}
}

@misc{PSCBridges,
    title        = {Bridges: {A HPC} Resource for Accelerating Computational Research},
    author       = {{Pittsburgh Supercomputing Center (PSC)}},
    year         = 2015,
    note         = {Funded by the National Science Foundation under Grant No. OCI-1445606. Retired in 2021.},
    howpublished = {\url{https://www.psc.edu/bridges}}
}

@misc{Stampede3,
    title        = {{Stampede3: A}dvanced Supercomputing for Open Science Research},
    author       = {{Texas Advanced Computing Center (TACC)}},
    year         = 2024,
    note         = {Funded by the National Science Foundation under Award No. 2320757.},
    howpublished = {\url{https://www.tacc.utexas.edu}}
}

@article{Expanse2021,
    title        = {Expanse: {C}omputing without Boundaries: Architecture, Deployment, and Early Operations Experiences of a Supercomputer Designed for the Rapid Evolution in Science and Engineering},
    author       = {Strande, Shawn and Cai, Haisong and Tatineni, Mahidhar and Pfeiffer, Wayne and Irving, Christopher and Majumdar, Amit and Mishin, Dmitry and Sinkovits, Robert and Norman, Mike and Wolter, Nicole and Cooper, Trevor and Altintas, Ilkay and Kandes, Marty and Perez, Ismael and Shantharam, Manu and Thomas, Mary and Sivagnanam, Subhashini and Hutton, Thomas},
    year         = 2021,
    journal      = {Practice and Experience in Advanced Research Computing (PEARC '21)},
    pages        = {1--8},
    doi          = {10.1145/3437359.3465588}
}

@misc{Comet2015,
    title        = {{Comet: A} Petascale Supercomputer for the Long Tail of Science},
    author       = {{San Diego Supercomputer Center (SDSC)}},
    year         = 2015,
    note         = {Accessed: 2025-02-14},
    howpublished = {\url{https://www.sdsc.edu/News\%20Items/PR20170201_Comet_10k.html}}
}

@misc{Stampede1,
    title        = {{Stampede: H}igh-Performance Computing Resource at {TACC}},
    author       = {{Texas Advanced Computing Center (TACC)}},
    year         = 2013,
    note         = {Supported by the National Science Foundation under Grant No. OCI-1134872.}
}

@article{Stampede2,
    title        = {{Stampede2: T}he Evolution of an {XSEDE} Supercomputer},
    author       = {Nystrom, N. A. and  {et al.}},
    year         = 2017,
    journal      = {Proceedings of the Practice and Experience in Advanced Research Computing (PEARC '17)},
    doi          = {10.1145/3093338.3093385}
}

@misc{DeltaAI,
    title        = {{DeltaAI: E}xpanding {AI}-Focused Computing Capacity at {NCSA}},
    author       = {{National Center for Supercomputing Applications (NCSA)}},
    year         = 2023,
    note         = {Accessed: 2025-02-14},
    howpublished = {\url{https://deltaai.ncsa.illinois.edu}}
}

@misc{Delta,
    title        = {{Delta: A}dvanced Computing Resource for Science and Engineering},
    author       = {{National Center for Supercomputing Applications (NCSA)}},
    year         = 2022,
    note         = {Accessed: 2025-02-14},
    howpublished = {\url{https://delta.ncsa.illinois.edu}}
}

@inproceedings{Anvil2022,
    title        = {{Anvil} -- {S}ystem Architecture and Experiences from Deployment and Early User Operations},
    author       = {X. Carol Song and Preston Smith and {others}},
    year         = 2022,
    booktitle    = {Practice and Experience in Advanced Research Computing (PEARC '22)},
    publisher    = {Association for Computing Machinery},
    address      = {New York, NY, USA},
    pages        = 23,
    doi          = {10.1145/3491418.3530766},
}

@article{Bridges2,
    title        = {{Bridges-2: T}ransforming Research with Advanced Computational Capabilities},
    author       = {Brown, S. T. and Buitrago, P. and Hanna, E. and Sanielevici, S. and Scibek, R.},
    year         = 2021,
    journal      = {Pittsburgh Supercomputing Center}
}

@misc{access-ci,
    title        = {{ACCESS: A}dvanced Cyberinfrastructure Coordination Ecosystem},
    author       = {{ACCESS (Advanced Cyberinfrastructure Coordination Ecosystem: Services and Support)}},
    year         = 2025,
    note         = {Accessed: 2025-02-14},
    howpublished = {\url{https://access-ci.org}}
}

@article{Towns2014-po,
    title        = {{XSEDE: A}ccelerating Scientific Discovery},
    author       = {Towns, J and Cockerill, T and Dahan, M and Foster, I and Gaither, K and Grimshaw, A and Hazlewood, V and Lathrop, S and Lifka, D and Peterson, G D and Roskies, R and Scott, J R and Wilkins-Diehr, N},
    year         = 2014,
    journal      = {Computing in Science \& Engineering},
    volume       = 16,
    number       = 5,
    pages        = {62--74},
    issn         = {1521-9615}
}

@misc{spechpc,
    title        = {{SPEChpc}},
    author       = {{SPEC}},
    year         = 2025,
    note         = {Accessed: 2025-02-14},
    howpublished = {\url{https://www.spec.org/products}}
}

@article{bryngelson21,
    title        = {{MFC: A}n open-source high-order multi-component, multi-phase, and multi-scale compressible flow solver},
    author       = {Bryngelson, S. H. and Schmidmayer, K. and Coralic, V. and Maeda, K. and Meng, J. and Colonius, T.},
    year         = 2021,
    journal      = {Computer Physics Communications},
    volume       = 266,
    pages        = 107396
}

@article{Coralic2014,
    title        = {Finite-volume {WENO} scheme for viscous compressible multicomponent flows},
    author       = {Vedran Coralic and Tim Colonius},
    year         = 2014,
    journal      = {Journal of Computational Physics},
    volume       = 274,
    pages        = {95--121}
}

@article{schmidmayer2017,
    title        = {A model and numerical method for compressible flows with capillary effects},
    author       = {Kevin Schmidmayer and Fabien Petitpas and Eric Daniel and Nicolas Favrie and Sergey Gavrilyuk},
    year         = 2017,
    journal      = {Journal of Computational Physics},
    volume       = 334,
    pages        = {468--496},
    doi          = {https://doi.org/10.1016/j.jcp.2017.01.001},
    issn         = {0021-9991},
}

@article{Gottlieb1998,
    title        = {Total variation diminishing {R}unge--{K}utta schemes},
    author       = {Gottlieb, Sigal and Shu, Chi-Wang},
    year         = 1998,
    journal      = {Mathematics of Computation},
    volume       = 67,
    number       = 221,
    pages        = {73--85}
}

@inbook{Toro1997,
    title        = {The {HLL} and {HLLC} {R}iemann Solvers},
    author       = {Toro, Eleuterio F.},
    year         = 1997,
    booktitle    = {Riemann Solvers and Numerical Methods for Fluid Dynamics: A Practical Introduction},
    publisher    = {Springer Berlin Heidelberg},
    address      = {Berlin, Heidelberg},
    pages        = {293--311},
    isbn         = {978-3-662-03490-3}
}

@article{thompson1990,
    title        = {Time-dependent boundary conditions for hyperbolic systems, {II}},
    author       = {Kevin W Thompson},
    year         = 1990,
    journal      = {Journal of Computational Physics},
    volume       = 89,
    number       = 2,
    pages        = {439--461},
    issn         = {0021-9991}
}

@article{andrianov2004,
    title        = {The {R}iemann problem for the {B}aer--{N}unziato two-phase flow model},
    author       = {Nikolai Andrianov and Gerald Warnecke},
    year         = 2004,
    journal      = {Journal of Computational Physics},
    volume       = 195,
    number       = 2,
    pages        = {434--464},
    issn         = {0021-9991}
}

@article{allaire2002,
    title        = {A five-equation model for the simulation of interfaces between compressible fluids},
    author       = {Grégoire Allaire and Sébastien Clerc and Samuel Kokh},
    year         = 2002,
    journal      = {Journal of Computational Physics},
    volume       = 181,
    number       = 2,
    pages        = {577--616},
    issn         = {0021-9991}
}

@article{menikoff1989,
    title        = {The {R}iemann problem for fluid flow of real materials},
    author       = {Menikoff, Ralph and Plohr, Bradley J.},
    year         = 1989,
    month        = {Jan},
    journal      = {Reviews of Modern Physics},
    publisher    = {American Physical Society},
    volume       = 61,
    pages        = {75--130},
    issue        = 1
}

@article{kapila2001,
    title        = {Two-phase modeling of deflagration-to-detonation transition in granular materials: {R}educed equations},
    author       = {Kapila, A. K. and Menikoff, R. and Bdzil, J. B. and Son, S. F. and Stewart, D. S.},
    year         = 2001,
    month        = 10,
    journal      = {Physics of Fluids},
    volume       = 13,
    number       = 10,
    pages        = {3002--3024},
    issn         = {1070-6631}
}

@article{schmidmayer2020,
    title        = {An assessment of multicomponent flow models and interface capturing schemes for spherical bubble dynamics},
    author       = {Schmidmayer, K. and Bryngelson, S. H. and Colonius, T.},
    year         = 2020,
    journal      = {Journal of Computational Physics},
    volume       = 402,
    pages        = 109080,
    doi          = {10.1016/j.jcp.2019.109080},
}

@article{saurel2008,
   author = {Richard Saurel and Fabien Petitpas and Remi Abgrall},
   doi = {10.1017/S0022112008002061},
   journal = {Journal of Fluid Mechanics},
   pages = {313-350},
   publisher = {Cambridge University Press},
   title = {Modelling phase transition in metastable liquids: {A}pplication to cavitating and flashing flows},
   volume = {607},
   year = {2008}
}

@article{saurel2009,
    title        = {Simple and efficient relaxation methods for interfaces separating compressible fluids, cavitating flows and shocks in multiphase mixtures},
    author       = {Richard Saurel and Fabien Petitpas and Ray A. Berry},
    year         = 2009,
    journal      = {Journal of Computational Physics},
    volume       = 228,
    number       = 5,
    pages        = {1678--1712},
    issn         = {0021-9991}
}

@article{hendrick2005,
    title        = {Mapped weighted essentially non-oscillatory schemes: {A}chieving optimal order near critical points},
    author       = {Andrew K. Henrick and Tariq D. Aslam and Joseph M. Powers},
    year         = 2005,
    journal      = {Journal of Computational Physics},
    volume       = 207,
    number       = 2,
    pages        = {542--567},
    issn         = {0021-9991}
}

@article{Maeda2017,
    title        = {A source term approach for generation of one-way acoustic waves in the {E}uler and {N}avier--{S}tokes equations},
    author       = {Kazuki Maeda and Tim Colonius},
    year         = 2017,
    journal      = {Wave Motion},
    volume       = 75,
    pages        = {36--49},
    issn         = {0165-2125}
}

@inproceedings{Wienke2012,
    title        = {{OpenACC} --- {F}irst Experiences with Real-World Applications},
    author       = {Wienke, Sandra and Springer, Paul and Terboven, Christian and an Mey, Dieter},
    year         = 2012,
    booktitle    = {Euro-Par 2012 Parallel Processing},
    publisher    = {Springer Berlin Heidelberg},
    address      = {Berlin, Heidelberg},
    pages        = {859--870},
    isbn         = {978-3-642-32820-6},
    editor       = {Kaklamanis, Christos and Papatheodorou, Theodore and Spirakis, Paul G.}
}

@phdthesis{johnson2008,
    title        = {Numerical Simulations of Non-Spherical Bubble Collapse with Applications to Shockwave Lithotripsy},
    author       = {Eric Johnsen},
    year         = 2008,
    school       = {California Institute of Technology}
}

@phdthesis{meng2016,
    title        = {Numerical simulations of droplet aerobreakup},
    author       = {Jomela C. Meng},
    year         = 2016,
    school       = {California Institute of Technology}
}

@article{mohseni2000,
    title        = {Numerical Treatment of Polar Coordinate Singularities},
    author       = {Kamran Mohseni and Tim Colonius},
    year         = 2000,
    journal      = {Journal of Computational Physics},
    volume       = 157,
    number       = 2,
    pages        = {787--795},
    issn         = {0021-9991}
}

@article{tseng2003ghost,
    title        = {A ghost-cell immersed boundary method for flow in complex geometry},
    author       = {Tseng, Yu-Heng and Ferziger, Joel H},
    year         = 2003,
    journal      = {Journal of Computational Physics},
    publisher    = {Elsevier},
    volume       = 192,
    number       = 2,
    pages        = {593--623}
}

@article{flaatten2011solutions,
    title        = {On solutions to equilibrium problems for systems of stiffened gases},
    author       = {Fl{\aa}tten, Tore and Morin, Alexandre and Munkejord, Svend Tollak},
    year         = 2011,
    journal      = {SIAM Journal on Applied Mathematics},
    publisher    = {SIAM},
    volume       = 71,
    number       = 1,
    pages        = {41--67}
}

@article{Strang1968,
    title        = {On the Construction and Comparison of Difference Schemes},
    author       = {Gilbert Strang},
    year         = 1968,
    month        = 9,
    journal      = {SIAM Journal on Numerical Analysis},
    volume       = 5,
    pages        = {506--517},
    doi          = {10.1137/0705041},
    issn         = {0036-1429},
    issue        = 3
}

@article{Zein2010,
    title        = {Modeling phase transition for compressible two-phase flows applied to metastable liquids},
    author       = {Ali Zein and Maren Hantke and Gerald Warnecke},
    year         = 2010,
    month        = 4,
    journal      = {Journal of Computational Physics},
    volume       = 229,
    pages        = {2964--2998},
    doi          = {10.1016/j.jcp.2009.12.026},
    issn         = {00219991},
    issue        = 8
}

@article{radhakrishnan24,
    title        = {Method for portable, scalable, and performant {GPU}-accelerated simulation of multiphase compressible flow},
    author       = {Radhakrishnan, A. and {Le Berre}, H. and Wilfong, B. and Spratt, J.-S. and {Rodriguez Jr.}, M. and Colonius, T. and Bryngelson, S. H.},
    year         = 2024,
    journal      = {Computer Physics Communications},
    volume       = 302,
    pages        = 109238,
    doi          = {10.1016/j.cpc.2024.109238}
}

@article{zhang1994,
    title        = {Ensemble phase‐averaged equations for bubbly flows},
    author       = {Zhang, D. Z. and Prosperetti, A.},
    year         = 1994,
    month        = {09},
    journal      = {Physics of Fluids},
    volume       = 6,
    number       = 9,
    pages        = {2956--2970},
    doi          = {10.1063/1.868122},
    issn         = {1070-6631}
}

@article{bryngelson19_IJMF,
    title        = {A quantitative comparison of phase-averaged models for bubbly, cavitating flows},
    author       = {Bryngelson, S. H. and Schmidmayer, K. and Colonius, T.},
    year         = 2019,
    journal      = {International Journal of Multiphase Flow},
    volume       = 115,
    pages        = {137--143},
    doi          = {10.1016/j.ijmultiphaseflow.2019.03.028},
}

@article{borges2008improved,
    title        = {An improved weighted essentially non-oscillatory scheme for hyperbolic conservation laws},
    author       = {Borges, Rafael and Carmona, Monique and Costa, Bruno and Don, Wai Sun},
    year         = 2008,
    journal      = {Journal of Computational Physics},
    publisher    = {Elsevier},
    volume       = 227,
    number       = 6,
    pages        = {3191--3211}
}

@article{fu2016family,
    title        = {A family of high-order targeted {ENO} schemes for compressible-fluid simulations},
    author       = {Fu, Lin and Hu, Xiangyu Y and Adams, Nikolaus A},
    year         = 2016,
    journal      = {Journal of Computational Physics},
    publisher    = {Elsevier},
    volume       = 305,
    pages        = {333--359}
}

@techreport{shu1997essentially,
    title        = {Essentially Non-Oscillatory and Weighted Essentially Non-Oscillatory Schemes for Hyperbolic Conservation Laws},
    author       = {Shu, Chi-Wang},
    year         = 1997,
    month        = {November},
    address      = {Hampton, VA},
    number       = {NASA/CR-97-206253},
    institution  = {Institute for Computer Applications in Science and Engineering, NASA Langley Research Center},
    type         = {ICASE Report No. 97-65}
}

@article{Guilaume2015,
    title        = {{Viscoelastic models with consistent hypoelasticity for fluids undergoing finite deformations}},
    author       = {Altmeyer, Guillaume and Rouhaud, Emmanuelle and Panicaud, Benoit and Roos, Arjen and Kerner, Richard and Wang, Mingchuan},
    year         = 2015,
    month        = aug,
    journal      = {Mechanics of Time-Dependent Materials},
    volume       = 19,
    number       = 3,
    pages        = {375--395},
    doi          = {10.1007/s11043-015-9269-5}
}

@article{guillaume2016,
    title        = {{Viscoelasticity behavior for finite deformations, using a consistent hypoelastic model based on Rivlin materials}},
    author       = {Altmeyer, Guillaume and Panicaud, Benoit and Rouhaud, Emmanuelle and Wang, Mingchuan and Roos, Arjen and Kerner, Richard},
    year         = 2016,
    month        = nov,
    journal      = {Continuum Mechanics and Thermodynamics},
    volume       = 28,
    number       = 6,
    pages        = {1741--1758}
}

@article{rodriguez2019,
    title        = {A high-order accurate five-equations compressible multiphase approach for viscoelastic fluids and solids with relaxation and elasticity},
    author       = {Mauro Rodriguez and Eric Johnsen},
    year         = 2019,
    journal      = {Journal of Computational Physics},
    volume       = 379,
    pages        = {70--90},
    issn         = {0021-9991}
}

@phdthesis{spratt2024,
    title        = {Numerical Simulations of Cavitating Bubbles in Elastic and Viscoelastic Materials for Biomedical Applications},
    author       = {Spratt, Jean-S{\'e}bastien},
    year         = 2024,
    school       = {California Institute of Technology}
}

@article{maeda2018eulerian,
    title        = {Eulerian--{La}grangian method for simulation of cloud cavitation},
    author       = {Maeda, Kazuki and Colonius, Tim},
    year         = 2018,
    journal      = {Journal of Computational Physics},
    publisher    = {Elsevier},
    volume       = 371,
    pages        = {994--1017}
}

@article{fuster2011modelling,
    title        = {Modelling bubble clusters in compressible liquids},
    author       = {Fuster, Daniel and Colonius, Tim},
    year         = 2011,
    journal      = {Journal of Fluid Mechanics},
    publisher    = {Cambridge University Press},
    volume       = 688,
    pages        = {352--389}
}

@article{menikoff89,
    title        = {The {R}iemann problem for fluid-flow of real materials},
    author       = {R. Menikoff and B. J. Plohr},
    year         = 1989,
    journal      = {Reviews of Modern Physics},
    volume       = 61,
    number       = 1,
    pages        = {75--130}
}

@article{plesset77,
    title        = {Bubble dynamics and cavitation},
    author       = {Plesset, Milton S and Prosperetti, Andrea},
    year         = 1977,
    journal      = {Annual Review of Fluid Mechanics.},
    volume       = 9,
    number       = 1,
    pages        = {145--185}
}

@article{ando09,
    title        = {Improvement of acoustic theory of ultrasonic waves in dilute bubbly liquids},
    author       = {Ando, K. and Colonius, T. and Brennen, C. E.},
    year         = 2009,
    journal      = {Journal of the Acoustical Society of America},
    volume       = 126,
    number       = 3,
    pages        = {El69-El74}
}

@article{Bryngelson2020,
    title        = {A {G}aussian moment method and its augmentation via {LSTM} recurrent neural networks for the statistics of cavitating bubble populations},
    author       = {Bryngelson, Spencer H. and Charalampopoulos, Alexis and Sapsis, Themistoklis P. and Colonius, Tim},
    year         = 2020,
    month        = {jun},
    journal      = {International Journal of Multiphase Flow},
    volume       = 127,
    pages        = 103262,
    issn         = {03019322}
}

@article{Fox2018,
    title        = {{Conditional hyperbolic quadrature method of moments for kinetic equations}},
    author       = {Fox, Rodney O. and Laurent, Fr{\'{e}}d{\'{e}}rique and Vi{\'{e}}, Aymeric},
    year         = 2018,
    month        = {jul},
    journal      = {Journal of Computational Physics},
    volume       = 365,
    pages        = {269--293},
    issn         = {00219991}
}

@article{pirozzoli2013generalized,
    title        = {Generalized characteristic relaxation boundary conditions for unsteady compressible flow simulations},
    author       = {Pirozzoli, Sergio and Colonius, Tim},
    year         = 2013,
    journal      = {Journal of Computational Physics},
    volume       = 248,
    pages        = {109--126}
}

@article{strang1968construction,
    title        = {On the construction and comparison of difference schemes},
    author       = {Strang, Gilbert},
    year         = 1968,
    journal      = {SIAM Journal on Numerical Analysis},
    volume       = 5,
    number       = 3,
    pages        = {506--517}
}

@book{hairer93solving,
    title        = {Solving ordinary differential equations {I. N}onstiff problems},
    author       = {Hairer, E. and Nørsett, S. P. and Wanner, G.},
    year         = 1993,
    publisher    = {Springer}
}

@article{guillard1999behaviour,
    title        = {On the behaviour of upwind schemes in the low {M}ach number limit},
    author       = {Guillard, Herv{\'e} and Viozat, C{\'e}cile},
    year         = 1999,
    journal      = {Computers \& Fluids},
    volume       = 28,
    number       = 1,
    pages        = {63--86}
}

@article{thornber2008improved,
    title        = {An improved reconstruction method for compressible flows with low Mach number features},
    author       = {Thornber, Ben and Mosedale, Andrew and Drikakis, Dimitris and Youngs, David and Williams, Robin JR},
    year         = 2008,
    journal      = {Journal of Computational Physics},
    volume       = 227,
    number       = 10,
    pages        = {4873--4894}
}

@article{chen2022anti,
    title        = {Anti-dissipation pressure correction under low {M}ach numbers for Godunov-type schemes},
    author       = {Chen, Shu-Sheng and Li, Jin-Ping and Li, Zheng and Yuan, Wu and Gao, Zheng-Hong},
    year         = 2022,
    journal      = {Journal of Computational Physics},
    volume       = 456,
    pages        = 111027
}

@phdthesis{weiler1986topological,
    title        = {Topological structures for geometric modeling (Boundary representation, manifold, radial edge structure)},
    author       = {Weiler, Kevin J},
    year         = 1986,
    school       = {Rensselaer Polytechnic Institute}
}

@article{devanna24,
    title        = {{URANOS-2.0: I}mproved performance, enhanced portability, and model extension towards exascale computing of high-speed engineering flows},
    author       = {Francesco {De Vanna} and Giacomo Baldan},
    year         = 2024,
    journal      = {Computer Physics Communications},
    volume       = 303,
    pages        = 109285,
    issn         = {0010-4655}
}

@article{sathyanarayana2023,
    title        = {High-speed turbulent flows towards the exascale: {STREAmS-2} porting and performance},
    author       = {Srikanth Sathyanarayana and Matteo Bernardini and Davide Modesti and Sergio Pirozzoli and Francesco Salvadore},
    year         = 2023,
    journal = {arXiv preprint arXiv 2304.05494},
}

@article{zhu2018,
    title        = {{AFiD-GPU: A} versatile {N}avier--{S}tokes solver for wall-bounded turbulent flows on GPU clusters},
    author       = {Xiaojue Zhu and Everett Phillips and Vamsi Spandan and John Donners and Gregory Ruetsch and Joshua Romero and Rodolfo Ostilla-Mónico and Yantao Yang and Detlef Lohse and Roberto Verzicco and Massimiliano Fatica and Richard J.A.M. Stevens},
    year         = 2018,
    journal      = {Computer Physics Communications},
    volume       = 229,
    pages        = {199--210},
    issn         = {0010-4655}
}

@article{jansson2024,
    title        = {{Neko: A} modern, portable, and scalable framework for high-fidelity computational fluid dynamics},
    author       = {Niclas Jansson and Martin Karp and Artur Podobas and Stefano Markidis and Philipp Schlatter},
    year         = 2024,
    journal      = {Computers \& Fluids},
    volume       = 275,
    pages        = 106243,
    issn         = {0045-7930}
}

@article{kamrin2012,
    title        = {Reference map technique for finite-strain elasticity and fluid–solid interaction},
    author       = {Ken Kamrin and Chris H. Rycroft and Jean-Christophe Nave},
    year         = 2012,
    journal      = {Journal of the Mechanics and Physics of Solids},
    volume       = 60,
    number       = 11,
    pages        = {1952--1969},
    issn         = {0022-5096}
}

@inproceedings{ctr24_mrj-mcb,
    title        = {High-fidelity numerical simulations of inertial microbubble collapse at a gel-water interface with finite elasticity and phase change},
    author       = {Mirelys Carcana Barbosa and Mauro Rodriguez Jr. and Jin Yang},
    year         = 2024,
    booktitle    = {Center for Turbulence Research, Proceedings of the Summer Program},
    pages = {349--359},
}

@misc{ref:gri30,
    author = {Gregory P. Smith and David M. Golden and Michael Frenklach and Nigel W. Moriarty and Boris Eiteneer and Mikhail Goldenberg and C. Thomas Bowman and Ronald K. Hanson and Soonho Song and William C. Gardiner and Vitali V. Lissianski and Zhiwei Qin},
    title        = {{GRI-Mech 3.0}},
    year         = 1999,
    howpublished = {\url{http://www.me.berkley.edu/gri_mech/}}
}

@article{ref:Saxena2006,
    title        = {Testing a small detailed chemical-kinetic mechanism for the combustion of hydrogen and carbon monoxide},
    author       = {Priyank Saxena and Forman A. Williams},
    year         = 2006,
    month        = 4,
    journal      = {Combustion and Flame},
    volume       = 145,
    pages        = {316--323},
    issn         = {00102180},
    issue        = {1-2}
}

@article{ref:cisneros2022,
    title        = {Flow and Combustion in a Supersonic Cavity Flameholder},
    author       = {Esteban Cisneros-Garibay and Carlos Pantano and Jonathan B Freund},
    year         = 2022,
    journal      = {AIAA Journal},
    pages        = {1--12},
    issn         = {0001-1452}
}

@article{ref:alkhateeb2009,
    title        = {One-dimensional slow invariant manifolds for spatially homogenous reactive systems},
    author       = {Ashraf N. Al-Khateeb and Joseph M. Powers and Samuel Paolucci and Andrew J. Sommese and Jeffrey A. Diller and Jonathan D. Hauenstein and Joshua D. Mengers},
    year         = 2009,
    month        = 7,
    journal      = {The Journal of Chemical Physics},
    volume       = 131,
    issn         = {0021-9606},
    issue        = 2
}

@book{ref:mcbride2002,
    title        = {{NASA G}lenn Coefficients for Calculating Thermodynamic Properties of Individual Species},
    author       = {Bonnie J McBride},
    year         = 2002,
    publisher    = {National Aeronautics and Space Administration, John H. Glenn Research Center}
}

@book{ref:Gardiner1984,
    title        = {Combustion Chemistry},
    author       = {W. Gardiner},
    year         = 1984,
    publisher    = {Springer New York},
    isbn         = {978-1-4684-0188-2},
    city         = {New York, NY},
}

@article{ref:Zeng2002,
    title        = {Review on Chemical Reactions of Burning Poly(methyl methacrylate) {PMMA}},
    author       = {W. R. Zeng and S. F. Li and W. K. Chow},
    year         = 2002,
    month        = 9,
    journal      = {Journal of Fire Sciences},
    volume       = 20,
    pages        = {401--433},
    issn         = {0734-9041},
    issue        = 5
}

@article{ref:Ren2021,
    title        = {Theoretical Single-Droplet Model for Particle Formation in Flame Spray Pyrolysis},
    author       = {Yihua Ren and Jinzhi Cai and Heinz Pitsch},
    year         = 2021,
    month        = 1,
    journal      = {Energy \& Fuels},
    volume       = 35,
    pages        = {1750--1759},
    issn         = {0887-0624},
    issue        = 2
}

@article{ref:Aier2020,
    title        = {Coupled population balance and large eddy simulation model for polydisperse droplet evolution in a turbulent round jet},
    author       = {Aditya Aiyer and Charles Meneveau},
    year         = 2020,
    month        = 11,
    journal      = {Physical Review Fluids},
    volume       = 5,
    pages        = 114305,
    issn         = {2469-990X},
    issue        = 11
}

@inproceedings{koen2013,
    title        = {{TestCase C3.5 - DNS} of the transition of the {T}aylor--{G}reen vortex},
    author       = {Hillewaert, Koen},
    year         = 2013,
    month        = {05},
    booktitle    = {1st International Workshop on High Order CFD Methods},
    pages = {1--5},
}

@misc{fypp,
	author = {Aradi, B},
	title = {{Fypp}: {P}ython-powered {F}ortran metaprogramming},
	url = {https://github.com/aradi/fypp},
    note = {{GitHub} Repository},
	year = {2025},
}

@article{martinez2014,
    title        = {A detailed verification procedure for compressible reactive multicomponent {N}avier--{S}tokes solvers},
    author       = {Pedro José {Martínez-Ferrer} and Romain Buttay and Guillaume Lehnasch and Arnaud Mura},
    year         = 2014,
    journal      = {Computers \& Fluids},
    volume       = 89,
    pages        = {88--110},
    issn         = {0045-7930}
}

@misc{cuTENSOR,
    author       = {{NVIDIA Corporation}},
    title        = {{cuTENSOR}: {A} High-Performance {CUDA} Library For Tensor Primitives},
    url          = {https://developer.nvidia.com/cutensor},
    year         = {2019}
}

@misc{cuFFT,
    author       = {{NVIDIA Corporation}},
    title        = {{cuFFT}: {H}igh-Performance {CUDA} {FFT} Library},
    url          = {https://developer.nvidia.com/cufft},
    year         = {2007}
}

@misc{hipBLAS,
    author       = {{A}dvanced {M}icro {D}evices {I}nc.},
    title        = {{hipBLAS}: {H}igh-Performance Basic Linear Algebra Subprograms Library},
    url          = {https://github.com/ROCmSoftwarePlatform/hipBLAS},
    note         = {{Github} Repository},
    year         = {2020}
}

@misc{hipFFT,
    author       = {{A}dvanced Micro Devices Inc.},
    title        = {{hipFFT}: {H}igh-Performance Fast {F}ourier Transform Library},
    url          = {https://github.com/ROCmSoftwarePlatform/hipFFT},
    note         = {{Github} Repository},
    year         = {2020}
}

@article{FFTW,
  author = {Matteo Frigo and Steven G. Johnson},
  title = {The design and implementation of {FFTW3}},
  journal = {Proceedings of the IEEE},
  volume = {93},
  number = {2},
  year = {2005},
  pages = {216--231},
}

@InProceedings{OpenACC,
    author="Wienke, Sandra
    and Springer, Paul
    and Terboven, Christian
    and an Mey, Dieter",
    editor="Kaklamanis, Christos
    and Papatheodorou, Theodore
    and Spirakis, Paul G.",
    title="OpenACC --- {F}irst Experiences with Real-World Applications",
    booktitle="Euro-Par 2012 Parallel Processing",
    year="2012",
    publisher="Springer Berlin Heidelberg",
    address="Berlin, Heidelberg",
    pages="859--870",
}

@article{mignone2006hllc,
  title={An {HLLC} Riemann solver for relativistic flows--{II. Magnetohydrodynamics}},
  author={Mignone, Andrea and Bodo, G},
  journal={Monthly Notices of the Royal Astronomical Society},
  volume={368},
  number={3},
  pages={1040--1054},
  year={2006},
  publisher={The Royal Astronomical Society}
}

@article{miyoshi2005multi,
  title={A multi-state {HLL} approximate {Riemann} solver for ideal magnetohydrodynamics},
  author={Miyoshi, Takahiro and Kusano, Kanya},
  journal={Journal of Computational Physics},
  volume={208},
  number={1},
  pages={315--344},
  year={2005},
  publisher={Elsevier}
}

@article{dai1994extension,
  title={Extension of the piecewise parabolic method to multidimensional ideal magnetohydrodynamics},
  author={Dai, Wenlong and Woodward, Paul R},
  journal={Journal of Computational Physics},
  volume={115},
  number={2},
  pages={485--514},
  year={1994},
  publisher={Elsevier}
}

@book{leveque2002finite,
  title={Finite volume methods for hyperbolic problems},
  author={LeVeque, Randall J},
  volume={31},
  year={2002},
  publisher={Cambridge university press}
}

@article{Giles1990,
  title        = {Nonreflecting boundary conditions for Euler equation calculations},
  author       = {Giles, Michael B.},
  journal      = {AIAA Journal},
  volume       = {28},
  number       = {12},
  pages        = {2050--2058},
  year         = {1990},
  doi          = {10.2514/3.10430}
}

@article{LodatoDomingoVervisch2008,
  title        = {Three-dimensional non-reflecting boundary conditions for direct and large-eddy simulation of compressible flows},
  author       = {Lodato, Guido and Domingo, Pierre and Vervisch, Luc},
  journal      = {Journal of Computational Physics},
  volume       = {227},
  number       = {10},
  pages        = {5105--5143},
  year         = {2008},
  doi          = {10.1016/j.jcp.2008.01.027}
}

@article{TamDong1996,
  title        = {Radiation and outflow boundary conditions for direct computation of acoustic and flow disturbances in a nonuniform mean flow},
  author       = {Tam, Christopher K. W. and Dong, Zhengping},
  journal      = {Journal of Computational Physics},
  volume       = {129},
  number       = {1},
  pages        = {164--180},
  year         = {1996},
  doi          = {10.1006/jcph.1996.0237}
}

@article{Jaensch2016,
  title        = {Time-domain impedance boundary conditions for compressible viscous flow},
  author       = {Jaensch, Stefan and Sovardi, Carlo and Polifke, Wolfgang},
  journal      = {Journal of Computational Physics},
  volume       = {314},
  pages        = {145--159},
  year         = {2016},
  doi          = {10.1016/j.jcp.2016.03.026}
}

@article{simoesmoreira1999evaporation,
   author = {J. R. Simões-Moreira and J. E. Shepherd},
   doi = {10.1017/S0022112098003796},
   issn = {0022-1120},
   journal = {Journal of Fluid Mechanics},
   month = {3},
   pages = {63-86},
   title = {Evaporation waves in superheated dodecane},
   volume = {382},
   year = {1999},
}

@article{saurel2005modelling,
   author = {O. Le Métayer and J. Massoni and R. Saurel},
   doi = {10.1016/j.jcp.2004.11.021},
   issn = {00219991},
   issue = {2},
   journal = {Journal of Computational Physics},
   month = {5},
   pages = {567-610},
   title = {Modelling evaporation fronts with reactive {R}iemann solvers},
   volume = {205},
   year = {2005},
}

@book{brennen1995, 
    place={New York}, 
    title={Cavitation and Bubble Dynamics}, 
    ISBN={0195094093}, 
    publisher={Oxford University Press}, 
    author={Brennen, Christopher Earls}, 
    year={1995}, 
    month={Jan} 
}

@article{charalampopoulos21,
  author = {Charalampopoulos, A. and Bryngelson, S. H. and Colonius, T. and Sapsis, T. P.},
  title = {Hybrid quadrature moment method for accurate and stable representation of non-{G}aussian processes and their dynamics},
  journal = {Philosophical Transactions of the Royal Society A},
  year = {2022},
  volume = {380},
  number = {2229},
}

@article{bryngelson_levin26,
  author = {Bryngelson, S. H.},
  title = {Fast integration method for averaging polydisperse bubble population dynamics},
  year = {2026},
  journal = {Computers \& Fluids},
  volume = {304},
  pages = {106877},
}

@article{gresho1990,
  title={On the theory of semi-implicit projection methods for viscous incompressible flow and its implementation via a finite element method that also introduces a nearly consistent mass matrix. {Part 2: I}mplementation},
  author={Gresho, Philip M and Chan, Stevens T},
  journal={International Journal for Numerical Methods in Fluids},
  volume={11},
  number={5},
  pages={621--659},
  year={1990},
}

@article{toth2000,
  title={The $\nabla \cdot B = 0$ constraint in shock-capturing magnetohydrodynamics codes},
  author={T{\'o}th, G{\'a}bor},
  journal={Journal of Computational Physics},
  volume={161},
  number={2},
  pages={605--652},
  year={2000},
  publisher={Elsevier}
}

@manual{OpenMP,
  title        = {{OpenMP} Application Programming Interface Version 5.2},
  author       = {{OpenMP Architecture Review Board}},
  year         = 2021,
  note         = {\url{https://www.openmp.org/wp-content/uploads/OpenMP-API-Specification-5-2.pdf}},
  organization = {{OpenMP Architecture Review Board}},
}

@inproceedings{wilfongGB25,
    author = {Wilfong, Benjamin and Radhakrishnan, Anand and Le Berre, Henry and Vickers, Daniel and Prathi, Tanush and Tselepidis, Nikolaos and Dorschner, Benedikt and Budiardja, Reuben and Cornille, Brian and Abbott, Stephen and Sch\"{a}fer, Florian and Bryngelson, Spencer},
    title = {Simulating many-engine spacecraft: {E}xceeding 1 quadrillion degrees of freedom via information geometric regularization},
    year = {2025},
    publisher = {Association for Computing Machinery},
    address = {New York, NY, USA},
    booktitle = {Proceedings of the International Conference for High Performance Computing, Networking, Storage and Analysis},
    pages = {14-24},
    series = {SC '25}
}

@inproceedings{wilfong-hpctests,
  title = {Testing and benchmarking emerging supercomputers via the {MFC} flow solver},
  author = {Wilfong, B. and Radhakrishnan, A. and {Le Berre}, H. A. and Prathi, T. and Abbott, S. and Bryngelson, S. H.},
  publisher = {Association for Computing Machinery},
  address = {New York, NY, USA},
  booktitle = {Proceedings of the SC '25 Workshops of the International Conference for High Performance Computing, Networking, Storage and Analysis},
  year = {2025},
  series = {SC '25}
}

\end{document}